\documentclass[showpacs,showkeys,preprintnumbers,prd,amsmath,amssymb,floats,floatfix,nofootinbib]{revtex4}
\usepackage{amsmath,amssymb,bm,natbib}
\usepackage[utf8]{inputenc}
\usepackage[colorlinks,citecolor=blue,urlcolor=blue,linkcolor=blue]{hyperref}
\usepackage[paperwidth=210mm, paperheight=297mm,centering,hmargin=3cm,vmargin=3.5cm]{geometry}
\usepackage{epstopdf}
\usepackage{lineno}
\usepackage{color}
\usepackage{booktabs}
\usepackage{graphicx}
\usepackage{multirow}
\usepackage{slashed}
\usepackage{comment}
\usepackage{float}
\makeatletter

\newcommand{\Rmnum}[1]{\expandafter\@slowromancap\romannumeral #1@}

\makeatother
\begin{document}
\def\bib{\bibitem}
\def\be{\begin{equation}}
\def\ee{\end{equation}}
\def\beq{\begin{equation}}
\def\eeq{\end{equation}}
\def\beqar{\begin{eqnarray}}
\def\eeqar{\end{eqnarray}}
\def\barr{\begin{array}}
\def\earr{\end{array}}
\def\dis{\displaystyle}
\def\lsim{\:\raisebox{-0.5ex}{$\stackrel{\textstyle<}{\sim}$}\:}
\def\gsim{\:\raisebox{-0.5ex}{$\stackrel{\textstyle>}{\sim}$}\:}
\def\tilh{\tilde{h}}
\def\and{\qquad {\rm and } \qquad}
\def\vev{\small \em {\it v.e.v. }}
\def\p{\partial}
\def\ga{\gamma^\mu}
\def\slp{p \hspace{-1ex}/}
\def\sleps{ \epsilon \hspace{-1ex}/}
\def\slk{k \hspace{-1ex}/}
\def\slq{q \hspace{-1ex}/\:}
\def\prd#1{Phys. Rev. {\bf D#1}}
\def\etal{ {\it et al.} }
\def\ie{ {\it i.e.} }
\def\eg{ {\it e.g.} }

\title{Measuring Higgs self couplings in the presence of VVH and VVHH at the ILC}

\author{\bf Satendra Kumar\textsuperscript{a, b}\footnote{satendrak@iiitu.ac.in}, \bf Poulose Poulose \textsuperscript{b}\footnote{poulose@iitg.ernet.in}, \bf Rafiqul Rahaman\textsuperscript{c}\footnote{rr13rs033@iiserkol.ac.in}, \bf Ritesh K. Singh\textsuperscript{d}\footnote{ritesh.singh@iiserkol.ac.in}}
% The "\note" macro will give a warning: "Ignoring empty anchor..."

\affiliation{\textsuperscript{a}
Indian Institute of Information Technology Una, 
Una, Himachal Pradesh 177 209, India}

\affiliation{\textsuperscript{b}
Indian Institute of Technology Guwahati,
Guwahati 781 039, India}

\affiliation{\textsuperscript{c, d}
IISER Kolkata, Mohanpur, West Bengal-741246, India}

\begin{abstract} 
The recent discovery of a Higgs boson at the LHC, while establishing the Higgs mechanism as the way  of electroweak symmetry breaking, started an era of precision measurements  involving the Higgs boson. In an effective Lagrangian framework, we consider the $e^+e^-\rightarrow ZHH$ process at the ILC running at a centre of mass energy of 500 GeV to investigate the effect of the $ZZH$ and $ZZHH$ couplings on the sensitivity of $HHH$ coupling in this process. Our results show that  the sensitivity of the trilinear Higgs self couplings on this process has somewhat strong dependence on the Higgs-gauge boson couplings. Single and two parameter reach of the ILC with an integrated luminosity of 1000 fb$^{-1}$ are obtained on all the effective couplings  indicating how these limits are affected by the presence of anomalous $ZZH$ and $ZZHH$ couplings. The kinematic distributions studied to understand the effect of the anomalous couplings, again, show a  strong influence of $Z$-$H$ couplings on the dependence of these distributions on $HHH$ coupling. Similar results are indicated in the case of the process, $e^+e^-\rightarrow \nu\bar \nu HH$, considered at a centre of mass energy of 2 TeV, where the cross section is large enough.  The effect of $WWH$ and $WWHH$ couplings on the sensitivity of $HHH$ coupling is clearly established through our analyses in  this process.
\end{abstract} 
\pacs{12.15.-y, 14.70.Fm, 13.66.Fg}
\keywords{electron positron collisions, anomalous couplings, effective Higgs Lagrangian}

%\begin{document}
\maketitle
\section{Introduction}\label{sec:introduction}
With the discovery of the new resonance of mass around 125 GeV at the LHC \cite{Chatrchyan:2012xdj,Aad:2012tfa,ATLAS:2013xla,ATLAS:2013nma,ATLAS:2013vla,CMS:xwa,CMS:bxa,Aad:2013wqa,Chatrchyan:2012xdj,Chatrchyan:2013lba,Aad:2012tfa,ATLAS:2016oum,ATLAS:2016pkl,ATLAS:2016zzs,Aad:2015vsa,ATLAS:2016hru}, a new era is open in the investigations of elementary particle dynamics. The new particle is so far consistent in every way with the long expected Higgs boson of the Standard Model (SM). All the expected SM decays are observed at the LHC, albeit some tension in the decay widths, which are still consistent within the statistical fluctuations. The spin and parity analysis favour a spin-zero, even-parity object \cite{ATLAS:2013xla,ATLAS:2013nma,ATLAS:2013vla,CMS:xwa,CMS:bxa,Aad:2015mxa}.
These measurements establish the role of Higgs mechanism in the electroweak symmetry breaking (EWSB) and the discovered resonance as the Higgs boson. So far the measurements are consistent with the properties of the Higgs boson expected in the SM Higgs mechanism. While we wait for further statistics to establish more detailed identity of this discovery, it is worth revisiting the role of new physics in the Higgs sector in the light of the new measurements. It is well accepted that, even if all the properties of the new particle meets the expectations of the SM, there still remain several questions on the SM. One of the serious issues within the Higgs sector is the difficulty with quadratically diverging quantum corrections to the mass of the Higgs boson, or the so called hierarchy problem. This itself should convince us that the SM is at the most  effective theory, highly successful at the electroweak scale. Another contentious issue is the stability of the vacuum. With the quantum corrections to the quartic Higgs coupling that the massive fermions and gauge bosons would induce, it is indicated that the minimum of the potential corresponds to a local metastable vacuum, with a possible global vacuum at much higher energies \cite{Degrassi:2012ry}.
Among the plethora of suggestions to look beyond the SM, one could indeed narrow down to scenarios that can accommodate a light Higgs boson, very likely an elementary one, with properties very close to that of the SM Higgs boson. One may need to wait till the LHC reveals further indications of new physics, if we are lucky, or perhaps even need to wait till the new generation lepton colliders, like the International Linear Collider (ILC) \cite{BrauJames:2007aa, Djouadi:2007ik,MoortgatPick:2005cw}, start exploring the TeV scale physics. Being a discovery machine, the LHC is capable of observing any direct production of new particle resonance at the energy scales explored, while the latter is more suitable to explore the new physics through detailed precision analysis, in the absence of any such direct observation of new physics.

Taking a cue from the observations so far, we are compelled to consider a case with new physics somewhat decoupled from the electroweak physics, which in turn is dictated by the SM. In that case, the effect of new physics will be reflected in the various couplings through the quantum corrections they acquire. The best way to study such effects is through an effective Lagrangian, which encodes the new physics effects in higher dimensional operators with anomalous couplings. Interesting phenomenological studied with effective Higgs couplings, including the possibility of CP violation in the Higgs sector is discussed in the literature \footnote{An important issue of the top quark Yukawa coupling in the context of CP-mixed Higgs boson is studies in Refs. \cite{Ananthanarayan:2013cia, Ananthanarayan:2014eea, Muhlleitner:2012jy, Godbole:2011hw}}.
The study of Higgs sector through an effective Lagrangian, and effective couplings goes back to Refs.\cite{Weinberg:1978kz,Weinberg:1980wa,Georgi:1994qn,Buchmuller:1985jz,Hagiwara:1993ck,Hagiwara:1993qt,Alam:1997nk,
Barger:2003rs,Giudice:2007fh,Grzadkowski:2010es,Contino:2010rs,Grzadkowski:2010es,Grzadkowski:2010es,
GutierrezRodriguez:2011gi,GutierrezRodriguez:2009uz,GutierrezRodriguez:2005fe,Rindani:2010pi,Rindani:2009pb}.
 More recently,  the Lagrangian including complete set of dimension-6 operators was studied in
Refs. \cite{Baak:2012kk, Einhorn:2013kja,Contino:2013kra,Amar:2014fpa,Masso:2014xra,Biekoetter:2014jwa,Willenbrock:2014bja}. For some of the recent reference discussing the constraints on the anomalous couplings within different approches, please see
\cite{Bonnet:2011yx,Corbett:2012dm,Chang:2013cia,Elias-Miro:2013mua,Banerjee:2013apa,Boos:2013mqa,Masso:2012eq,Han:2004az,Corbett:2012ja,Dumont:2013wma,Pomarol:2013zra,Ellis:2014dva,Belusca-Maito:2014dpa,Gupta:2014rxa,Ellis:2017kfi,Denizli:2017pyu,DiVita:2017vrr,Ellis:2018gqa,Liu:2018peg,Rindani:2018ubx,Hesari:2018ssq}.
Ref. \cite{Ellis:2014dva} studied the H+V, where V= Z, W, associated production at the LHC and the Tevatron to discuss the bounds obtainable from the global fit to the presently available data, whereas Refs.~\cite{Belusca-Maito:2014dpa} and  \cite{Ellis:2018gqa}  discuss the constraint on the parameters coming from the LHC results as well as other precision data from LEP, SLC and the Tevatron.
Experimental studies on the Higgs couplings at the LHC are presented in, for example, \cite{ Aad:2013wqa, Teyssier:2014hta, ATLAS:2018otd, Khachatryan:2016vau, CMS:2013xfa}.

Higgs self couplings give direct information about the scalar potential, and therefore, very important to understand the nature of the EWSB. The process, $e^+ e^-\rightarrow ZHH$ is one of the best suited to study the Higgs  trilinear coupling \cite{DeRujula:1991ufe,GutierrezRodriguez:2008nk,Takubo:2009ws,Tian:2010np,Battaglia:2001nn,Barger:2003rs,Killick:2013mya,Djouadi:1999gv,Baer:2013cma,Castanier:2001sf,Liu:2018peg}.  At the same time, this process also depends on the Higgs-Gauge boson couplings, $ZZH$ and $ZZHH$, which will affect the determination of the the $HHH$ coupling. Another process that could probe the $HHH$ couplings is $e^+e^- \rightarrow \nu \bar{\nu} HH$ following the WW fusion \cite{Battaglia:2001nn,Barger:2003rs,Killick:2013mya,Djouadi:1999gv,Rindani:2018ubx}, which is also affected by the $WWH$ and $WWHH$ couplings. With the first phase of the ILC expected to run at 250 GeV centre of mass energy, efforts to probe the $ZH$ production process, where the influence of the quartic couplings would be felt at one loop, are seriously studied \cite{Rindani:2018ubx, DiVita:2017vrr}.
In this report we will focus on the Higgs pair production processes within the framework of the effective Lagrangian. One goal of this study is to investigate how significant is the effect of $VVH$ coupling, where $V=Z, ~W$, in the extraction of  the $HHH$ coupling. In contrast to other similar studies, we perform a multi-parameter analysis employing the Markov-Chain-Monte-Carlo method to obtain simultaneous constraints in hyperspace spanned by all relevant parameters.

The report is presented in the following way. In Section~\ref{sec:setup} the effective Lagrangian will be presented, with the currently available constraint on the parameters. In Section~\ref{sec:discussions} the processes under consideration will be presented, with details. In Section~\ref{sec:summary} the results will be summarized.

\section{General Setup}\label{sec:setup}
The effective Lagrangian with full set of dimension-6 operator involving the Higgs bosons is described in Ref. ~\cite{Giudice:2007fh,Grzadkowski:2010es,Contino:2010rs,Grzadkowski:2010es,Contino:2013kra,Ellis:2014dva}. In this report, we shall restrict our discussion to the processes $e^+e^- \rightarrow ZHH$, and $e^+e^- \rightarrow \nu\bar{\nu}WW\rightarrow \nu\bar{\nu}HH$. Relevant to these processes, part of the Lagrangian is given by
\begin{eqnarray}
{\cal L}_{\rm Higgs}^{\rm anom} &=&    \frac{\bar c_{H}}{2 v^2} \partial^\mu\big(\Phi^\dag \Phi\big) \partial_\mu \big( \Phi^\dagger \Phi \big) +  \frac{\bar c_6}{v^2}\lambda~\big(\Phi^\dag\Phi\big)^3+\frac{\bar c_{\gamma}}{m_W^2} g'^2~\Phi^\dag \Phi B_{\mu\nu} B^{\mu\nu}
   + \frac{\bar  c_{g}}{m_W^2}g_s^2 ~\Phi^\dag \Phi G_{\mu\nu}^a G_a^{\mu\nu} \nonumber\\
  &&+ \frac{\bar c_{HW}}{m_W^2} ig~\big(D^\mu \Phi^\dag \sigma_{k} D^\nu \Phi\big) W_{\mu \nu}^k 
  + \frac{\bar c_{HB}}{m_W^2} ig'~ \big(D^\mu \Phi^\dag D^\nu \Phi\big) B_{\mu \nu} \nonumber\ \\
  && + \frac{\bar c_{W}}{2m_W^2} ig~\big( \Phi^\dag \sigma_{k} \overleftrightarrow{D}^\mu \Phi \big)  D^\nu  W_{\mu \nu}^k 
  +\frac{\bar c_{B}}{2 m_W^2} ig'~\big(\Phi^\dag \overleftrightarrow{D}^\mu \Phi \big) \partial^\nu  B_{\mu \nu},%  \nonumber\ \\
%  & + & \frac{\bar c_{t}}{v^2} y_t    \Phi^\dag \Phi\ \Phi^\dag\cdot{\bar Q}_L t_R + \frac{\bar c_{b}}{v^2} y_b     %\Phi^\dag \Phi\ \Phi \cdot {\bar Q}_L b_R + \frac{\bar c_{\tau}}{v^2} y_\tau\ \Phi^\dag \Phi\ \Phi \cdot %{\bar L}_L \tau_R 	\, 
\label{eq:Leff}
\end{eqnarray}
where
\(
  \Phi^\dag {\overleftrightarrow D}_\mu \Phi = 
    \Phi^\dag D^\mu \Phi - D_\mu\Phi^\dag \Phi \ ,
\) $D^\mu$ being the appropriate covariant derivative operator, and $\Phi$, the usual Higgs doublet in the SM. Also, $G_{\mu\nu}^a$, $W_{\mu\nu}^k$ and $B_{\mu\nu}$ are the field tensors corresponding to the $SU(3)_C$, $SU(2)_L$ and $U(1)_Y$ of the SM gauge groups, respectively, with gauge couplings $g_s$, $g$ and $g'$, in that order. $\sigma_k$ are the Pauli matrices, and $\lambda$ is the usual (SM) quadratic coupling constant of the Higgs field.
The above Lagrangian leads to the following in the unitary gauge and mass basis
~\cite{Alloul:2013naa}
\begin{eqnarray}
  {\cal L}_{H,Z,W} ^{\rm anom}=&& -v \lambda g_{HHH}^{(1)} H^3 + \frac{1}{2} g_{HHH}^{(2)} H \partial_{\mu} H \partial^{\mu} H -\frac{1}{4} g_{HZZ}^{(1)} Z_{\mu\nu} Z^{\mu\nu} H-
 \frac{1}{4} g_{HZZ}^{(2)} Z_{\nu}\partial_{\mu} Z^{\mu\nu} H \nonumber\\
 &&+\frac{1}{2} g_{HZZ}^{(3)} Z_{\mu} Z^{\mu} H -\frac{1}{2} g_{HAZ}^{(1)} Z_{\mu\nu} F^{\mu\nu} H-g_{HAZ}^{(2)}Z_{\nu} \partial_{\mu} F^{\mu\nu} H \nonumber\\ 
 && - \frac{1}{8} g_{HHZZ}^{(1)} Z_{\mu\nu} Z^{\mu\nu} H^2 - \frac{1}{2} g_{HHZZ}^{(2)}Z_{\nu} \partial_{\mu} Z^{\mu\nu} H^2 - \frac{1}{4} g_{HHZZ}^{(3)} Z_{\mu} Z^{\mu} H^2 \nonumber\\
 && - \frac{1}{2} g_{HWW}^{(1)} W^{\mu\nu} W_{\mu\nu}^{\dagger} H -\left [g_{HWW}^{(2)} W^{\nu} \partial^{\mu} W_{\mu\nu}^{\dagger} H + h.c.\right] + g~m_{W} W_{\mu}^{\dagger} W^{\mu} H \nonumber\\ 
 && - \frac{1}{4} g_{HHWW}^{(1)} W^{\mu\nu} W_{\mu\nu}^{\dagger} H^2 - \frac{1}{2} \left[g_{HHWW}^{(2)} W^{\nu} \partial^{\mu} W_{\mu\nu}^{\dagger} H^2 + h.c.\right] + \frac{1}{4} g^2 W_{\mu}^{\dagger} W^{\mu} H^2 
 \nonumber\\
\label{eq:LagPhys}
\end{eqnarray}

Various physical couplings present in the Lagrangian in Eq.~(\ref{eq:LagPhys}) are given in terms of the parameters of the effective Lagrangian in Eq.~(\ref{eq:Leff}) as

\begin{eqnarray}
&& g_{HHH}^{(1)} = 1+ \frac{5}{2} \bar{c_6},\hspace{1cm} g_{HHH}^{(2)} = \frac{g}{m_W} \bar{c}_H \nonumber\\
&& g_{HZZ}^{(1)}  = \frac{2g}{c_W^2 m_W} \left[ \bar{c}_{HB} s_W^2 - 4\bar{c}_{\gamma} s_W^4 + c_W^2 \bar{c}_{HW}  \right]\nonumber\\  
&& g_{HZZ}^{(2)}  = \frac{g}{c_W^2 m_W} \left[ (\bar{c}_{HW}+ \bar{c}_W) c_W^2 + (\bar{c}_B + \bar{c}_{HB}) s_W^2  \right],~~~~~g_{HZZ}^{(3)}  = \frac{g m_Z}{c_W} \left[ 1-2 \bar{c}_T  \right] \nonumber\\
&& g_{HAZ}^{(1)}  = \frac{g s_W}{c_W m_W} \left[ \bar{c}_{HW} -\bar{c}_{HB} + 8\bar{c}_{\gamma} s_W^2   \right]\nonumber\\ 
&& g_{HAZ}^{(2)}  = \frac{g s_W}{c_W m_W} \left[ \bar{c}_{HW} -\bar{c}_{HB} -\bar{c}_B + \bar{c}_W   \right] \nonumber\\
%\end{eqnarray}
%\end{minipage}
%\begin{minipage}{2.5in}
%\begin{eqnarray}
&&g_{HHZZ}^{(1)}  = \frac{g^2}{c_W^2 m_W^2} \left[ \bar{c}_{HB} s_W^2 - 4\bar{c}_{\gamma} s_W^4 + \bar{c}_{HW} c_W^2   \right]\nonumber\\
&&g_{HHZZ}^{(2)}  = \frac{g^2}{2 c_W^2 m_W^2} \left[ (\bar{c}_{HW} + \bar{c}_{W}) c_W^2 + (\bar{c}_{B} + \bar{c}_{HB}) s_W^2  \right],~~~~~   g_{HHZZ}^{(3)}  = \frac{g^2}{2 c_W^2} \left[1 - 6 \bar{c}_{T} \right]\nonumber\\
&&g_{HWW}^{(1)}  = \frac{2g}{m_W} \bar{c}_{HW}, ~~~~~~~~ g_{HWW}^{(2)}  = \frac{g}{2m_W} \left[\bar{c}_{W} + \bar{c}_{HW}\right] \nonumber\\
&&g_{HHWW}^{(1)}  = \frac{g^2}{m_W^2} \bar{c}_{HW}, ~~~~~~ g_{HHWW}^{(2)}  = \frac{g^2}{4 m_W^2} \left[\bar{c}_{W} + \bar{c}_{HW}\right] 
\end{eqnarray}

In total eight coefficients, namely, $\bar{c}_6, ~\bar{c}_H,~ \bar{c}_T,~ \bar{c}_{\gamma},~ \bar{c}_B, ~\bar{c}_W,~ \bar{c}_{HB},~ \bar{c}_{HW} $, govern the dyanmics of $ZHH$ and $\nu\bar\nu HH$ productions at the ILC. Coming to the experimental constraints on these parameters, the first two, $\bar{c}_6$ and $\bar{c}_H$ influence only the Higgs self couplings, and therefore, practically, do not have any experimental constraints on them. 
Electroweak precision tests constrain $\bar c_T$, $\bar c_W$ and $\bar c_B$ as ~\cite{Baak:2012kk}

\begin{eqnarray}
 &&\bar{c}_T(m_Z) \in [-1.5, 2.2] \times 10^{-3}, \nonumber \\
&&(\bar{c}_W(m_Z) + \bar{c}_B(m_Z))\in [-1.4, 1.9]\times 10^{-3}.
\end{eqnarray}

Note that,  $\bar{c}_W$ and $\bar{c}_B$ are not independently constrained, leaving a possibility of having  large values with a cancellation between them as per the above constraint. $\bar c_W$ itself, along with $\bar c_{HW}$ and $\bar c_{HB}$ is constrained from the LHC observations on the associated production of Higgs along with $W$ in Ref. ~\cite{Ellis:2014dva}.
Consideration of the Higgs associated production along with W, ATLAS and CMS along with D0 put a limit of  \( \bar{c}_W \in \big[-0.05, 0.04\big] \), when all other parameters are set to zero. A global fit using various information from ATLAS and CMS, including signal-strength information constrains the region in $\bar c_W-\bar c_{HW}$ plane, leading to a slightly more relaxed limit on $\bar c_W$, and a limit of about \( \bar{c}_{HW} \in \big[-0.1, 0.06\big] \).  The limit on $\bar{c}_{HB}$ estimated using a global fit in Ref.~\cite{Ellis:2014dva} is about \(\bar{c}_{HB} \in [-0.05, 0.05]\) with a one parameter fit.

The purpose of this study is to understand how to exploit a precision machine like the ILC to investigate suitable processes so as to derive information regarding these couplings. In the next section we shall explain the processes of interest in the present case, and discuss the details to understand the influence of one or more of the couplings mentioned above.

\section{Discussion of the processes considered}\label{sec:discussions}

\begin{figure}[h]
\begin{center}
\begin{tabular}{c c}
%\hspace{-10mm}
\includegraphics[width=50mm]{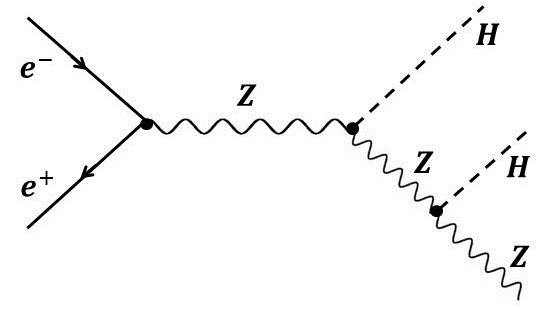}&
\hspace{18mm}
\includegraphics[width=50mm]{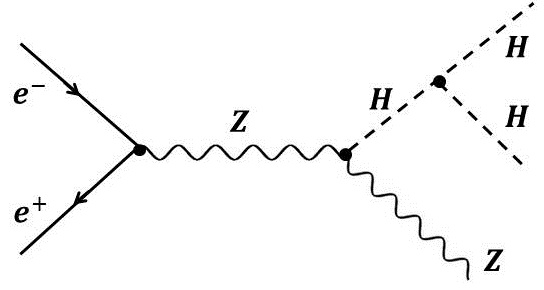}\\[5mm]
\includegraphics[width=50mm]{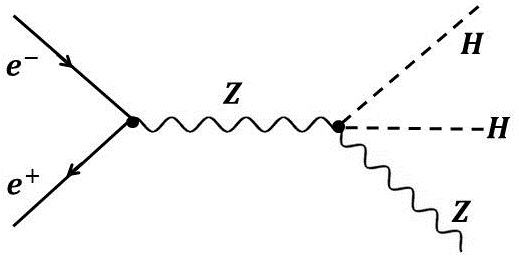}
\hspace{-80mm}
\end{tabular}
\end{center}
\caption{Feynman diagrams contributing to  the process $e^{-}e^+ \rightarrow ZHH$ in Standard Model. }
\label{fig:fdzhh}
\end{figure}

It is generally expected that the ILC, with its clean environment, fixed centre of mass energy, and additional features like availability of beam polarization, will be able to do the precision studies much more efficiently than what the LHC could do.  This is especially so in the case of Higgs self couplings. One of the best suited process to study the trilinear (self) coupling of the Higgs boson is $e^+e^- \rightarrow ZHH$, the phenomenological analysis of which is studied in detail within the context of the SM. The Feynman diagrams corresponding to this process in the SM are given in  Fig.~\ref{fig:fdzhh}.

\begin{figure}[h]
\begin{center}
\begin{tabular}{c c}
%\hspace{5mm}
\includegraphics[width=53mm]{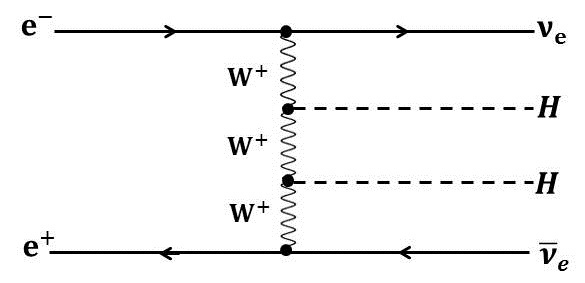}&
\hspace{15mm}
\includegraphics[angle=0,width=53mm]{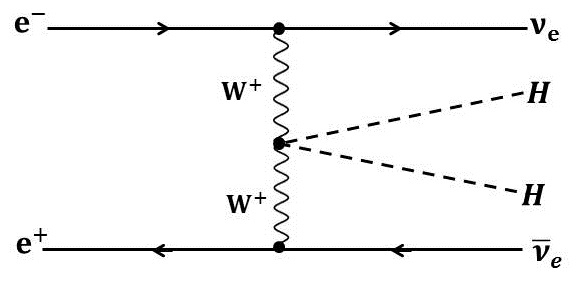} \\[5mm] 
\includegraphics[width=53mm]{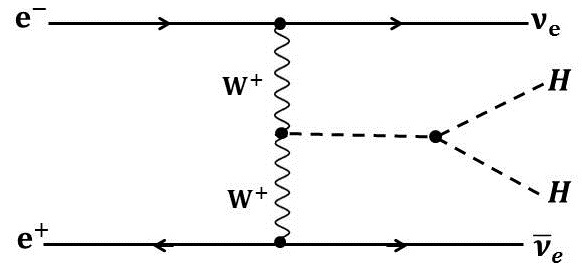}
\hspace{-80mm}
\end{tabular}
\end{center}
\caption{Feynman diagrams contributing to  the process $e^{-}e^+ \rightarrow  \nu\bar\nu HH$ in Standard Model, without considering $e^+e^-\rightarrow ZHH \rightarrow \nu\bar\nu HH $}
\label{fig:fdnnhh}
\end{figure}

Another process that is relevant to the study of $HHH$ coupling is $e^+e^- \rightarrow \nu_e \bar \nu_e HH$. The earlier process, $e^+e^- \rightarrow ZHH$, with the invisible decay of $Z\rightarrow \nu_e\bar\nu_e$ also leads to the same final state. However, this can be easily reduced by considering the missing invariant mass. The rest of the process goes through the Feynman diagrams presented in Fig.~\ref{fig:fdnnhh}.

Apart from the $HHH$ coupling, these processes are influenced by gauge-Higgs couplings like $ZZH$, $ZZHH$, $WWH$ and $WWHH$. Keeping in mind the above discussion of the effective couplings deviating from the SM due to the influence of the BSM at some higher energies, one must understand how such a scenario would affect the phenomenology, in order to draw any conclusion regarding these couplings. In the rest of this report, we shall revisit these processes, with a specific purpose of understanding the correlation between the  gauge-Higgs coupling and the trilinear Higgs couplings. 

For our analyses we use \texttt{MadGraph5}~\cite{Alwall:2014hca}, with the Effective Lagrangian implemented through \texttt{FeynRules}~\cite{Alloul:2013bka} as given by~\cite{Alloul:2013naa}.

\subsection{$e^+e^-\rightarrow ZHH$ Process}

\begin{figure}[h] \centering
\begin{tabular}{c c}
\hspace{-5mm}
\includegraphics[angle=0,width=100mm]{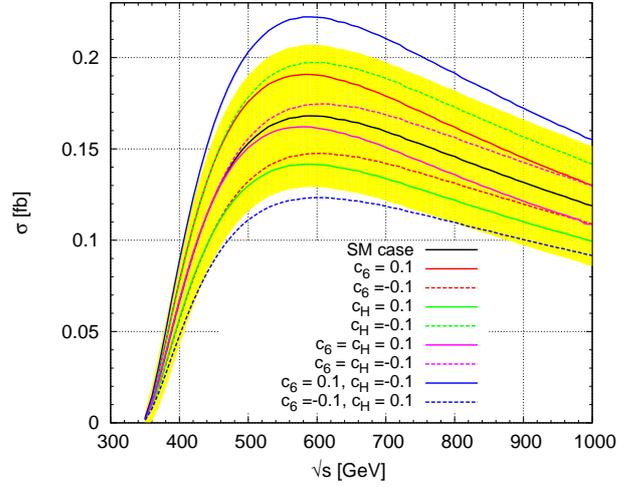} 
\end{tabular}
\caption{Cross section against $\sqrt{s}$ for  the process $e^{-}e^+ \rightarrow ZHH$, for different values of the parameters $c_6$ and $c_H$, with all others kept to zero. }
\label{fig:cs_roots_zhh}
\end{figure}

We shall first consider $e^+e^- \rightarrow ZHH$ process. In Fig.~\ref{fig:cs_roots_zhh} the cross section is plotted against the centre of mass for the SM case as well as for some selected $(c_6, c_H)$ points. The cross section peaks around a centre of mass energy of 600 GeV with  a value of about 0.17 fb, which slides down to about 0.16 fb at 500 GeV. We perform our analysis for the ILC running at a centre of mass energy of 500 GeV.  
%One may note that the cross section is small compared to many other typical electroweak processes, and one will need %to be patient to reach a very high luminosity of a few 1000 fb$^{-1}$ to make any meaningful study. This would be the %case with any process involving trilinear Higgs couplings, and therefore, the process under scruitiny is not at any %particular disadvantage because of this. 

\begin{figure}[h]\centering
\begin{tabular}{c c}
\hspace{-5mm}
\includegraphics[angle=0,width=75mm]{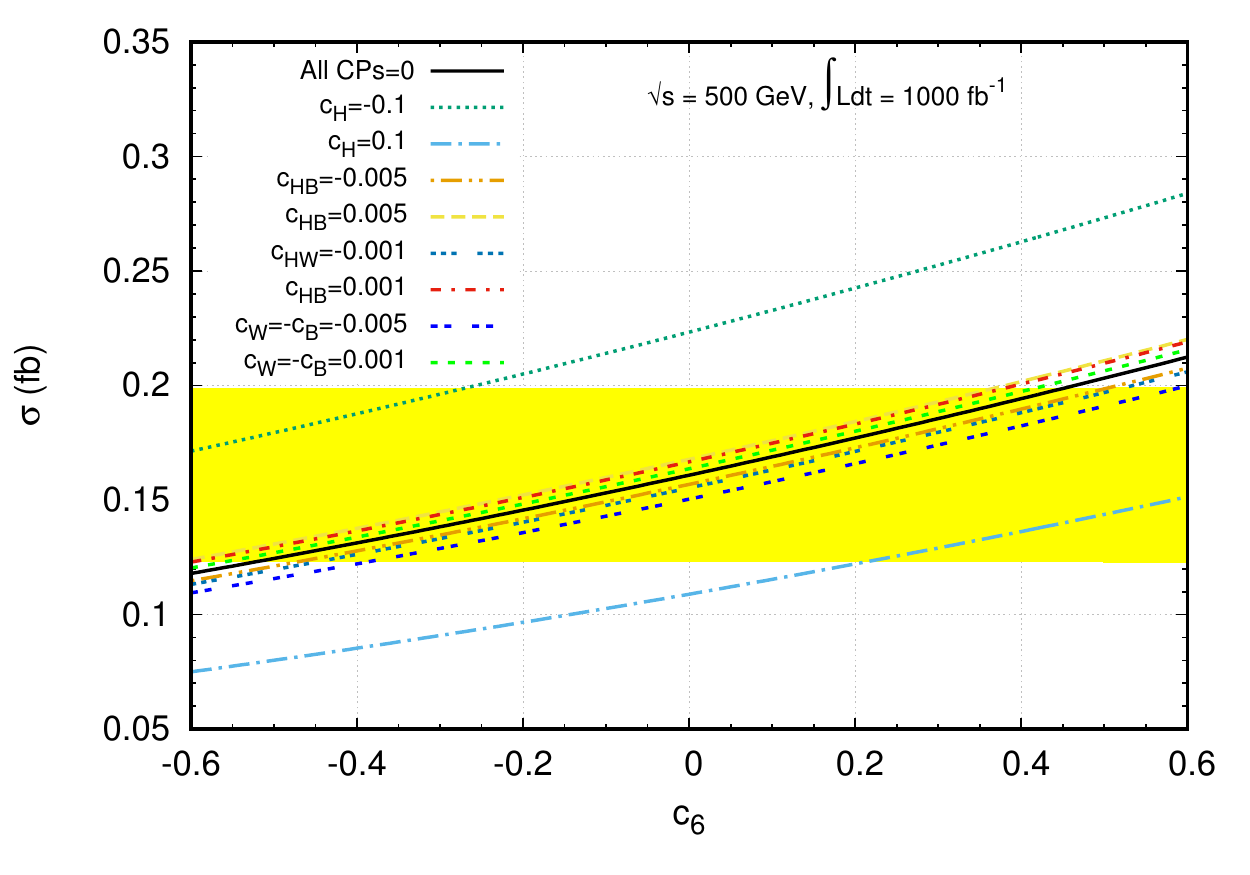}&
\includegraphics[angle=0,width=75mm]{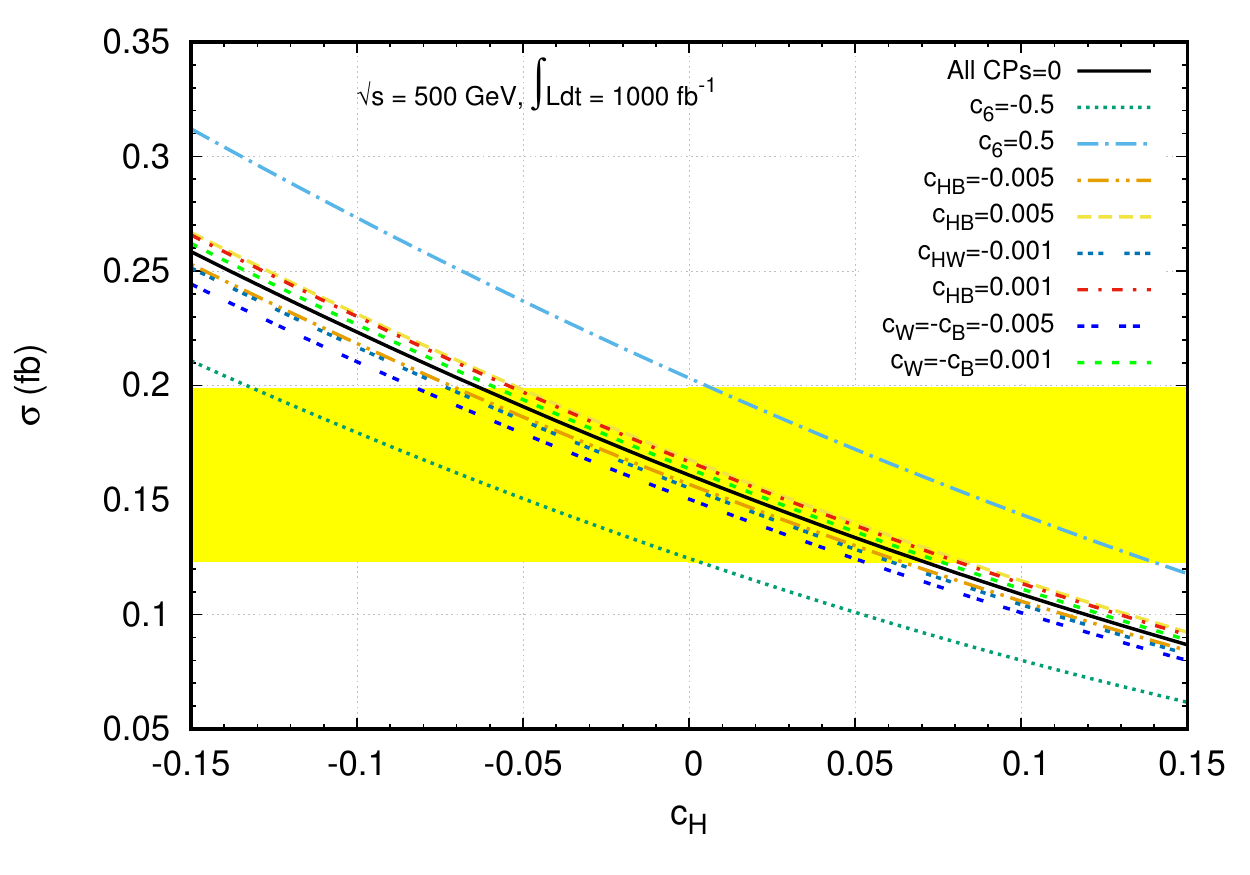}  \\
\end{tabular}
%\vspace*{-7cm}
\caption{
Cross section of $ZHH$ production against $c_6$ (left) and $c_H$ (right), when some of the other selected relevant parameters assume typical values is compared against the case when only $c_6$ or $c_H$  is present. The black solid lines corresponds to the case when all parameters other than $c_6$ (left) or $c_H$ (right) vanish. The centre of mass energy is assumed to be 
$\sqrt{s}=500$ GeV. In each case, all other parameters are set to zero. The yellow band indicates the $3\sigma$ limit 
of the SM cross section, with integrated luminosity of $1000$ fb$^{-1}$.}
\label{fig:cs_c6cH}
\end{figure}

To seek a functional form of cross section as a function of the anomalous couplings
we generated events in \texttt{MadGraph5} for the process $e^+e^-\to hhZ$  with and without  decaying the particles   
with $100$ random set of couplings $f_i=(c_6,c_H,c_W,c_B,c_{HB},c_{HW})$ within a given range  at $\sqrt{s}=500$ GeV
and fitted the cross section of the process with the formula
\begin{equation}
	\sigma=\sigma_{0} + f_i\times \sigma_i + f_i\times f_j \times \sigma_{ij}.
\end{equation}
Here $\sigma_{0}$ is the  cross section when $f_i=0$, i.e., the SM cross section. The production cross section of  $e^+e^-\to hhZ$ at $\sqrt{s}=500$ GeV as a function of the anomalous higgs couplings ($c_6,c_H,c_W,c_B,c_{HB},c_{HW}$) is obtained to be

\begin{eqnarray}\label{eq:sigma-prod}
	\sigma%(c_6,c_H,c_W,c_B,c_{HB},c_{HW})
	&=&0.160868 
	+0.0787057  \times c_6 -0.571945  \times c_H +2.64522 \times c_W -0.0588928  \times c_B  \nonumber\\
	&+& 1.09184 \times c_{HB}+5.69175 \times c_{HW} 
	+0.0119329 \times c_6^2+0.523069 \times c_H^2
	+121.355 \times c_W^2\nonumber\\&
	-&0.0636082 \times c_B^2+57.9799 \times c_{HB}^2+110.68 \times c_{HW}^2 
	-0.150628 \times c_6 \times c_H \nonumber\\&
	+&0.683334 \times c_6 \times c_W -0.000703205  \times c_6 \times c_B
	+0.266043 \times c_6 \times c_{HB}  +1.39575 \times c_6 \times c_{HW}\nonumber\\&
	-&4.89070 \times c_H \times c_W -0.00250087 \times c_H \times c_B -1.9789 \times c_H \times c_{HB} -10.1718 \times c_H \times c_{HW}\nonumber\\&
	+&0.00489508 \times c_W \times c_B+165.005 \times c_W \times c_{HB} -107.031 c_W \times c_{HW}  \nonumber\\&
	+&0.0343358 \times c_B \times c_{HB}+0.0649916 \times c_B \times c_{HW}
	-89.6786 \times c_{HB} \times c_{HW}~~~\text{fb}.
\end{eqnarray}
The total cross section including the decay of $Z\to l^+l^-$ and $h\to b\bar{b}$ would be
\begin{eqnarray}\label{eq:sigma-tot}
	\sigma_{tot}=\sigma\times Br(Z\to l^+l^-)
	\times Br(h\to b\bar{b})^2~~~\text{fb}.
\end{eqnarray}
%where $k(p_T^l)$ is the factor due $p_T$ cut of leptons. For $p_T>10$ GeV of leptons $k(p_T^l)=0.931966$.
%We have $Br(Z\to l^+l^-)=2\times 0.03363$ ($2$ is due to presence of $\mu$ and $e$) and 
%$Br(h\to b\bar{b})=0.58$.
%So the SM total cross section is $\sigma_{tot}(SM)=3.3922\times 10^{-3}$ fb.

The influence of  $c_6$ and $c_H$ on the production  cross section given in Eq.~(\ref{eq:sigma-prod})
are shown in Fig.~\ref{fig:cs_c6cH} in the {\em left-panel} and in the {\em right-panel}, respectively.  
We  compare the variation of cross section  with $c_6$ keeping all other parameters to the SM value, with the cases when some of the relevant parameters having non-standard values. The $3\sigma$ region (yellow band) of the SM value of the cross section, considering an integrated luminosity of 1000 fb$^{-1}$, is presented in these plots so as to make an estimate of the reach on the $c_6$. The plots clearly indicate the correlation between the influence of different parameters on the cross section. For example, assuming only $c_6$ takes a non-zero value, the reach at $3\sigma$ level is approximately $-0.5<c_6<0.45$, as indicated by the black solid line. However, as indicated by the red solid line, if we assume a typical value of $c_W=-c_B=-0.005$, the lower limit is considerably relaxed, with some moderate change in the upper bound to 0.5. On the other hand, for the case with $c_W=-c_B=0.001$, where the sign is reversed, the upper bound becomes more stringent, whereas the lower bound is more relaxed. A similar story can be read out for the cases with the presence of other parameters as well. The effect of all the parameters $c_W$, $ c_{HW}$ and $c_{HB}$, which contribute to the $ZZH$ and $ZZHH$ couplings are found to be significant. Strong dependence of the sensitivity of $c_6$ on the presence of $c_H$ is somewhat expected, for both parameters contribute to the $HHH$ coupling. The dependence on all the parameters on the sensitivity of $c_H$ on the cross section is also found to be significant for chosen typical values of the parameters.
\begin{figure}[h]\centering
\begin{tabular}{c}
\hspace{-8mm}
\includegraphics[angle=0,width=82mm]{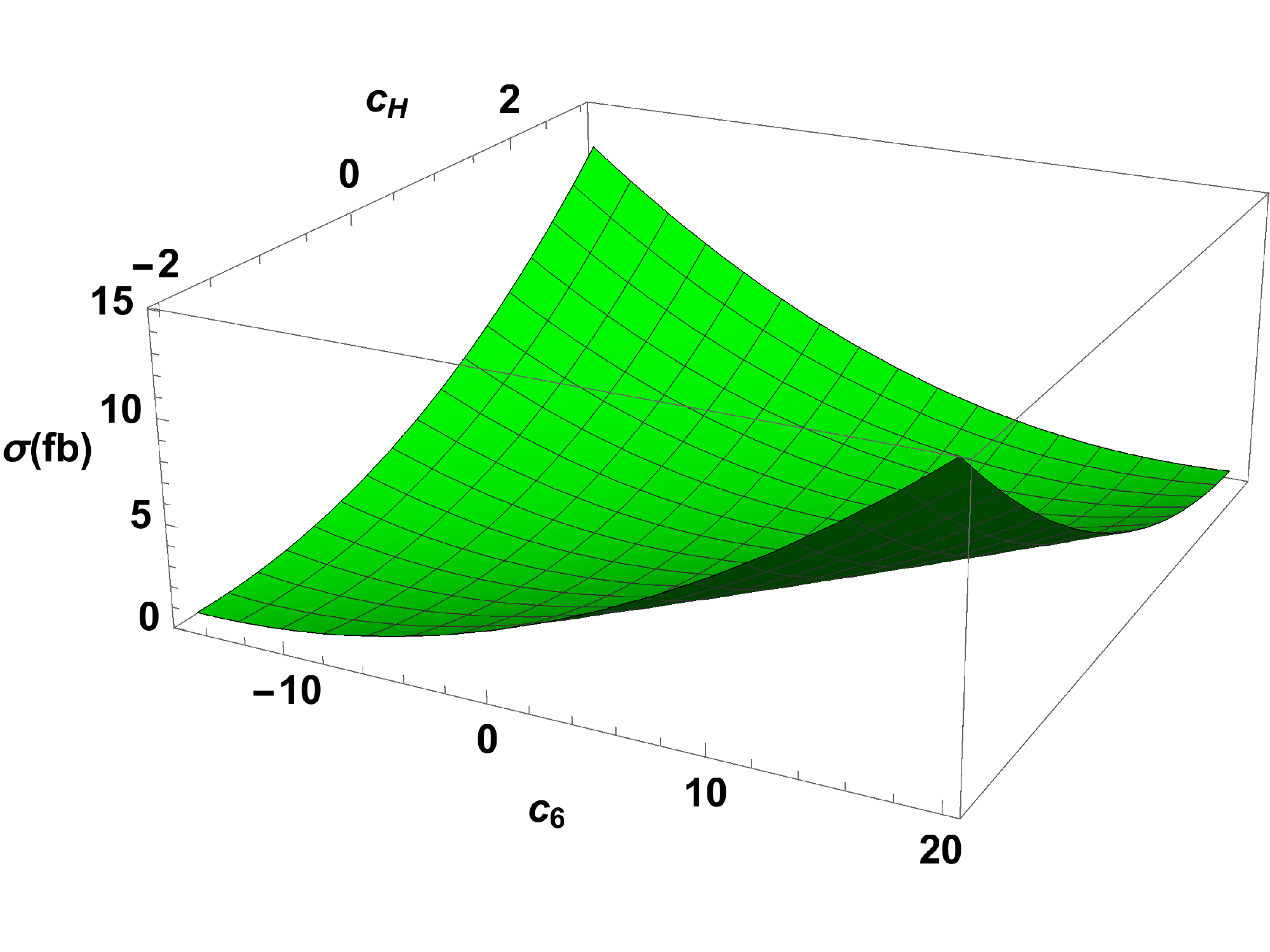}
\end{tabular}
%\vspace*{-7cm}
\caption{Cross section of $ZHH$ production plotted against $c_6$ and $c_H$ at $\sqrt{s}=500$ GeV, with all other parameters set to zero.}
\label{fig:sig_2param_c6cH}
\end{figure}
To see the simultaneous effect of $c_6$ and $c_H$,  the production cross section given in  is plotted against $c_6$ and $c_H$ 
In Fig.~\ref{fig:sig_2param_c6cH}. The correlation of the sensitivity between the two parameters is clear. The opposite sign combination seems to be more sensitive to the cross section, and therefore more stringent constraints could be drawn in this case compared to the same sign case.
\begin{figure}[h]\centering
\begin{tabular}{c c c}
\hspace{-8mm}
\includegraphics[angle=0,width=50mm]{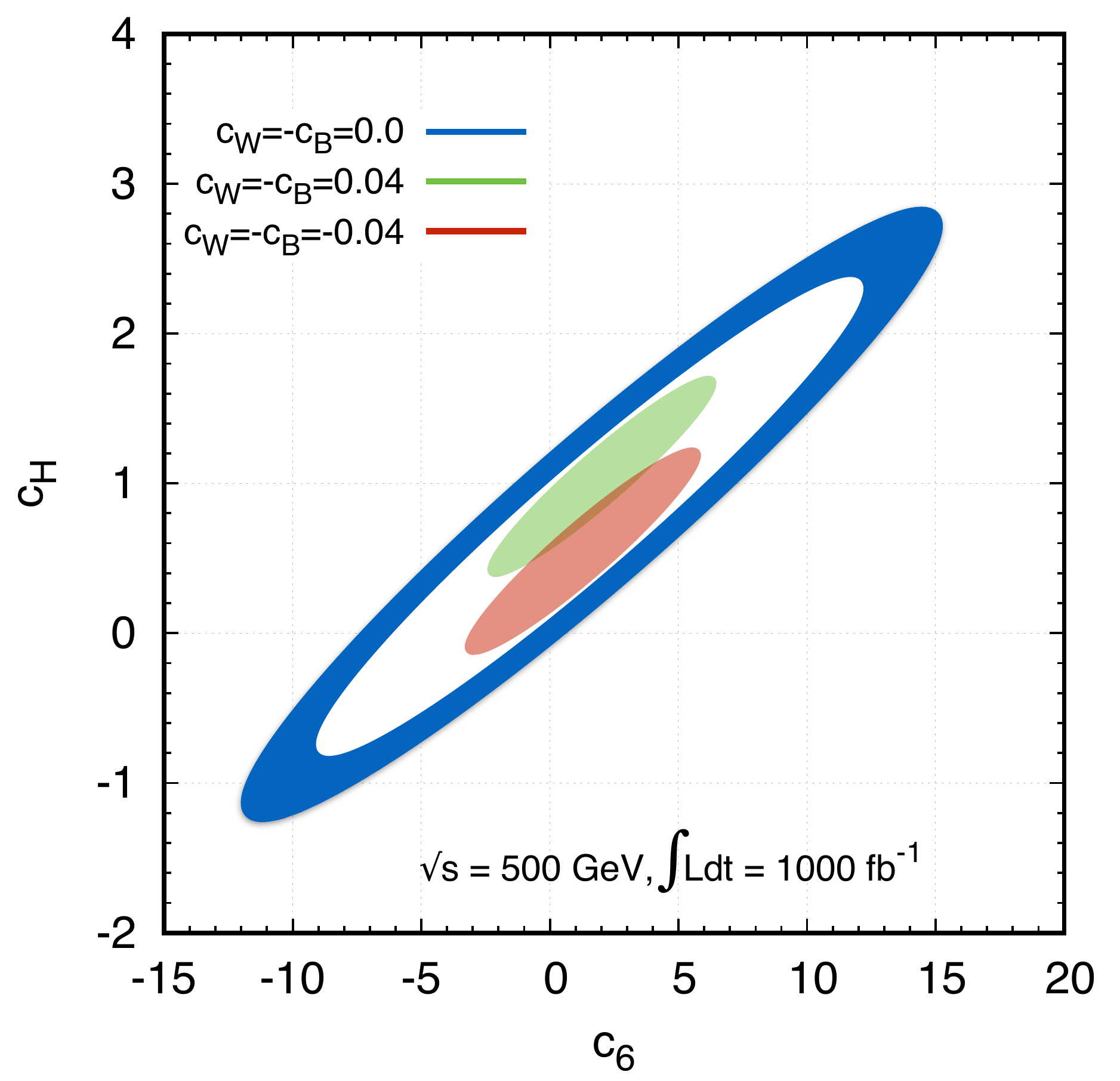} &
\includegraphics[angle=0,width=49.5mm]{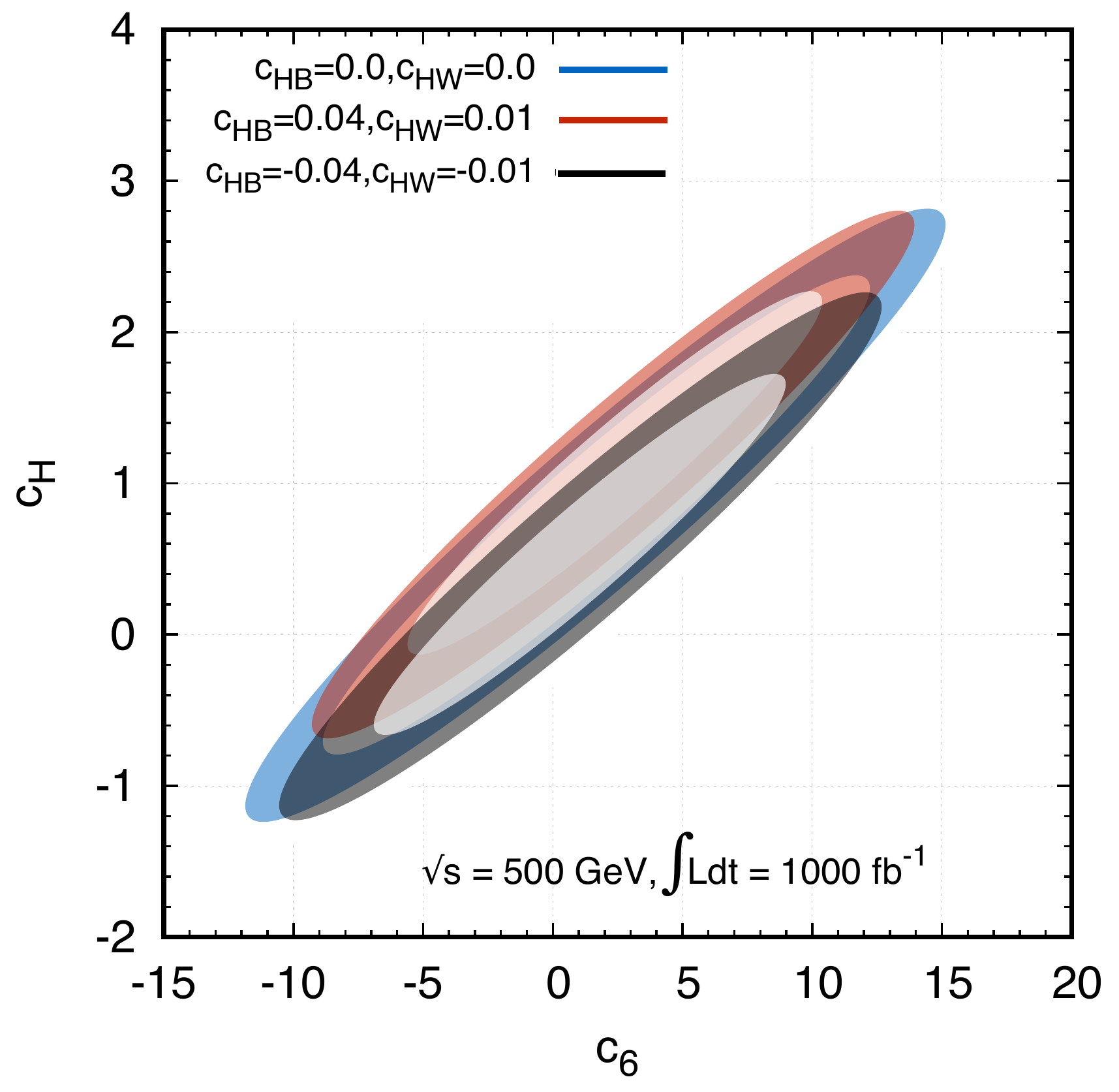} &
\includegraphics[angle=0,width=50mm]{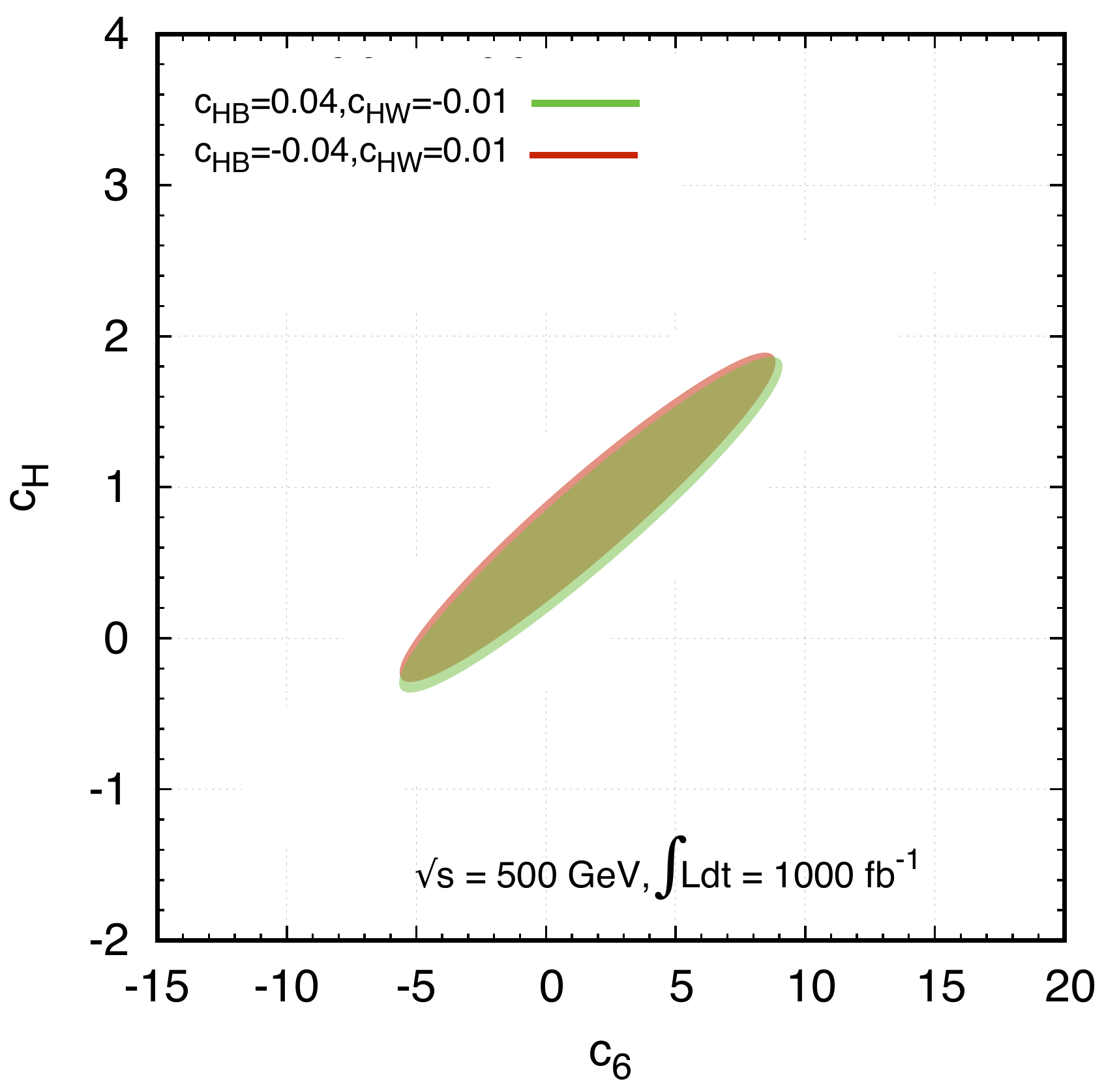}\\
\end{tabular}
%\vspace*{-7cm}
\caption{
The shaded regions correspond to regions in the $c_6$-$c_H$ plane with the total cross section is within the $3\sigma$ limit when $c_6=c_H=0$ in each case, for an integrated luminosity of $1000$ fb$^{-1}$ at a centre of mass energy of $500$ GeV. Values of the other anomalous couplings are as indicated in the figure, with all other couplings set to zero.}
\label{fig:lt_c6_cH}
\end{figure}

The reach of the ILC on the trilinear Higgs coupling through the process being considered can be established by considering the $3\sigma$ limit of the cross section at an integrated luminosity of 1000 $fb^{-1}$ as presented in Fig.~\ref{fig:lt_c6_cH}, for the case of SM, and cases with non-vanishing anomalous $ZZH$ and $ZZHH$ couplings. Please note that, when cross section is considered as a function of $c_6$ and  $c_H$,  the result is a second order polynomial with these two parameters (see Eq.~(\ref{eq:sigma-prod})). With this, the $3\sigma$ limit of the cross section leads to an elliptic equation corresponding to the relation between these two parameters. This result in an elliptic band in the $c_6 - c_H$ plane respecting the $3\sigma$ limit of the cross section. As is evident from the plots, these allowed bands of the parameters move in the parameter space, depending on the values of the other parameters, as illustrated by the cases of $c_W=-c_B$, $c_{HW}$ and $c_{HB}$. These results also illustrate how important the signs of different couplings are in a study of the sensitivity of the trilinear Higgs couplings. What we may learn from the above is  that the limits drawn with assuming the absence of all other parameters may not depict the actual situation.  %To derive limit on the couplings one needs to consider a decay channel of the process, which is discussed in the  sub-sub-section~\ref{subsect:limit}.

It is important to know the behaviour of the kinematic distributions, and how the anomalous parameters influence these, to derive any useful and reliable conclusions from the experimental results. This is so, even in cases where the fitting to obtain the reach of the parameters is done with the total number of events, as the reconstruction of events and the reduction of the background depend crucially on the kinematic distributions of the decay products. In the following, we shall present some illustrative cases of distributions at the production level, in order to understand the effect of different couplings on these. The changes in the kinematic distributions at the production level will also be carried over to the distributions of their decay products. Presently we would like to be content with the analysis at the production level, considering the limited scope of this work. As mentioned earlier we shall focus on the ILC running at a centre of mass energy of 500 GeV for our study.

\begin{figure}[h]\centering
\begin{tabular}{c c }
\hspace{-10mm}
\vspace*{0.1cm}
\includegraphics[angle=0,width=70mm]{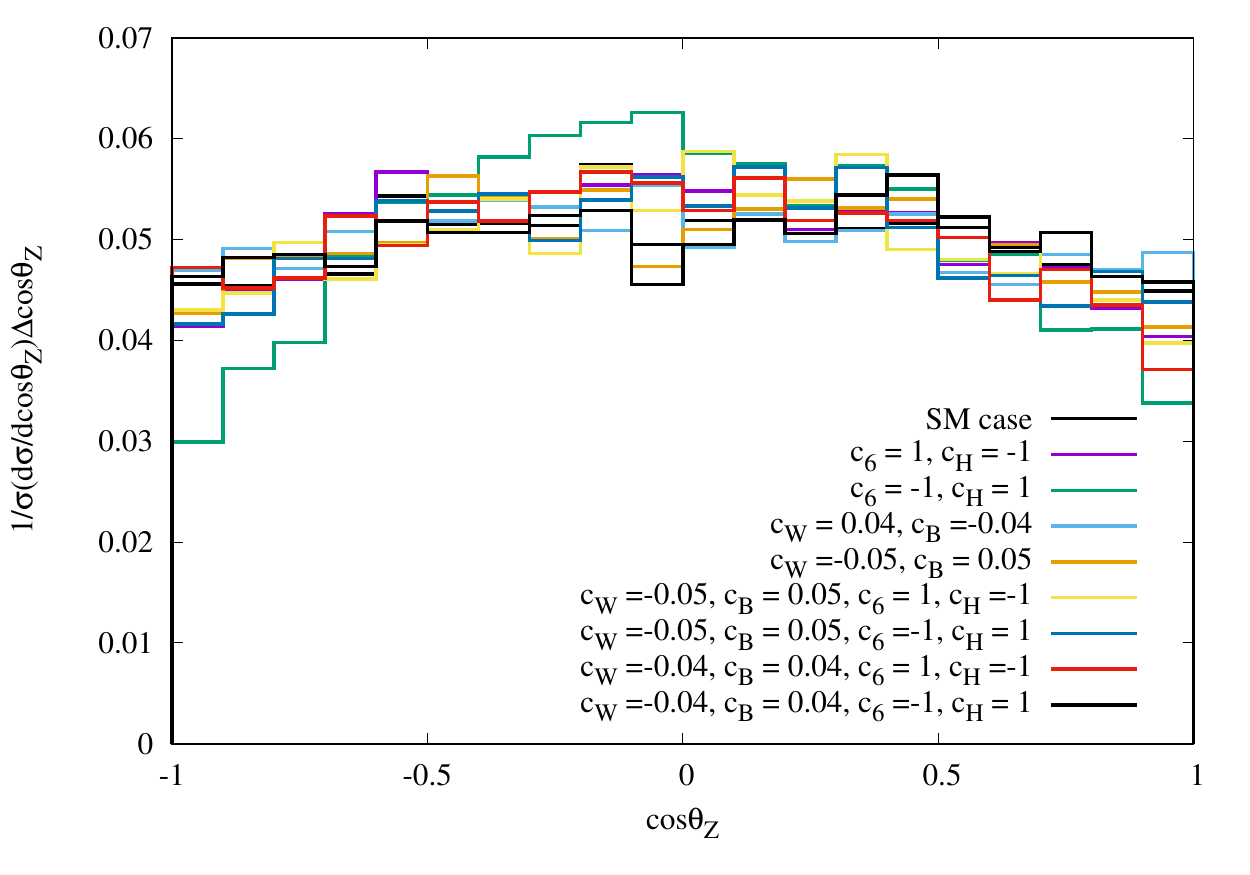} &
\includegraphics[angle=0,width=70mm]{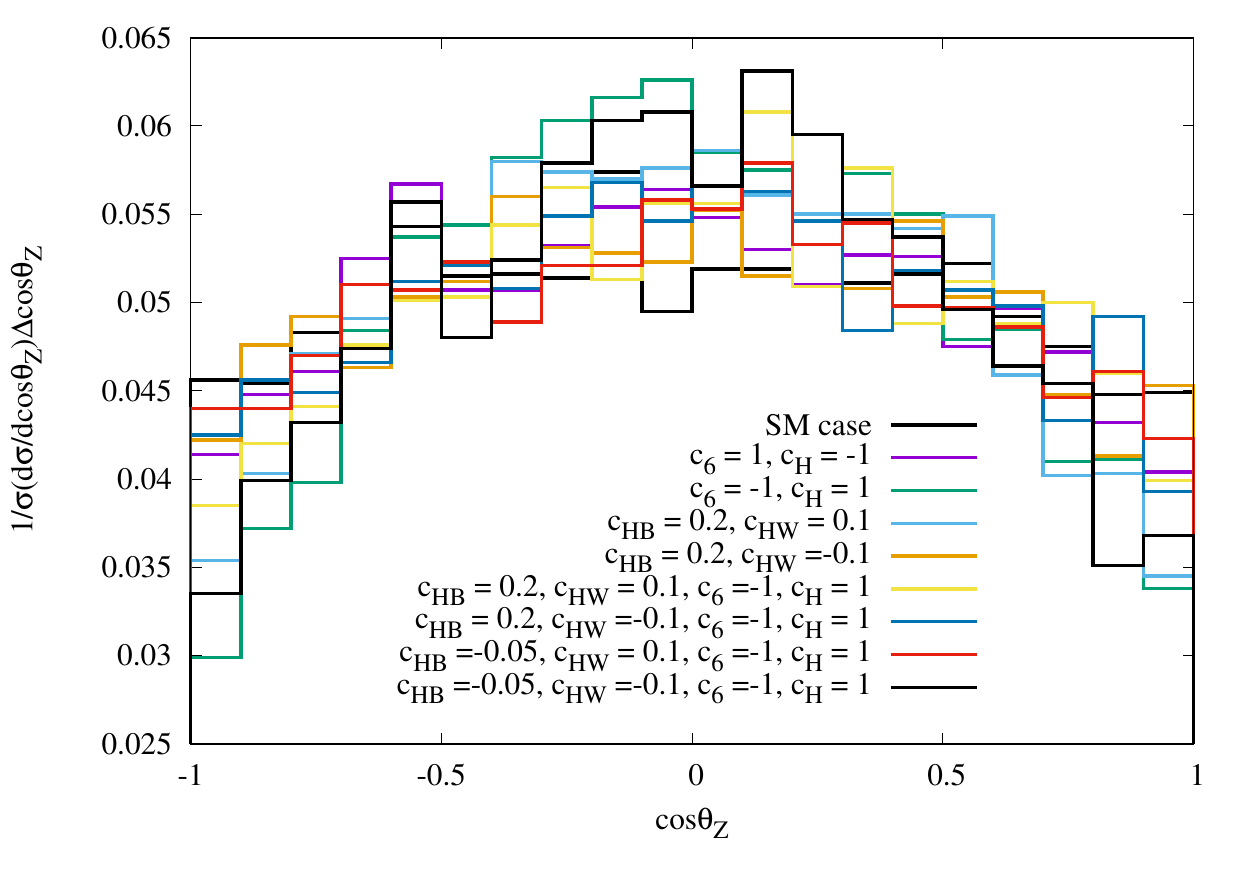}\\
\vspace*{0.1cm}
\hspace{-10mm}
\includegraphics[angle=0,width=70mm]{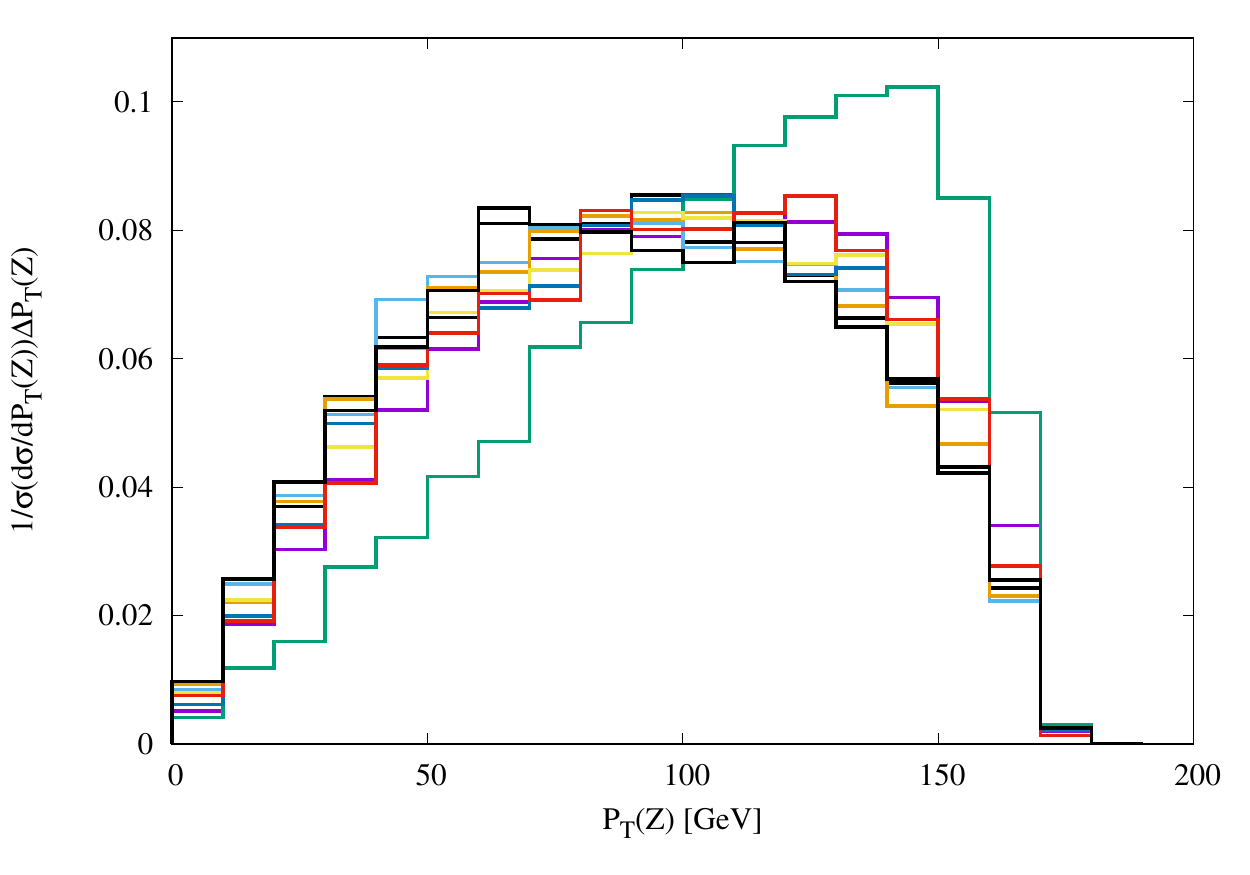} &
\includegraphics[angle=0,width=70mm]{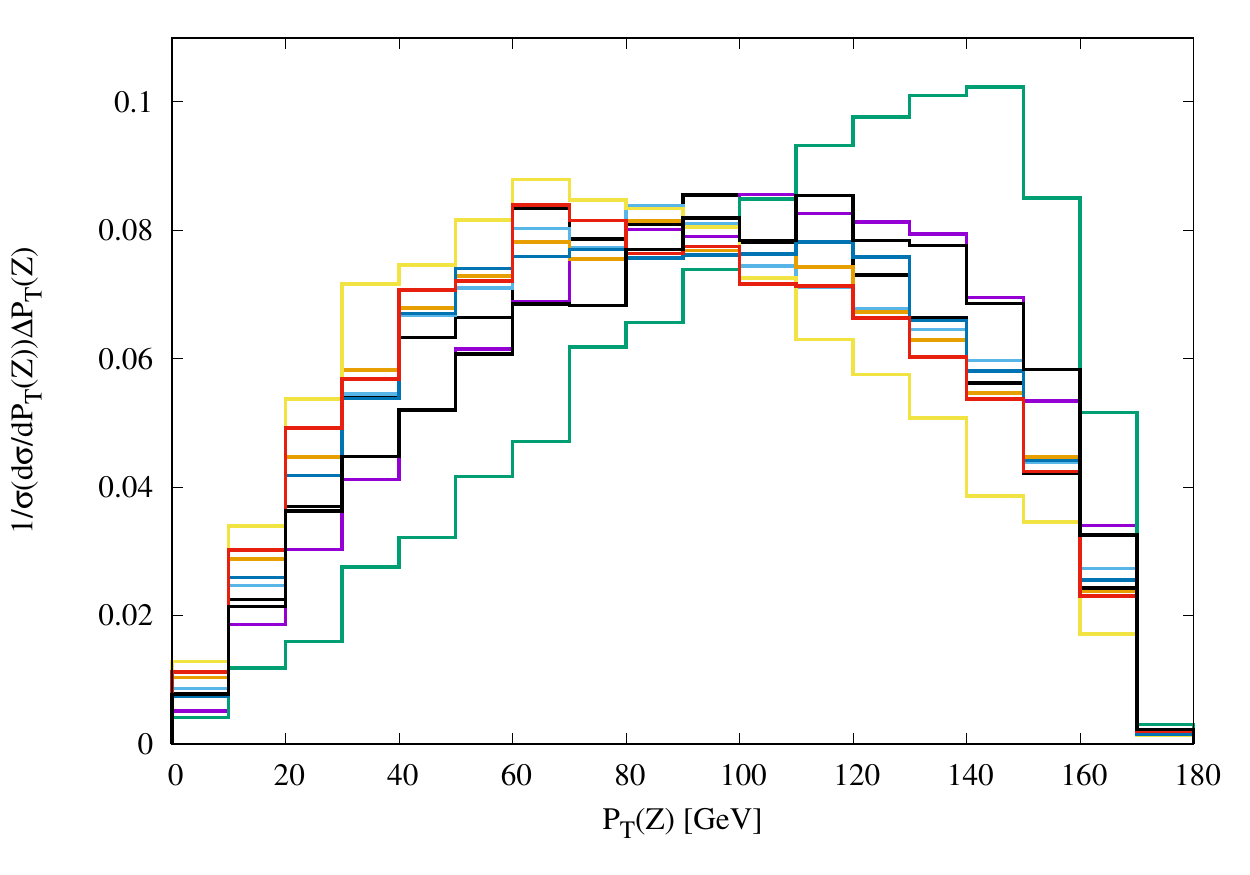}\\
\hspace{-10mm}
\includegraphics[angle=0,width=70mm]{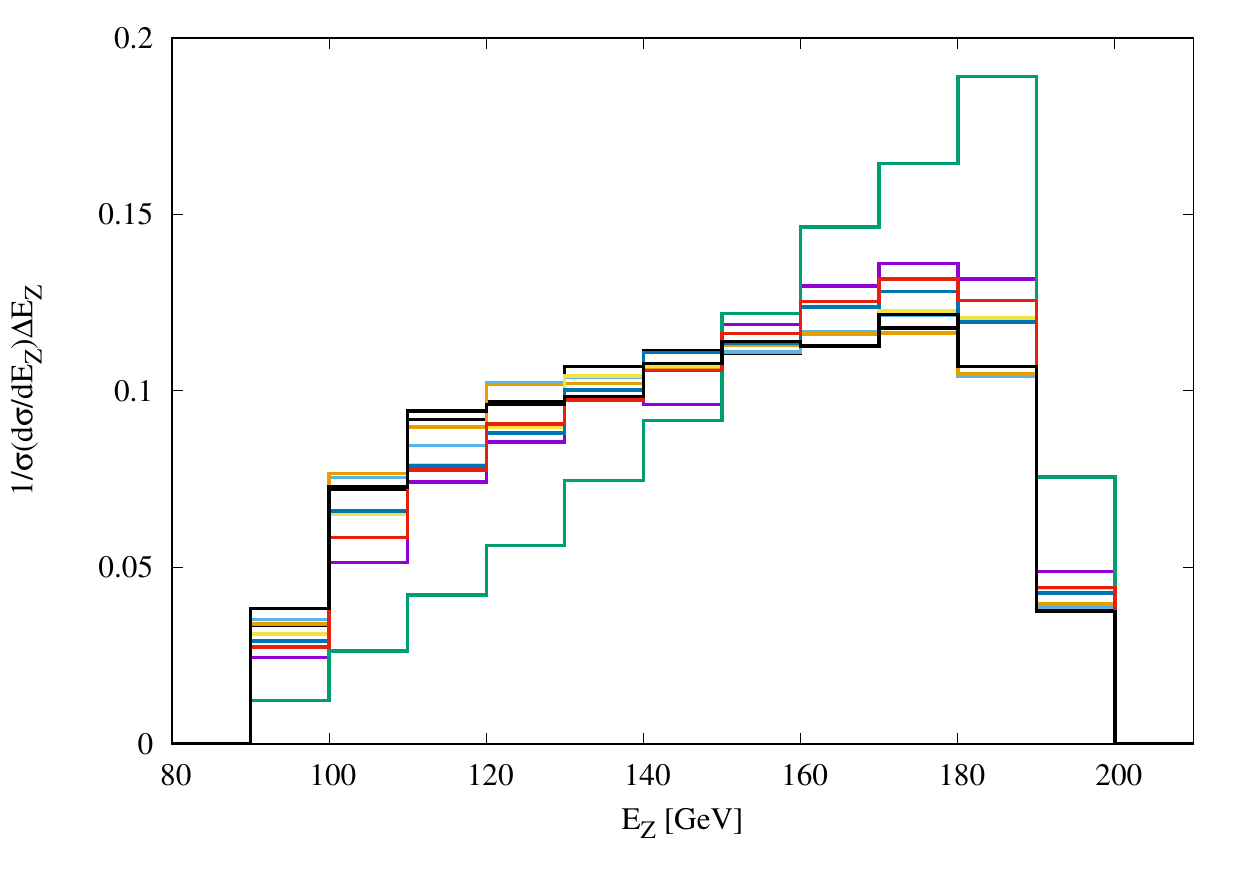} &
\includegraphics[angle=0,width=70mm]{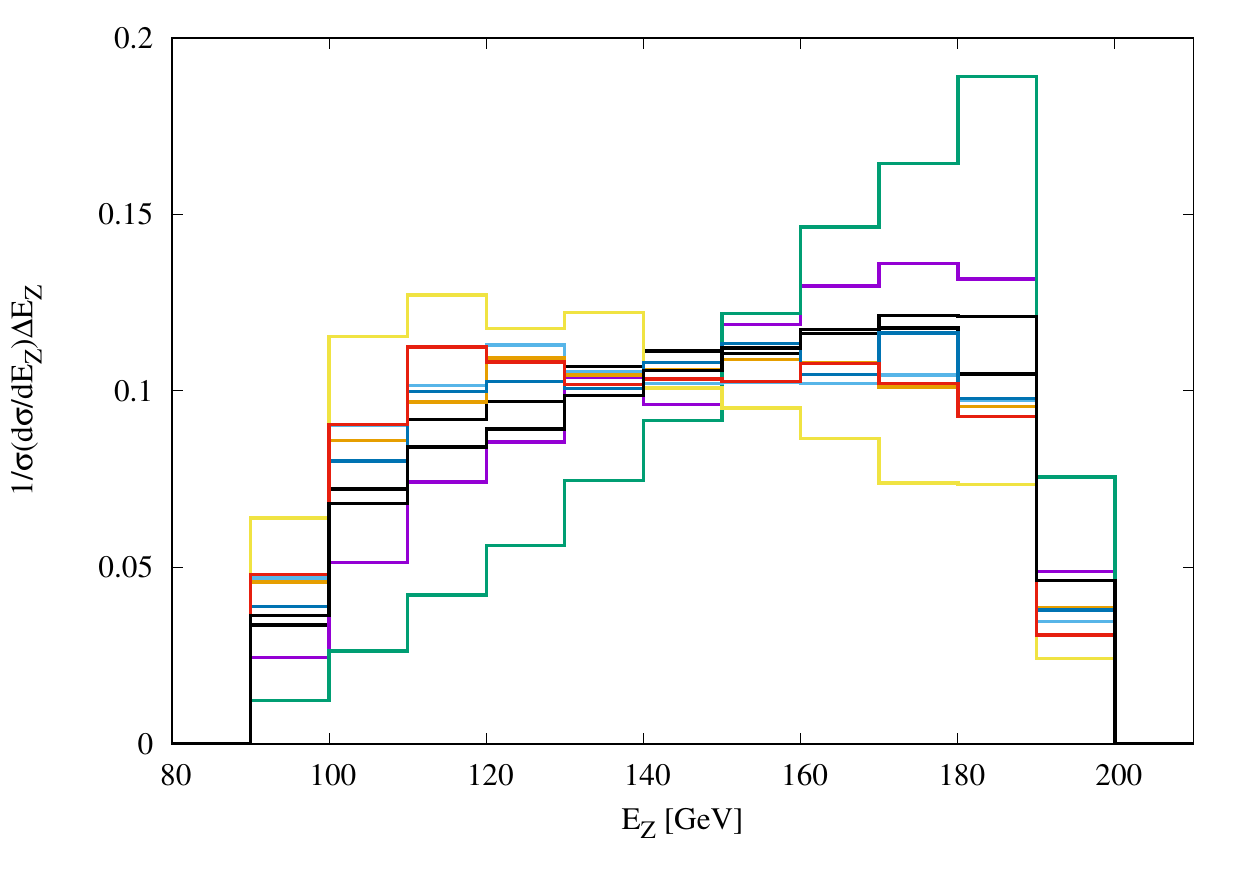} \\
\end{tabular}
\vspace*{0.1cm}
\caption{Distributions of the 
$\cos\theta_Z$, Transverse Momentum, and Energy of the $Z$ boson for the anomalous coupling values as in the inset, illustrating how the presence of $c_W$ (first column), and  $c_{HW}$ and $c_{HB}$ (second column) affect the influence of $c_6$ and $c_H$. A centre of mass energy of $500$ GeV is assumed.}
\label{fig:ctZ_ptZ_EZ}
\end{figure}

We first consider in Fig.\ref{fig:ctZ_ptZ_EZ} (top row), the normalized $\cos\theta_Z$ distributions of the $Z$ boson for the case of SM, as well as for different cases with anomalous couplings. The normalized distributions present the difference in the shape, which brings out the qualitative difference in a more visible manner. The figure on the left shows the case with $c_{W}=-c_B$  taking typical values, while the other parameters set to zero, whereas the figure on the right considers $c_{HW}$ and $c_{HB}$ non-zero, while setting other parameters to zero. The case with only $c_6$ and $c_H$ taking non-zero values, when compared with the SM case shows a perceivable  change in the distribution with more number of events piling in the small $\cos\theta_Z$ region. Such  experimental observations could, therefore, be considered as an indication of the anomalous $HHH$ coupling. On the other hand, the presence of anomalous $c_W$ and $c_B$ couplings does not affect the distribution much. More importantly, in their presence, the non-zero $c_W$ and $c_B$, the distribution remains close to the SM distribution, even with non-zero $c_6$ and $c_H$. Thus, a conclusion regarding the presence or otherwise of the $HHH$ coupling drawn from the $\cos\theta_Z$ distribution will depend on the values of $c_W$ and $c_B$. The figure on the right tells a similar story for the case of $c_{HW}$ and $c_{HB}$ replacing $c_W$.  In Fig.\ref{fig:ctZ_ptZ_EZ} (second row) and (third row), the $p_T$ and energy distributions of the $Z$ boson are plotted. Here too, we see that if only $c_6$ and $c_H$ are considered to be non-zero, events with high $p_T$ and high energy $Z$ bosons are preferred much more in comparison with the SM case. This conclusion is upset with the simultaneous presence of other parameters related to  $ZZH$ coupling. 

\begin{figure}[h]\centering
\begin{tabular}{c c}
\hspace{-10mm}
\vspace*{0.1cm}
\includegraphics[angle=0,width=70mm]{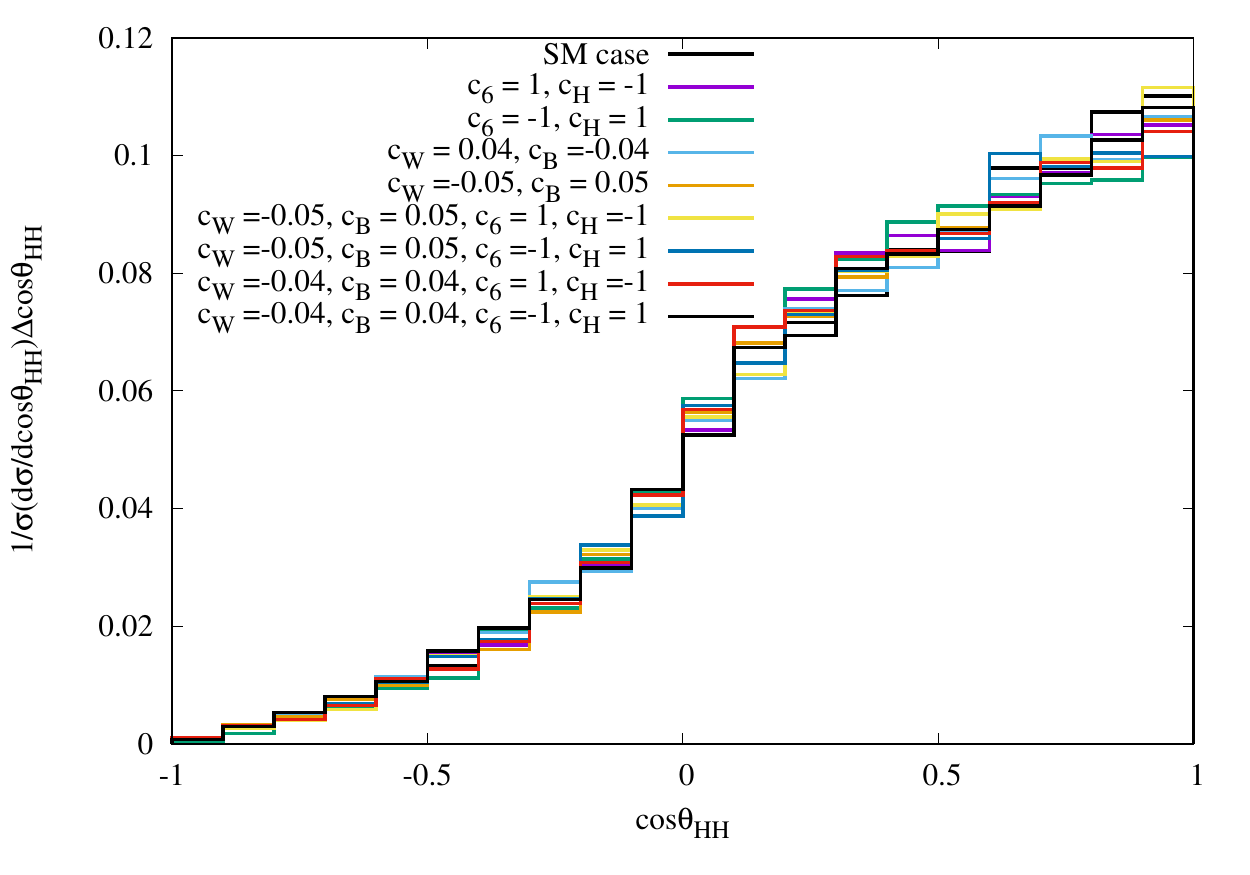}&
\includegraphics[angle=0,width=70mm]{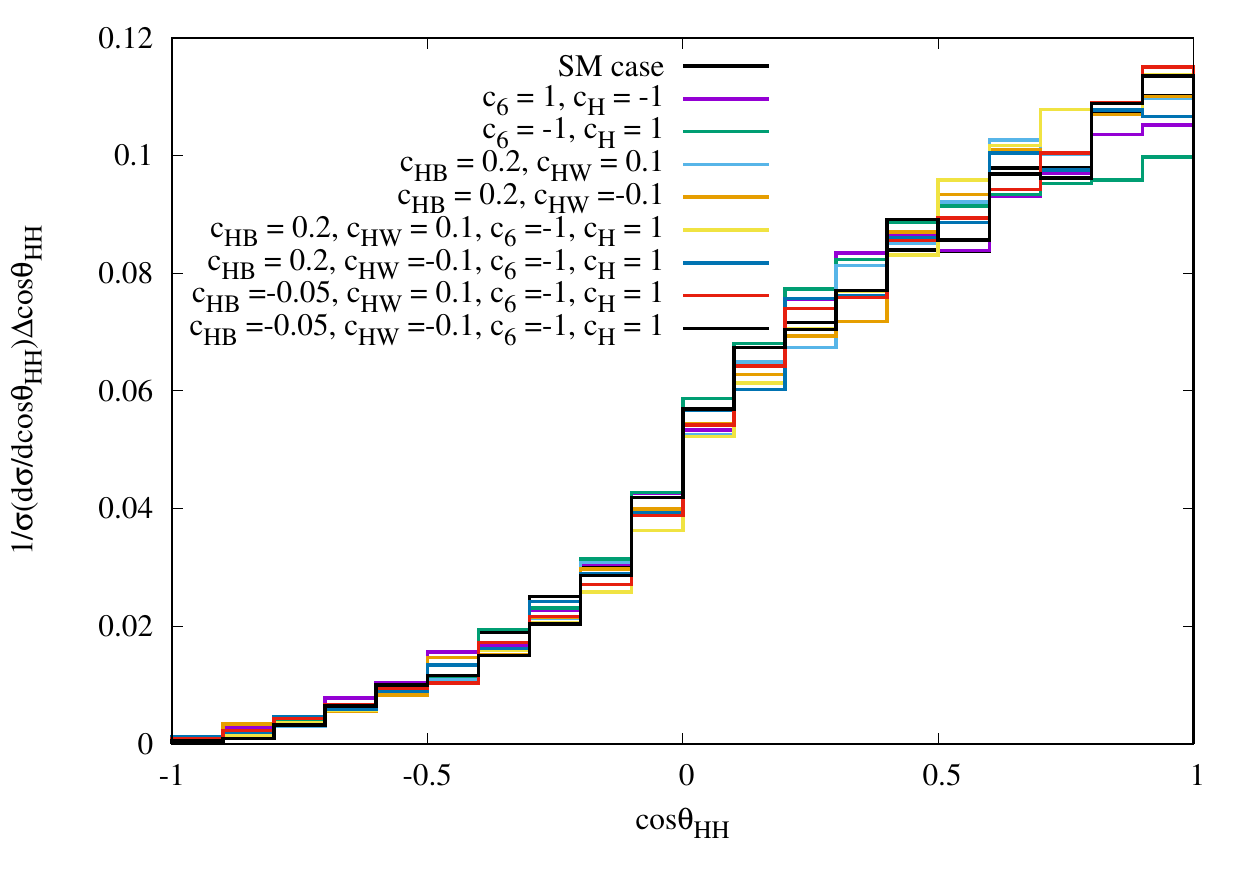}  \\
\hspace{-10mm}
\includegraphics[angle=0,width=70mm]{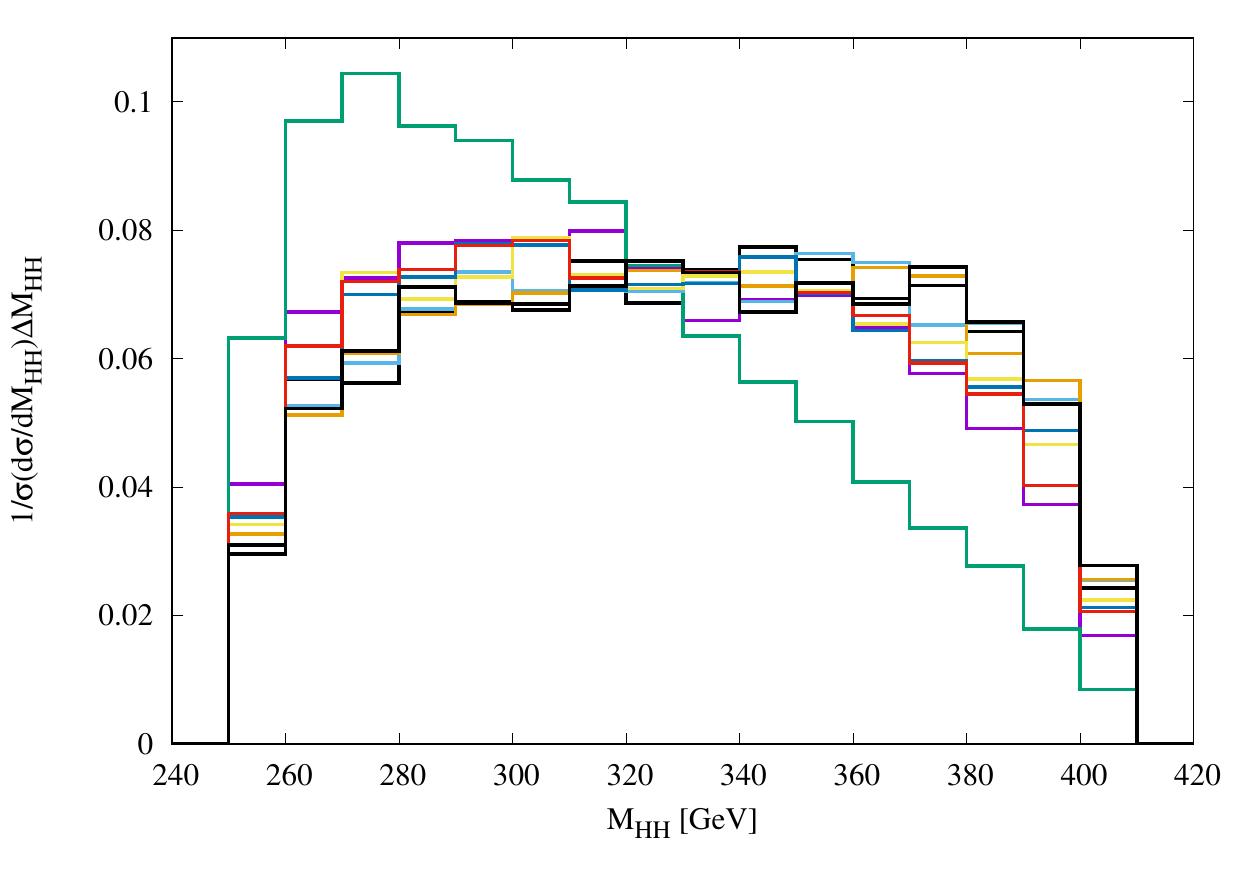} &
\includegraphics[angle=0,width=70mm]{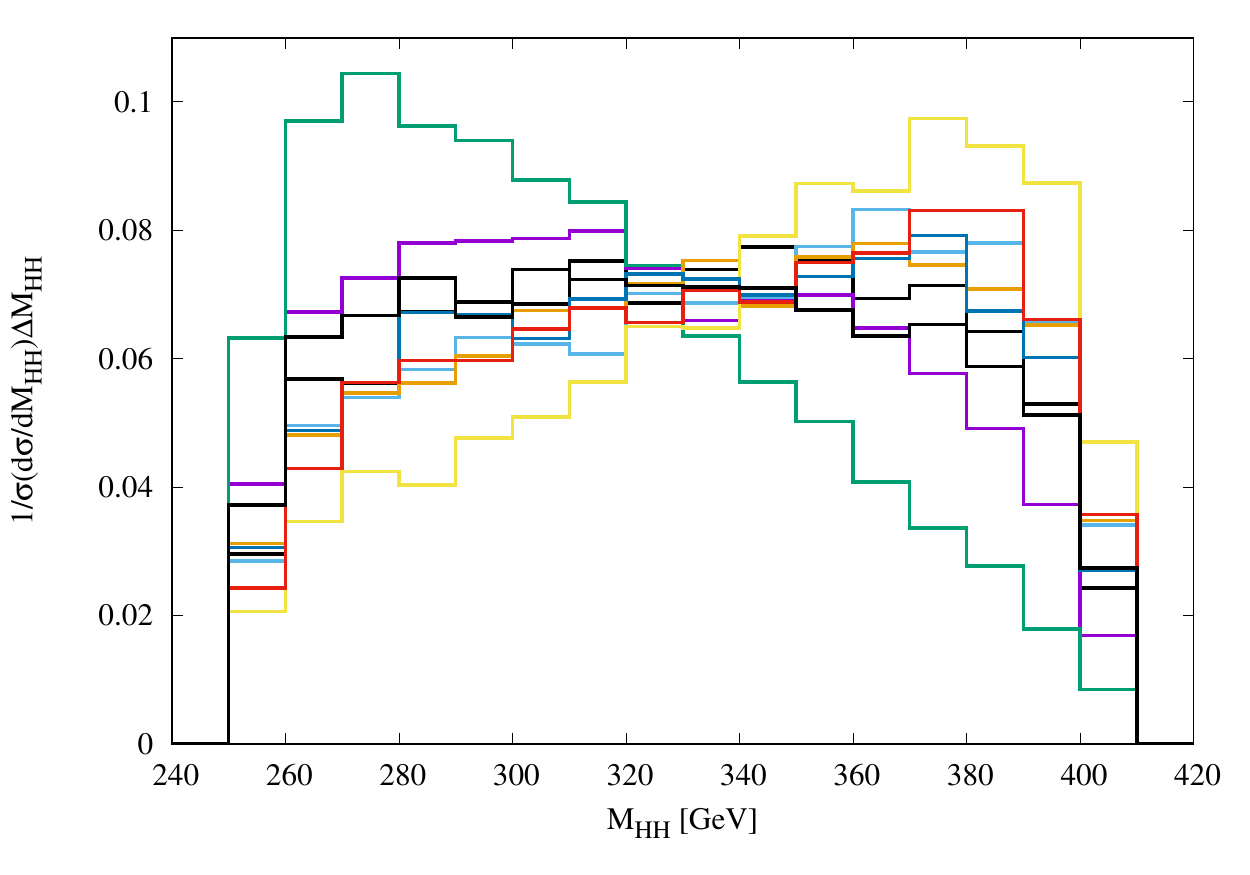}
\end{tabular}
\vspace*{0.1cm}
\caption{$\cos\theta_{HH}$ and the invariant mass of $HH$ distributions for the anomalous coupling values as in the inset, illustrating how the presence of $c_W$ (first column), and  $c_{HW}$ and $c_{HB}$ (second column) affect the influence of $c_6$ and $c_H$. A centre of mass energy of $500$ GeV is assumed.}
\label{fig:ct_m_HH}
\end{figure}

The distribution of the opening angle between the two Higgs bosons as well as their invariant mass distribution presented in Fig.~\ref{fig:ct_m_HH} indicate the same feature captured in the various distributions of the $Z$ bosons. While in all cases including the SM case, most of the events are in the forward hemisphere, in the presence of non-vanishing $c_6$ and $c_H$, but with $c_W=c_{HW}=c_{HB}=0$, the events are more evenly distributed within the forward hemisphere, compared to the rest of the cases including the SM case. The $HH$ invariant  mass demonstrate an even more dramatic difference in the different cases mentioned above. 

The conclusions that we draw from the above considerations is that single parameter considerations to understand the effect of $HHH$ coupling will not be realistic if other relevant gauge-Higgs couplings receive anomalous contributions. Our preliminary investigation clearly indicates that the correlations can be rather strong, for all the relevant parameters, and one needs to consider  a careful analysis to obtain realistic limits on the parameters.

%\subsubsection{Limits on anomalous couplings}\label{subsect:limit}
The reach on the parameter $c_6$ and $c_H$ discussed above uses the production
cross section, while the inclusion of decay will loosen the  limit on them. Here we use the
total cross section including the decay of the $Z$ bosons and the $H$ given in Eq.~(\ref{eq:sigma-tot}) and study the sensitivity to all the couplings.
The sensitivity of the total cross section $\sigma_{tot}$ to a couplings $f$ is defined as
\begin{equation}
{\cal S}(f)= \frac{\sigma_{tot}(f)-\sigma_{tot}(f=0)}{\delta\sigma_{tot}},
\end{equation}
$\delta\sigma_{tot}$ is the estimated error  in $\sigma_{tot}$.
\begin{figure}[h]\centering
\begin{tabular}{c c}
\includegraphics[angle=0,width=64mm]{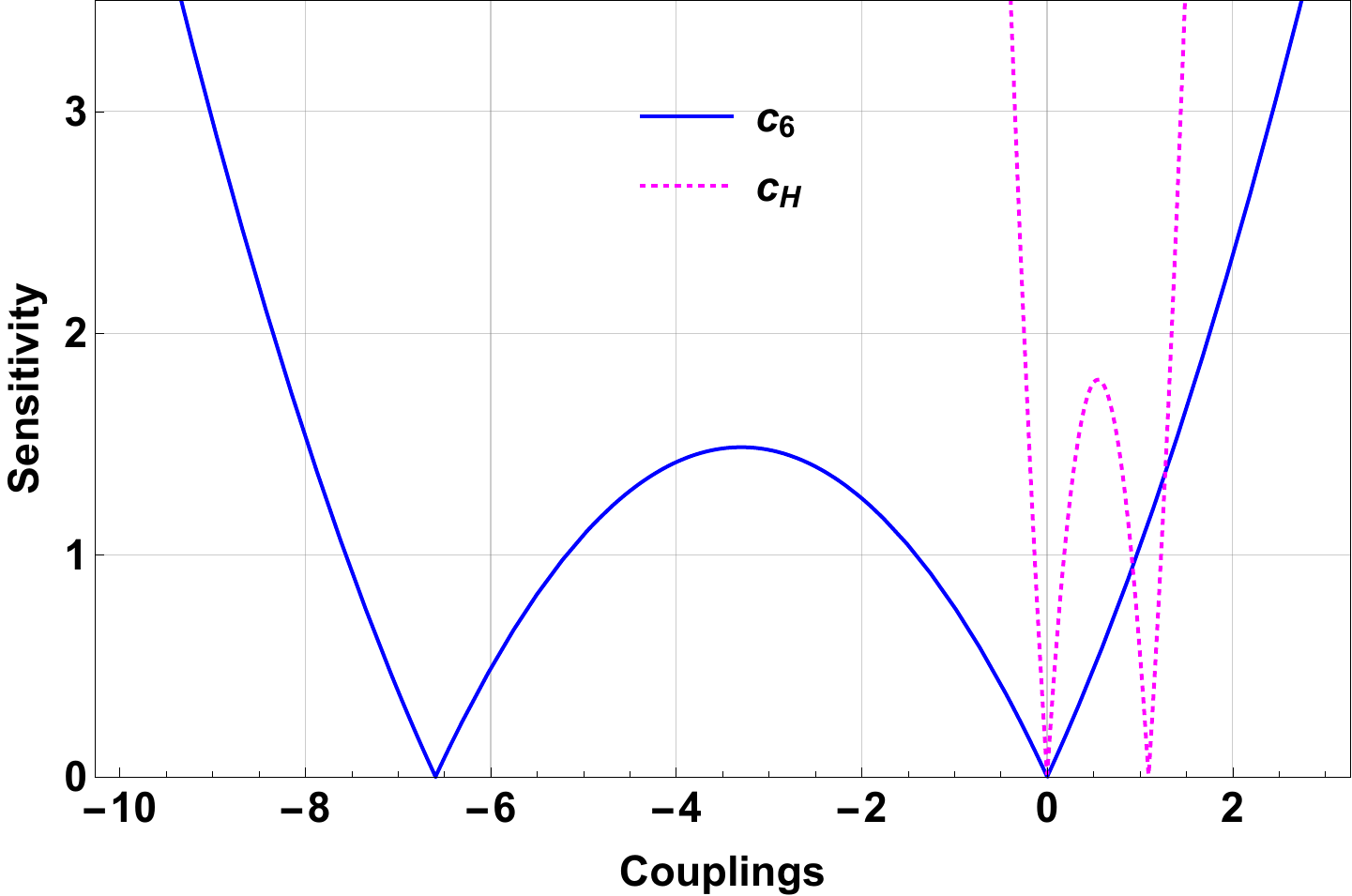} &
\hspace{15mm}
\includegraphics[angle=0,width=64mm]{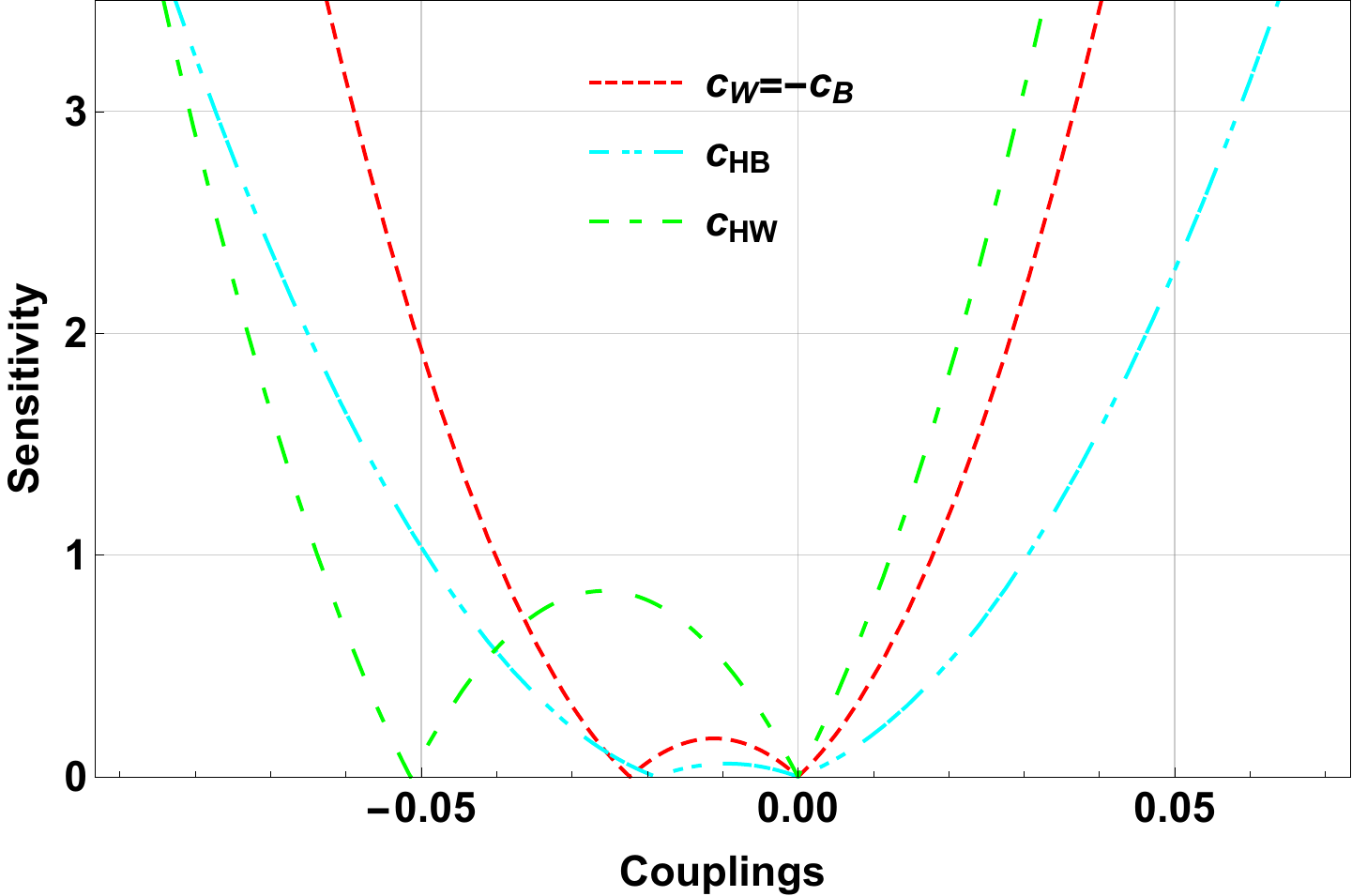} \\
\end{tabular}
\caption{Sensitivity of cross-section on anomalous couplings in $e^+e^-\to hhZ$ with $h \rightarrow b \bar b,~Z\to l^+l^-$ for $\sqrt{s}=500$ GeV, ${\cal L}=1000$ fb$^{-1}$  for $c_6$, $c_H$ (left) and $c_W=-c_B$, $c_{HB}$, $c_{HW}$ (right). The numerical value of one parameter limits on the couplings at $3\sigma$ sensitivity can be read in TABLE~\ref{tab:mcmclimits}.}
\label{fig:sensitivity}
\end{figure}
\begin{table}\caption{\label{tab:mcmclimits} Simultaneous limits on anomalous couplings from MCMC in  $e^+e^-\to hhZ$ with $h\rightarrow b\bar{b},~Z\to l^+l^-$ for $\sqrt{s}=500$ GeV, ${\cal L}=1000$ fb$^{-1}$}
	\renewcommand{\arraystretch}{1.50}
	\begin{center}
		\begin{tabular}{|c|l|l|l|c|}\hline	
			Parameter &  ~~~~~68~\% BCI&  ~~~~~95~\% BCI&  ~~~~~99~\% BCI& One param. $3\sigma$ limit\\\hline
 $c_6     $  &	$\in[-5.4 , +10]$ & $\in[ -12, +15 ]$ & $\in[ -15 , +18]$& $\in[-9.2,+2.4]$\\
 $c_H     $  &	$\in[-1.3 , +3.1]$ & $\in[ -3.1, +4.6 ]$ & $\in[ -4.1 , +5.6]$& $\in[-0.34,+1.4]$\\
 $c_W=-c_B$  &	$\in[-0.19 , +0.17]$ & $\in[ -0.34, +0.32 ]$ & $\in[ -0.43 , +0.41]$& $\in[-0.05,+0.03]$\\
 $c_{HB}  $  &	$\in[-0.21 , +0.23]$ & $\in[ -0.39, +0.41 ]$ & $\in[ -0.48 , +0.51]$& $\in[-0.07,+0.05]$\\
 $c_{HW}  $  &	$\in[-0.10 , +0.10]$ & $\in[ -0.18, +0.17 ]$ & $\in[ -0.23 , +0.22]$& $\in[-0.08,+0.02]$\\
			\hline
		\end{tabular}
	\end{center}
\end{table}
\begin{figure}[h]
\centering	
\begin{tabular}{c c c c c}
\includegraphics[angle=0,width=28mm]{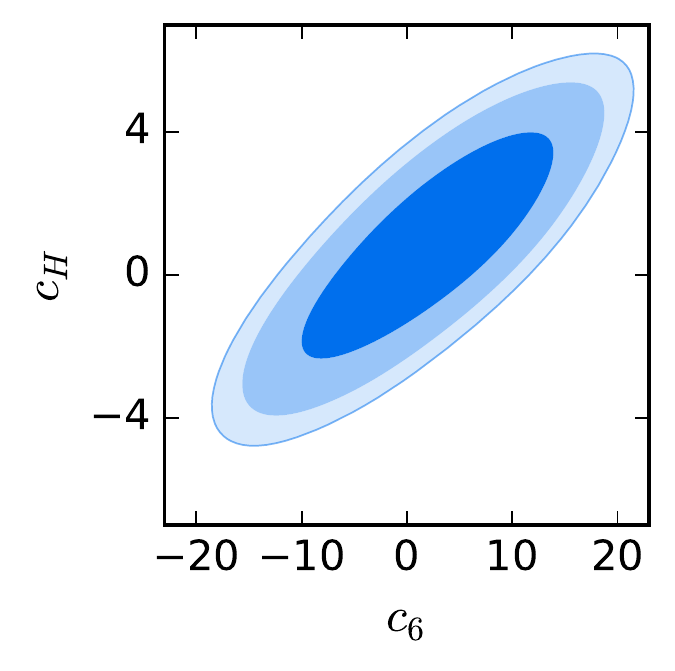}&  
\includegraphics[angle=0,width=28mm]{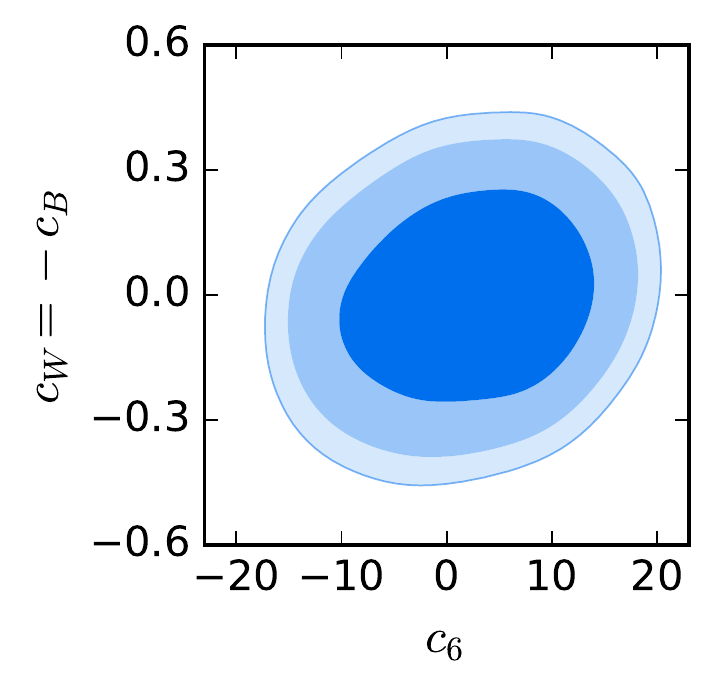} &
\includegraphics[angle=0,width=28mm]{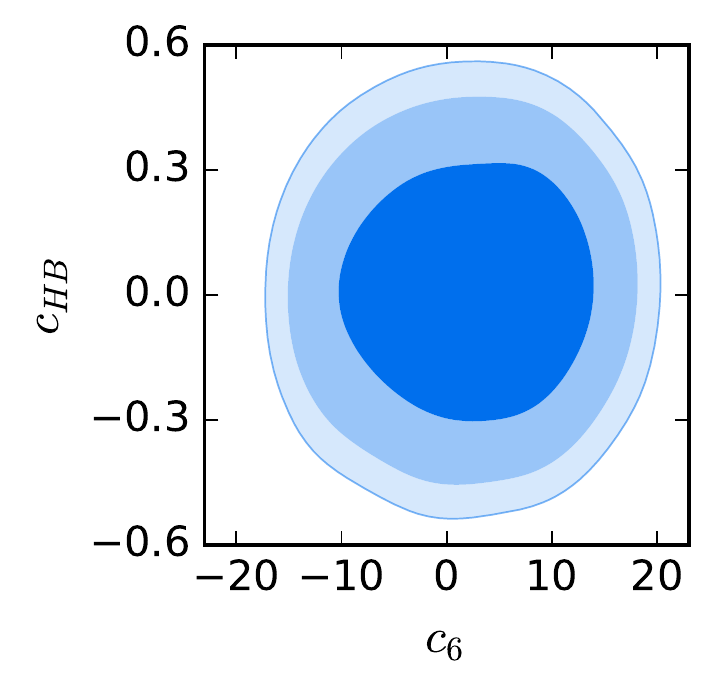} &
\includegraphics[angle=0,width=28mm]{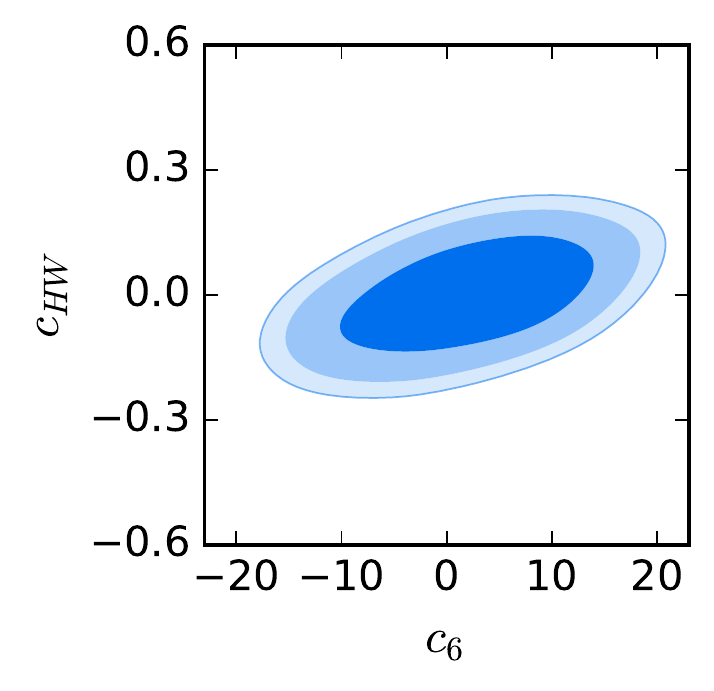}&
\includegraphics[angle=0,width=28mm]{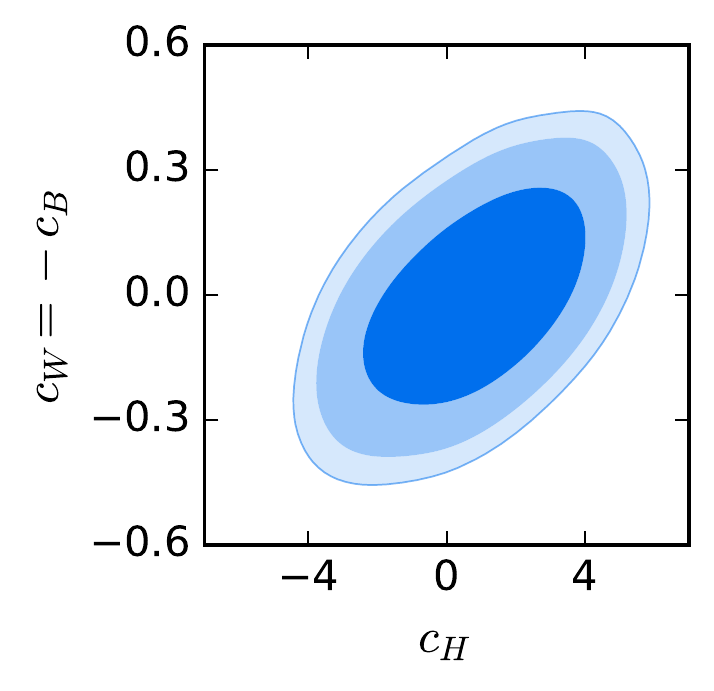}\\
\includegraphics[angle=0,width=28mm]{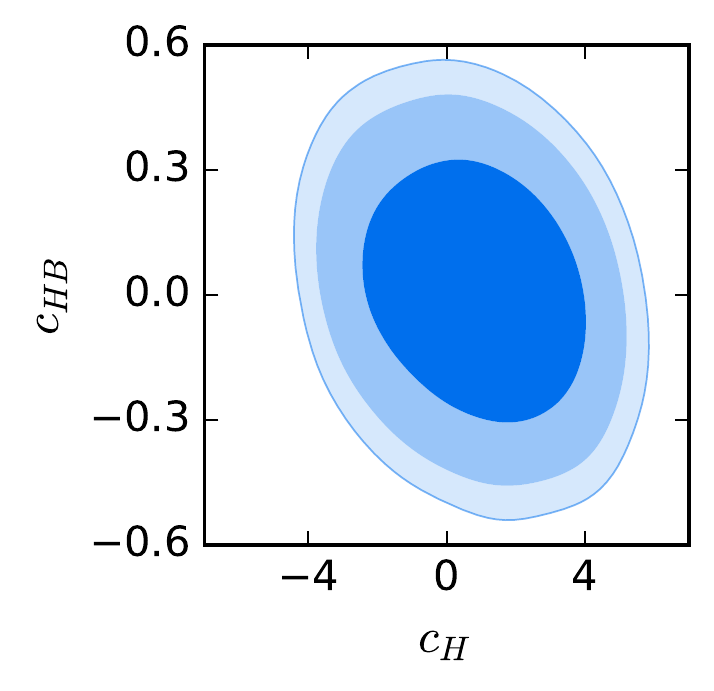}&
\includegraphics[angle=0,width=28mm]{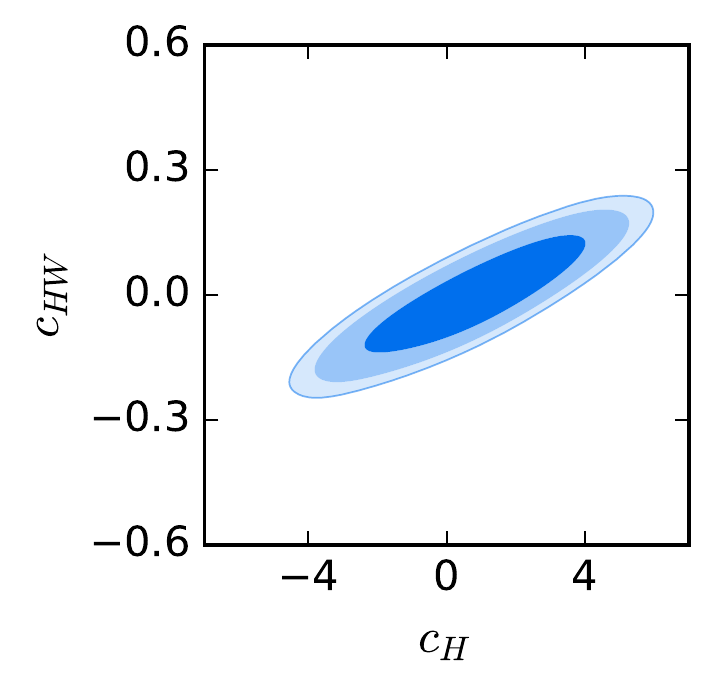}&
\includegraphics[angle=0,width=28mm]{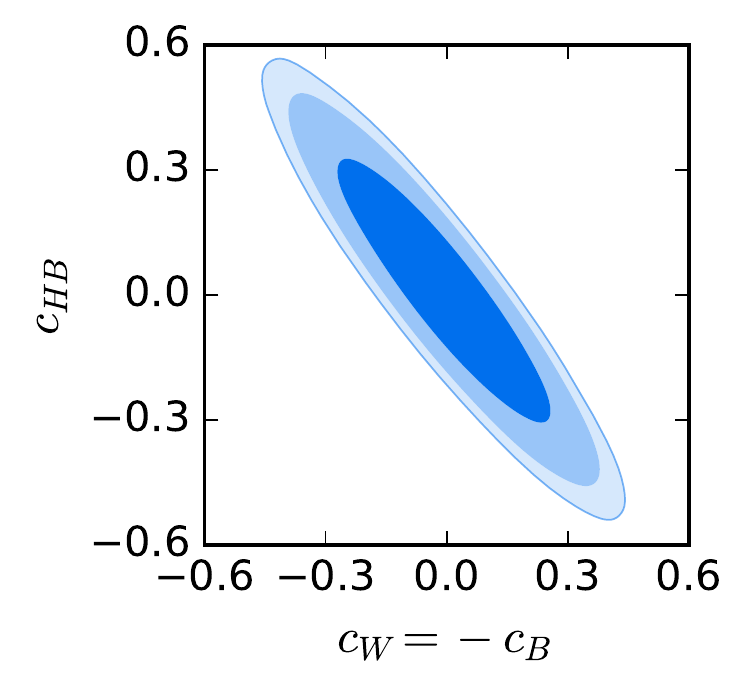} &
\includegraphics[angle=0,width=28mm]{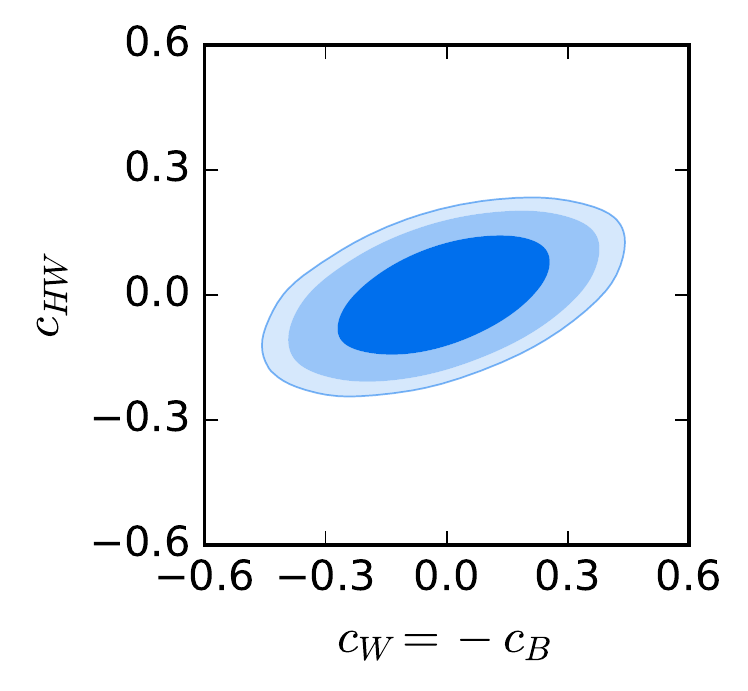} &
\includegraphics[angle=0,width=28mm]{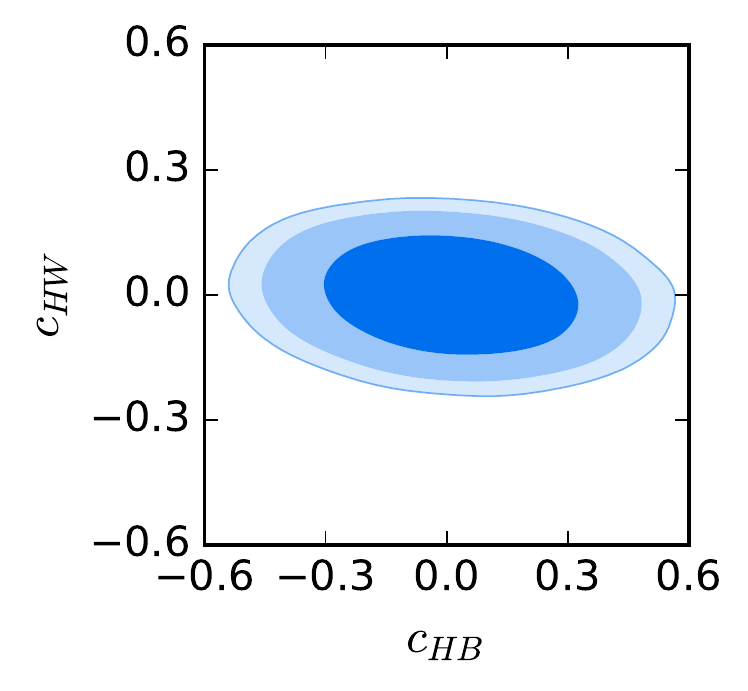} \\
\end{tabular}
\caption{ The MCMC $68~\%$ ({\em darkest-shaded}), $95~\%$ ({\em less darker-shaded}), $99~\%$ ({\em lighter-shaded}) BC contours on the parameter space  in $e^+e^-\to hhZ$ with $h\rightarrow b\bar{b},~Z\to l^+l^-$ for $\sqrt{s}=500$ GeV, ${\cal L}=1000$ fb$^{-1}$}
\label{fig:mcmcplots}
\end{figure}
In Fig.\ref{fig:sensitivity}, we present the sensitivity of the total cross section given in Eq.~(\ref{eq:sigma-tot}) to all the couplings  at  center of mass energy of $500$ GeV and Integrated luminosity of  $1000$ $fb^{-1}$.  
Due to the presence of linear piece along with quadratic piece, the sensitivity of $c_6$ and $c_H$ in the {\em left-panel} 
has double hump nature. Both   
$c_6$ and $c_H$ has roughly symmetric limits at $1\sigma$ ($c_6\in[-1.4,1.0]$) but these get asymmetric limits at $3\sigma$ 
($c_6\in[-9.20,+2.43]$) due to the double hump nature. The couplings $c_W=-c_B$, $c_{HB} $ and $c_{HW}$ shown in the {\em right-panel}  
 posses asymmetric limit at $1\sigma$ as well as at $3\sigma$ sensitivity. The one parameter limit at $3\sigma$ sensitivity
on all the couplings are shown in the last column of   TABLE~\ref{tab:mcmclimits}.

We perform a simultaneous multi-parameter  analysis with the total  cross section using the Markov-Chain--Monte-Carlo (MCMC) method.
We obtain simultaneous limits by varying all the couplings simultaneously using \texttt{GetDist}~\cite{Antony:GetDist} package with the MCMC chain. 
The $68~\%$, $95~\%$ and  $99~\%$ Bayesian-Confidence-Interval (BCI) on the couplings  are shown in the TABLE~\ref{tab:mcmclimits}. The $99~\%$ BCI 
on the couplings can be compared with the $3\sigma$ one parameter limit from sensitivity on the fifth column of the same table.
It can be seen that the simultaneous limits on $c_6$ are much tighter than its one parameter limit by its parametric dependence,
while simultaneous limits on all other couplings are less tighter than their one parameter limit. The correlation among all parameter after marginalization in the remaining parameters  are studied and they are shown in Fig.~\ref{fig:mcmcplots}. In the figure, the {\em darkest-shaded} contours are for $68~\%$ BC, less  {\em darker-shaded} contours are for  $95~\%$ BC and  {\em lightest-shaded} contours are for $99~\%$ BC.  The (anti) correlations among the couplings have emerged 
in the multi-parameter analysis.  A mild correlation can be seen in the panel $c_6$--$c_H$ and $c_W$--$c_{HW}$, while a strong anti-correlation is observed in the panel $c_W=-c_B$--$c_{HB}$.

\subsection{$e^+e^-\rightarrow HH\nu\bar \nu$ process}

\begin{figure}[!t]\centering
\begin{tabular}{c c}
%\hspace{-18mm}
\includegraphics[angle=0,width=80mm]{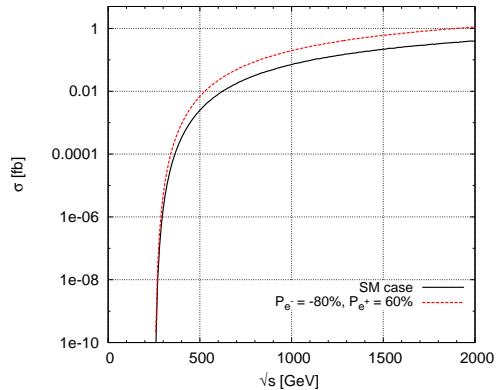} 
\end{tabular}
\vspace*{-1cm}
\caption{Total cross section of  $e^- e^+ \rightarrow \nu_e \bar\nu_e H H$ in the case of unpolarized and polarized beams, as indicated.}
\label{fig:sig_rs_nnhh}
\end{figure}

We shall now turn our attention to the second process involving $HHH$ couplings, as well as gauge-Higgs couplings. We consider the two Higgs production with missing energy through the process $e^+ e^- \rightarrow HH\nu \bar \nu$. The previous process, $e^+ e^- \rightarrow HHZ$, with $Z\rightarrow \nu \bar \nu$ has the same final state. But, this can be easily separated from the rest of the contributions in the SM, to the channels presented in the Feynman diagrams given in Fig.~\ref{fig:fdnnhh}, through, for example considering the missing invariant mass. The cross section for the process is plotted against the centre of mass energy for the case of polarized as well as unpolarized beams in Fig.~\ref{fig:sig_rs_nnhh}. The advantage of very high energy collider is evident here. We shall consider a centre of mass energy of 2 TeV, for which the cross section is close to 0.4 fb in case of unpolarized beams, and slightly more than 1 fb for $e^-$ beam of $-80~\%$ polarization and $e^+$ beam with $+60~\%$ polarization \cite{Behnke:2013xla}. This study will complement the  study of the $ZHH$ production in the sense that the physical couplings involved are $HHH$ along with $WWH$ and $WWHH$ instead of the ones involving the neutral gauge bosons. Although in the language of the effective Lagrangian, the couplings involved are similar to the ones in the previous process, their involvement in the current process is expected to be different.

\begin{figure}[h]\centering
\begin{tabular}{c c}
\hspace{-5mm}
\includegraphics[angle=0,width=80mm]{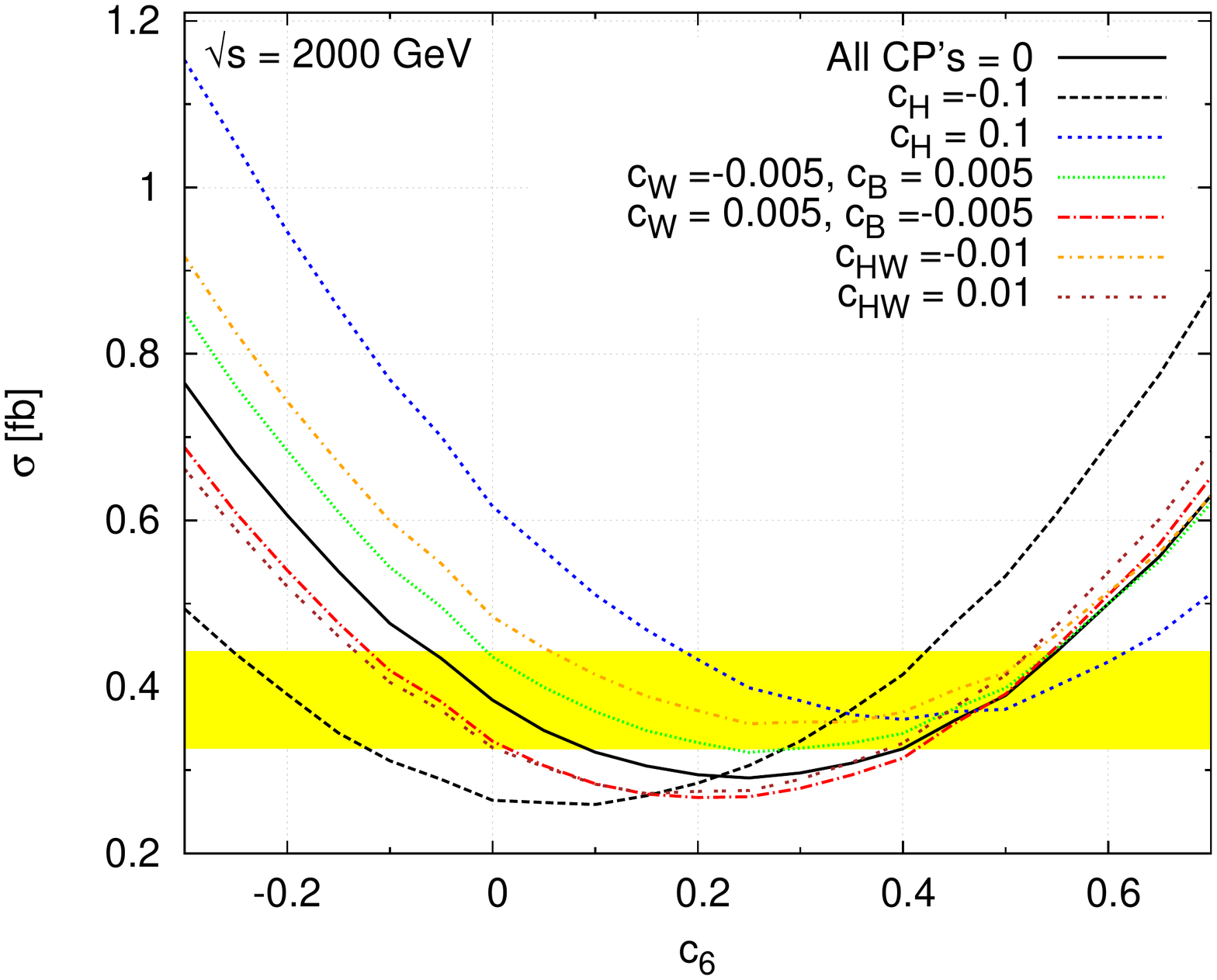}&
\includegraphics[angle=0,width=80mm]{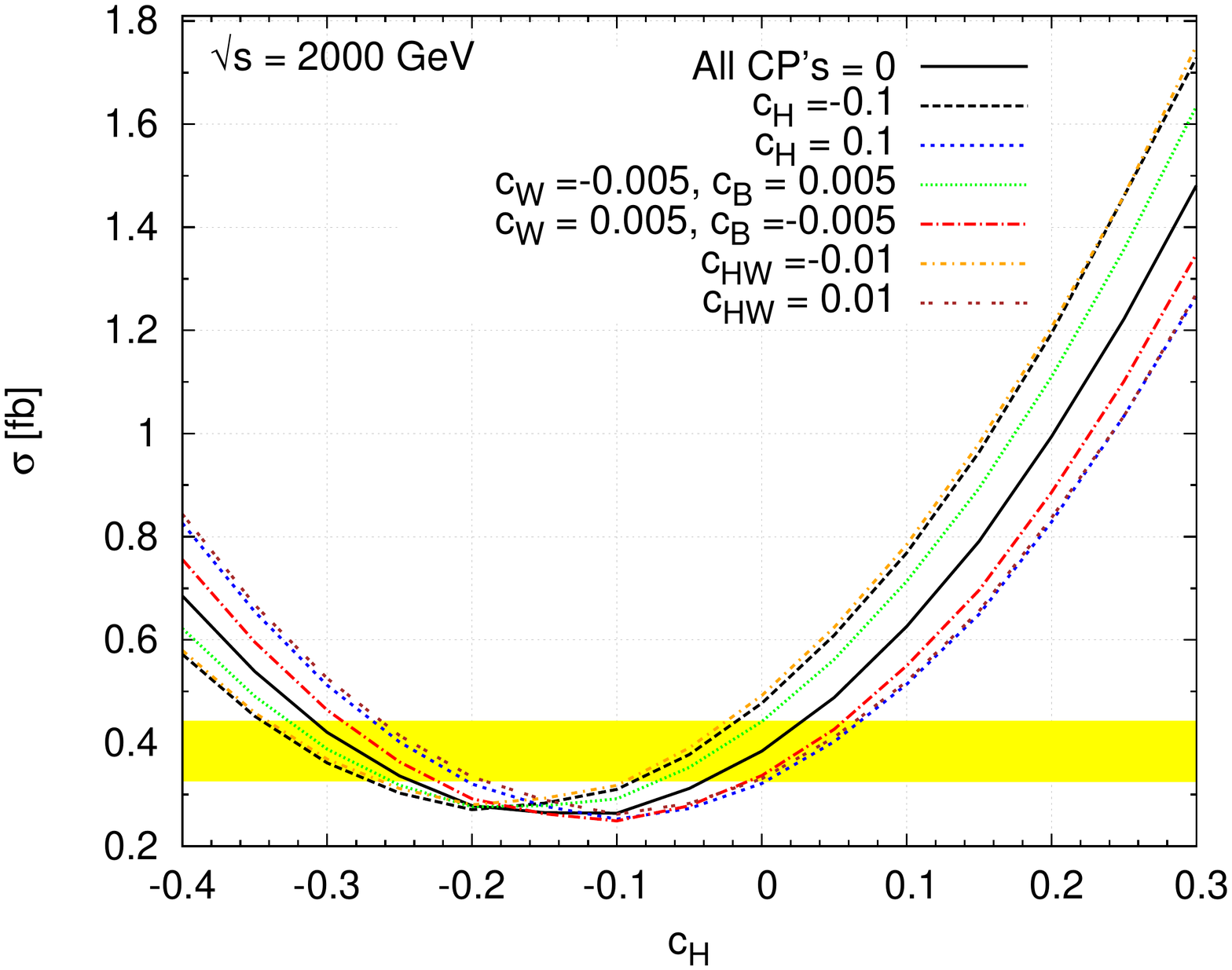}  \\
\end{tabular}
\vspace*{-1cm}
\caption{
Cross section of $\nu\bar\nu HH$ production against $c_6$ (left) and $c_H$ (right), when some of the other selected relevant parameters assume typical values is compared against the case when only $c_6$ or $c_H$  is present. The black solid lines corresponds to the case when all parameters other than $c_6$ (left) or $c_H$ (right) vanish. The centre of mass energy is assumed to be 
$\sqrt{s}=2$ TeV. In each case, all other parameters are set to zero. The yellow band indicates the $3\sigma$ limit 
of the SM cross section.}
\label{fig:cs_c6_nnhh}
\end{figure}

As in the earlier case, the sensitivity of $c_6$ and $c_H$ on the total cross section at the centre of mass energy of 2 TeV is presented in Figs.~\ref{fig:cs_c6_nnhh}, where all other parameters are set to zero, as well as in the presence of some of the relevant parameters. We have included the $3\sigma$ band of the SM cross section assuming 1000 fb$^{-1}$ luminosity. Clearly, the correlation is perceivable, and the conclusions are similar to the case of  $ZHH$ production, that the sensitivity of $HHH$ coupling on the process considered strongly depend on the values of other parameters relevant to $WWH$ and $WWHH$ couplings. 

\begin{figure}[!t]\centering
\begin{tabular}{c c}
\hspace{-8mm}
\vspace{-5mm}
\includegraphics[angle=0,width=80mm]{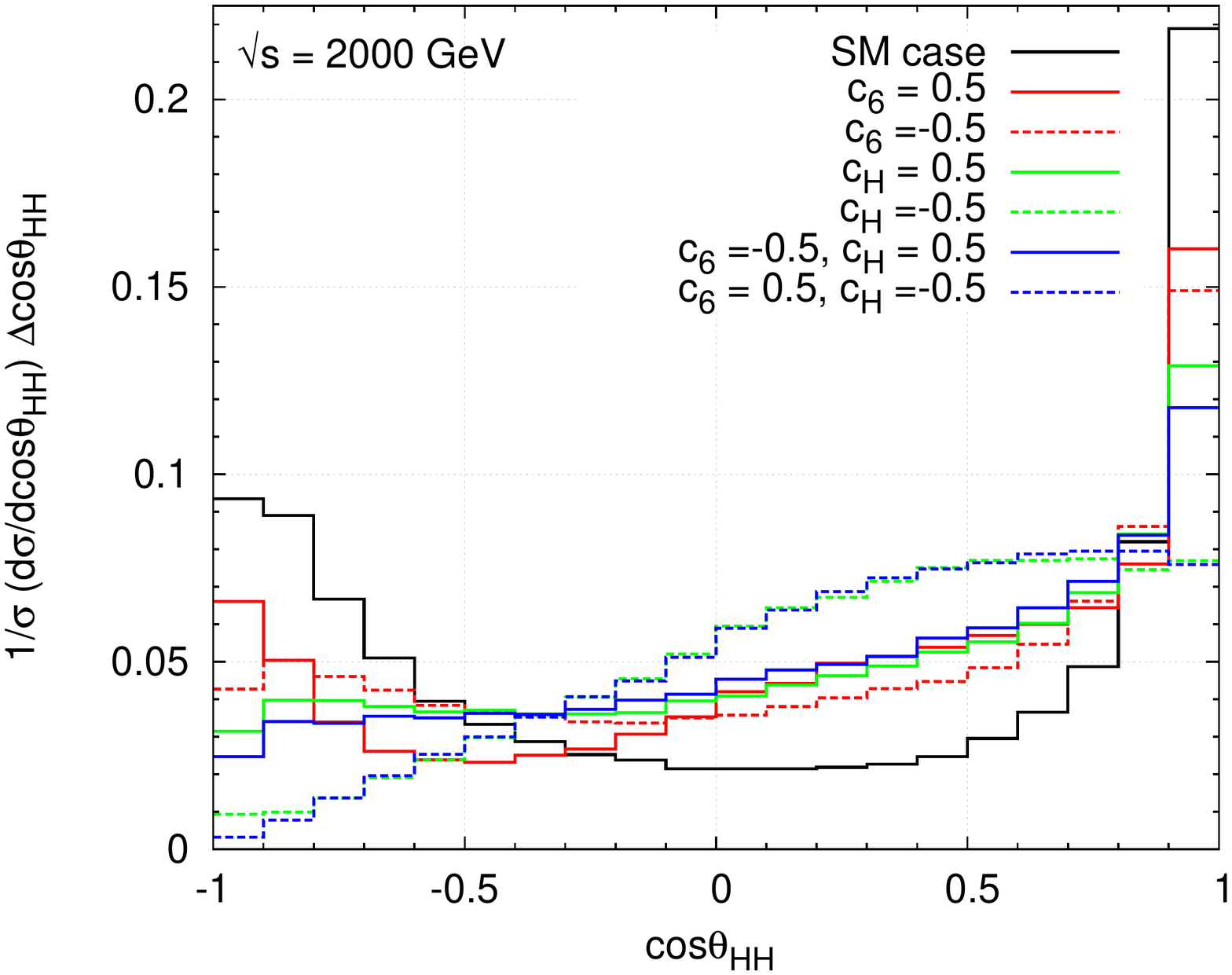} &
\hspace{-5mm}
\vspace{-5mm}
\includegraphics[angle=0,width=80mm]{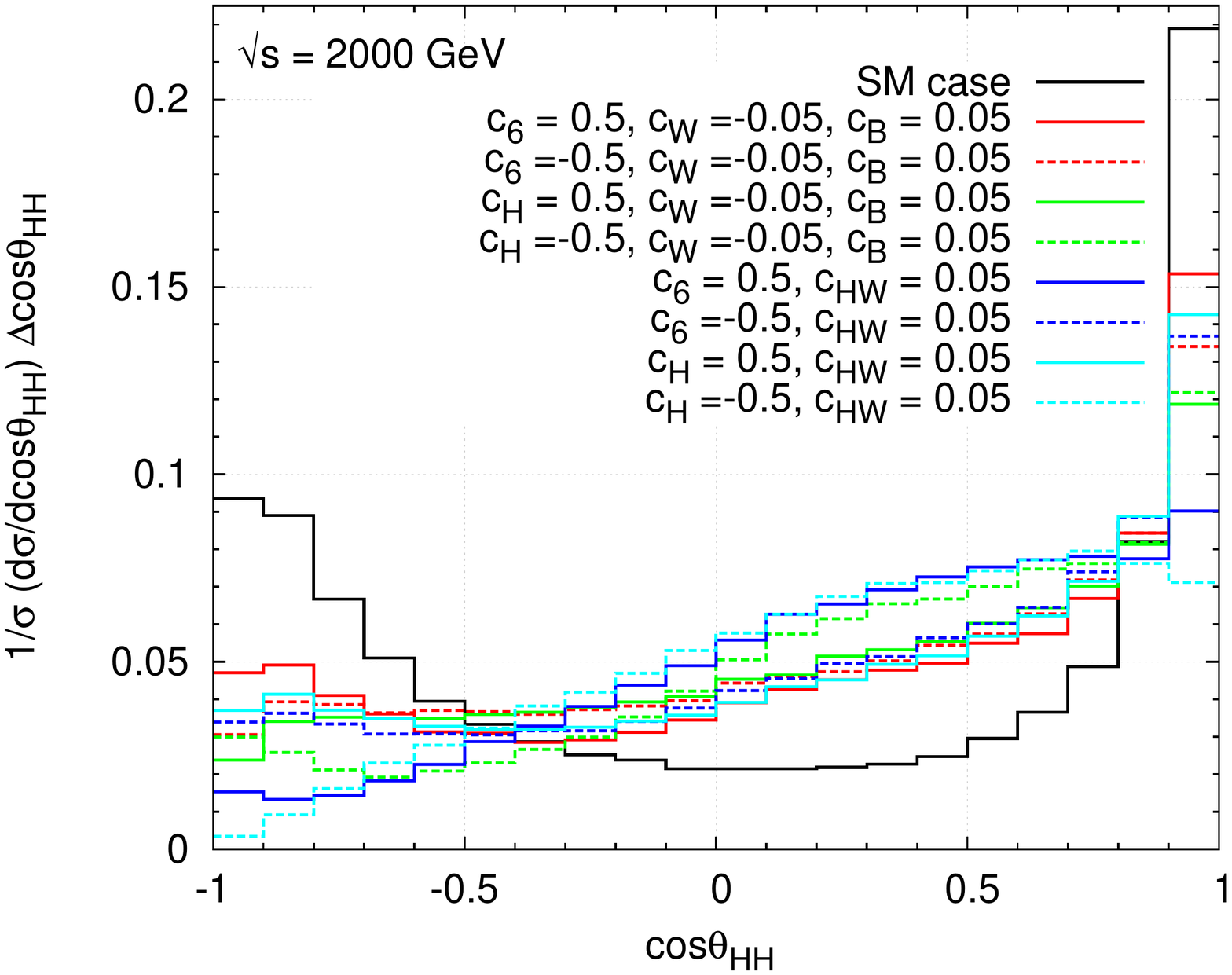} \\
\hspace{-8mm}
\vspace{-5mm}
\includegraphics[angle=0,width=80mm]{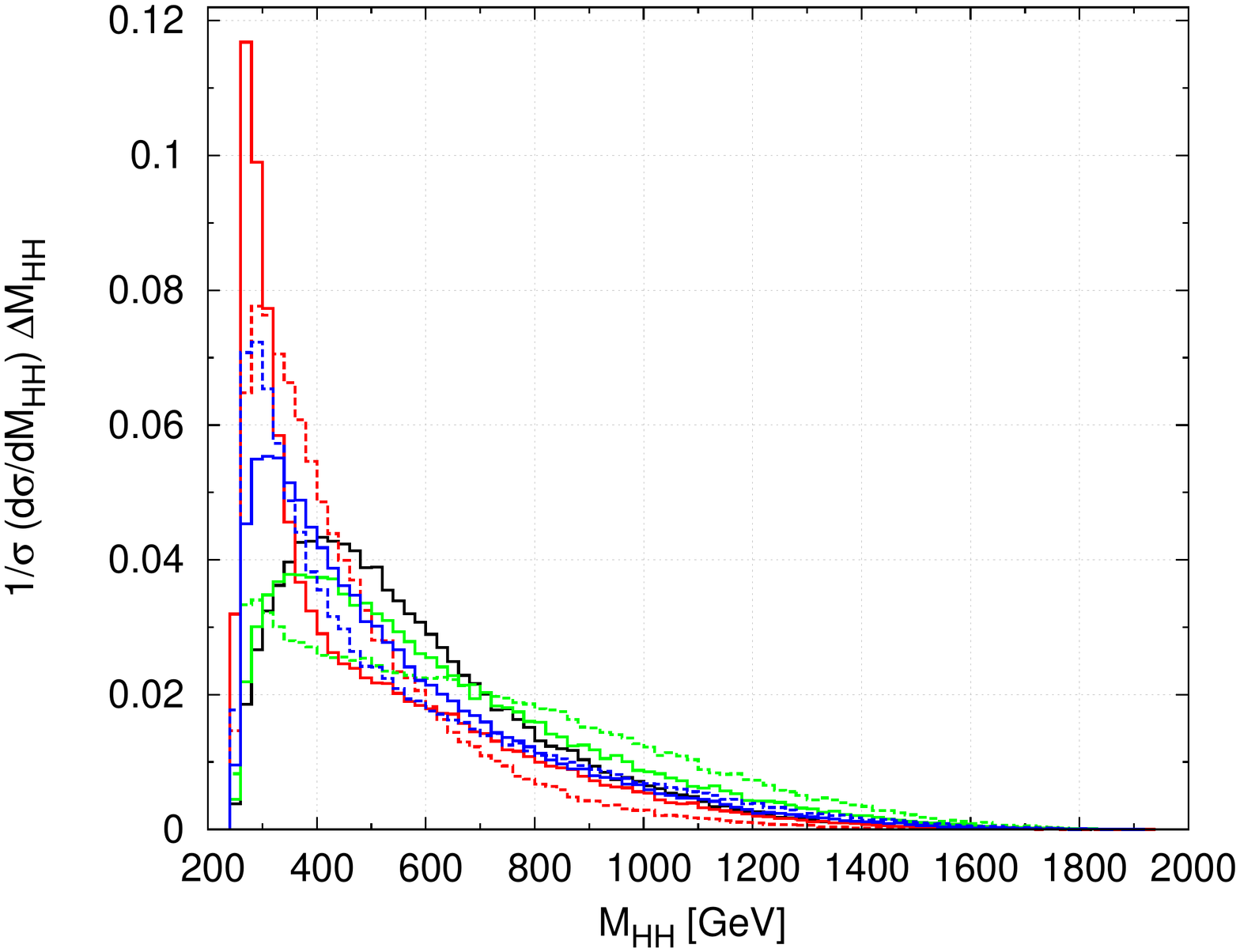} &
\hspace{-5mm}
\vspace{-5mm}
\includegraphics[angle=0,width=80mm]{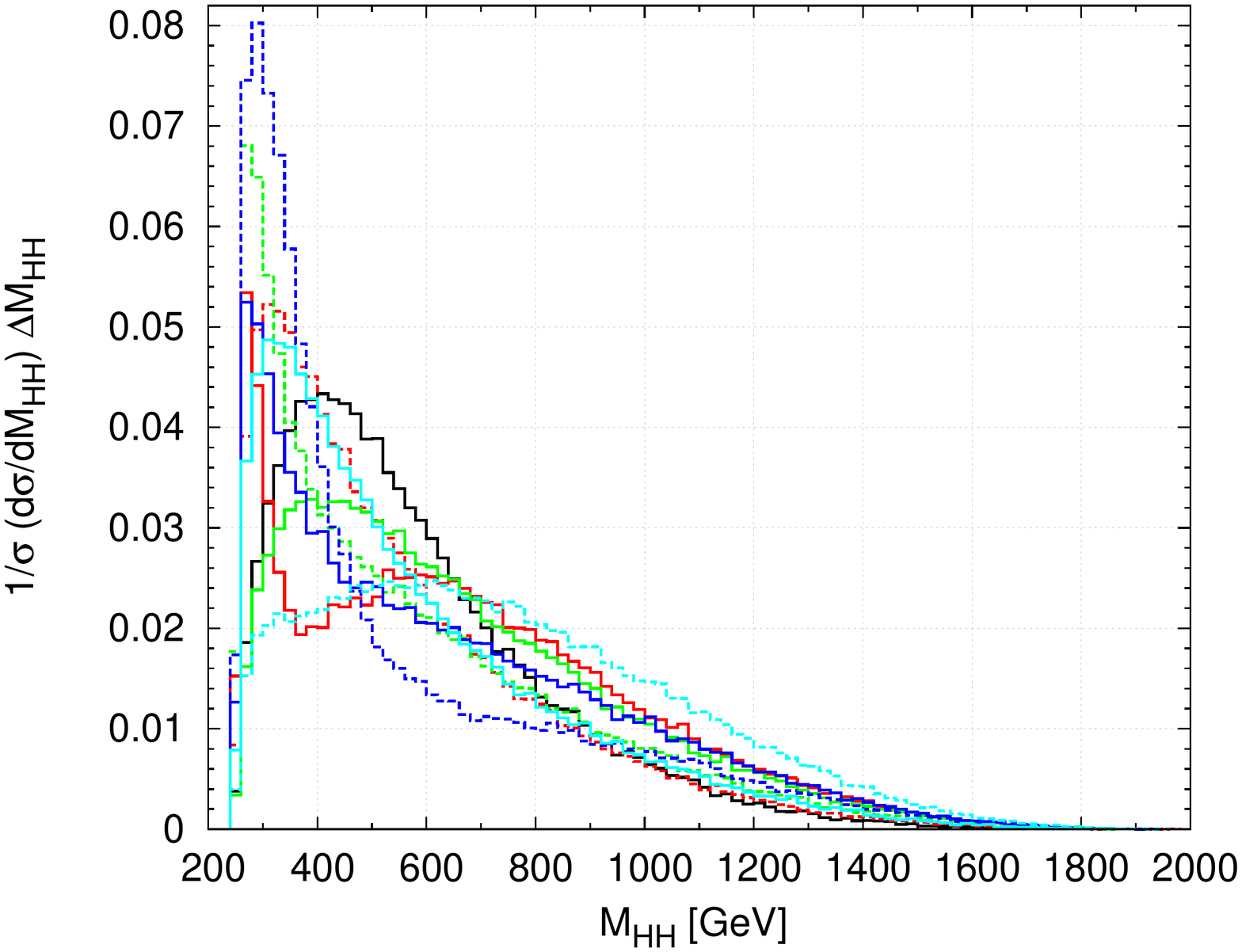}\\
\hspace{-5mm}
\vspace{-5mm}
\includegraphics[angle=0,width=80mm]{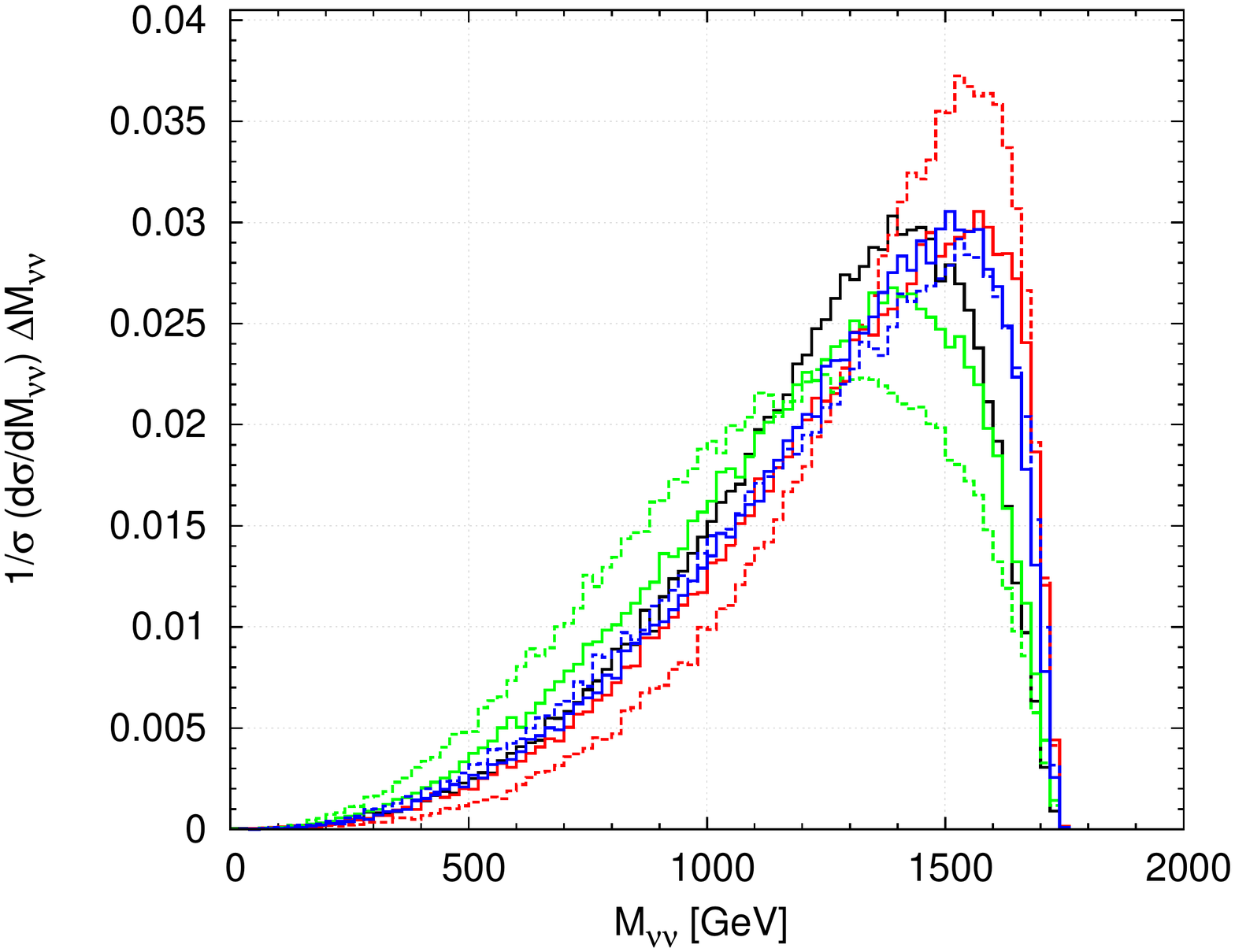} &
\hspace{-8mm}
\vspace{-5mm}
\includegraphics[angle=0,width=80mm]{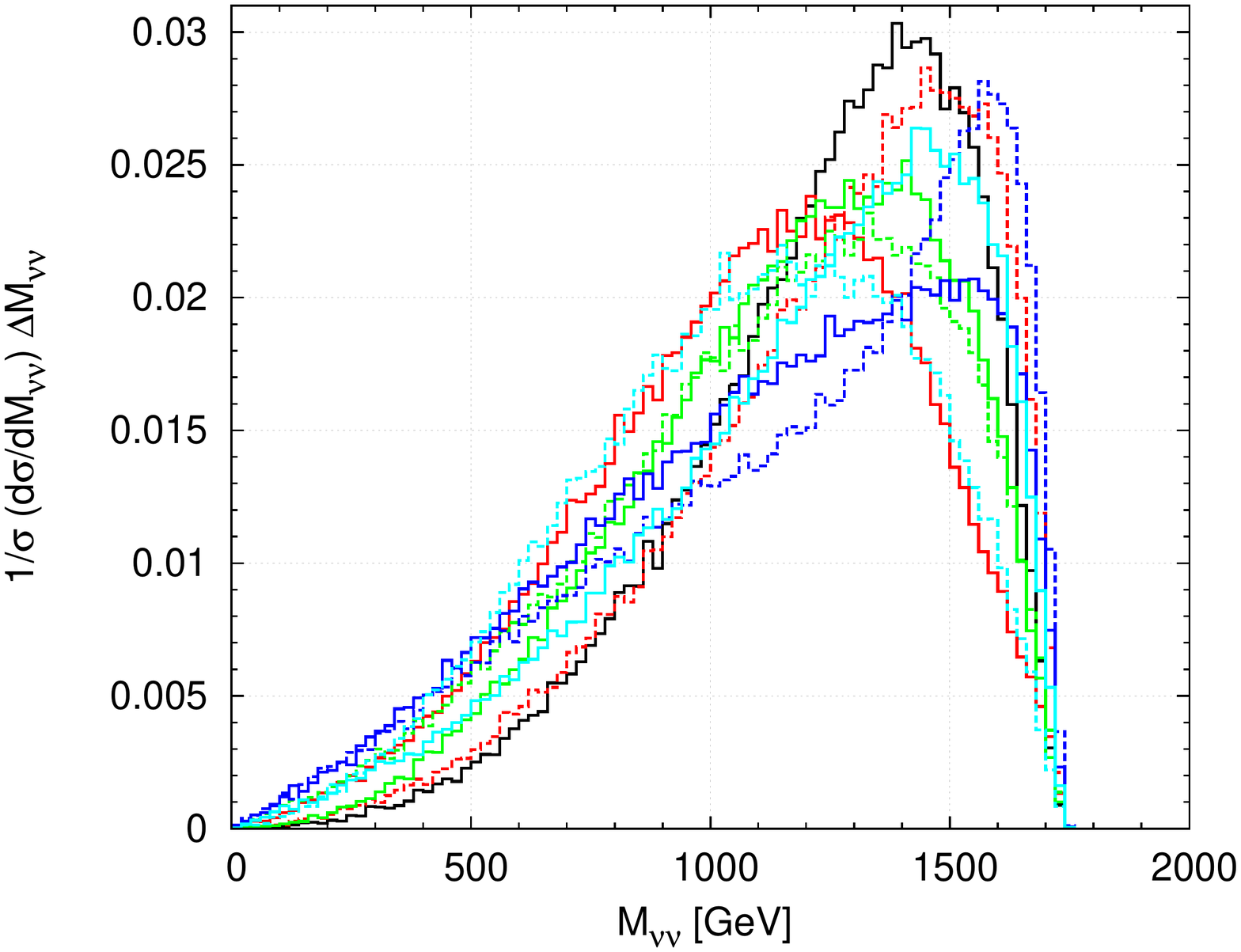}\\
\hspace{-8mm}
\vspace{-5mm}
\includegraphics[angle=0,width=80mm]{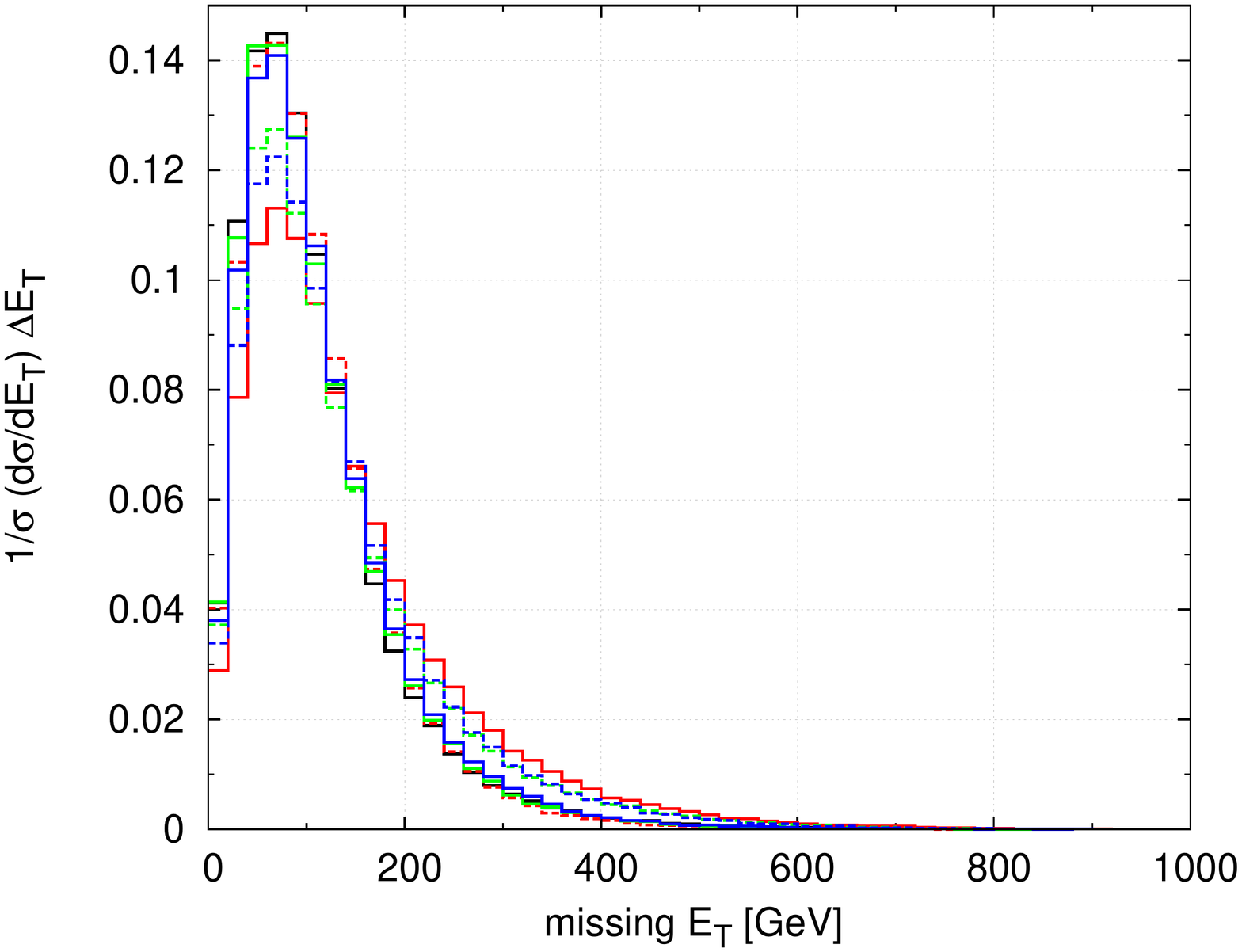} &
\hspace{-5mm}
\vspace{-5mm}
\includegraphics[angle=0,width=80mm]{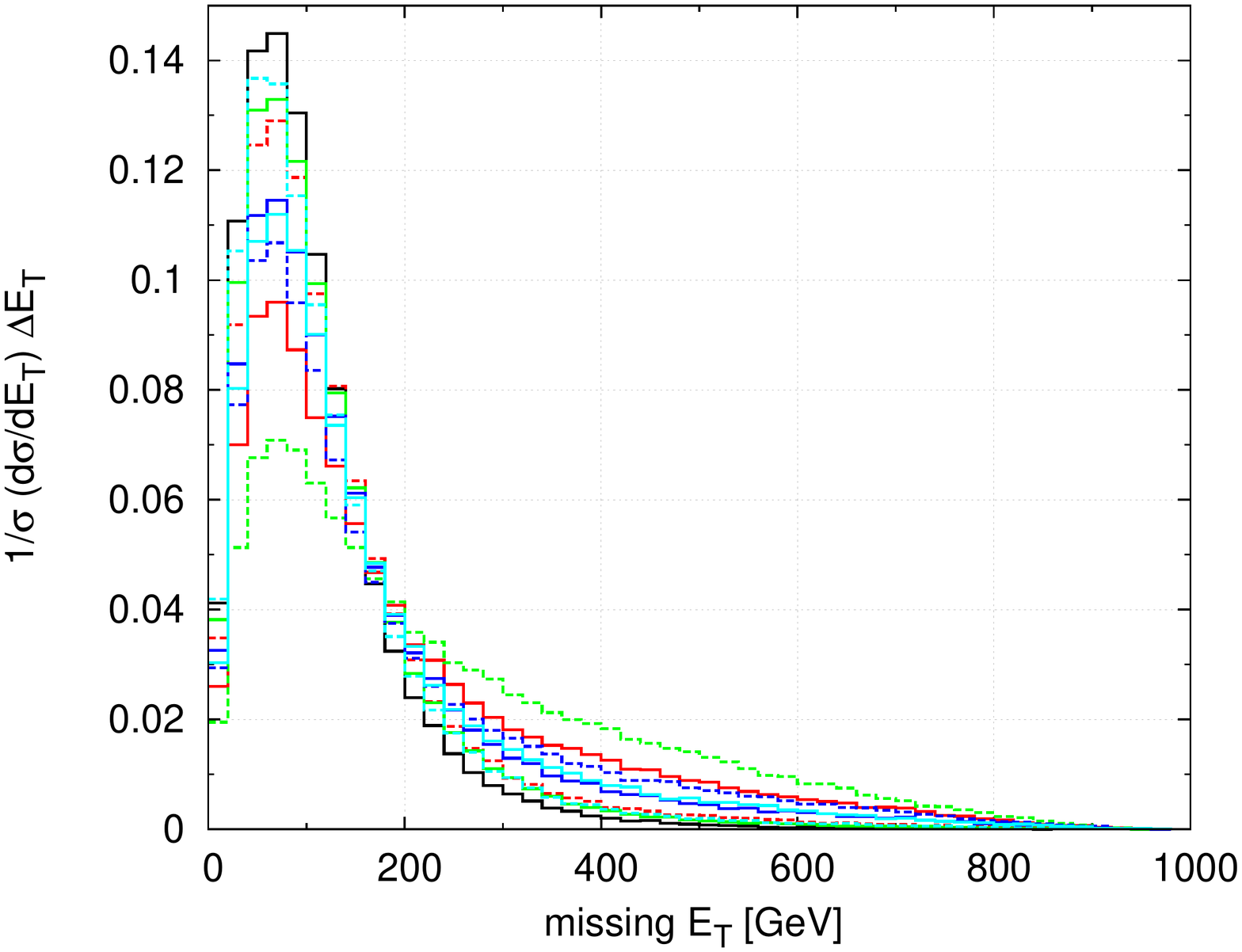}
\end{tabular}
%\vspace*{-7cm}
\caption{Kinematic distributions with  the anomalous coupling values as in the inset, illustrating how the presence of $c_W$ (first column), and  $c_{HW}$ and $c_{HB}$ (second column) affect the influence of $c_6$ and $c_H$. A centre of mass energy of $2000$ GeV is assumed.}
\label{fig:ctHH_Mhh_Mnn}
\end{figure}

Moving on to the kinematic distributions, we shall present the distributions of the opening angle between the two Higgs bosons is presented in Fig.\ref{fig:ctHH_Mhh_Mnn} (first row).  The effect of $c_W$, and $c_{HW}$ and $c_{HB}$ are presented separately in the first column and the second column, respectively. In both cases, the case with only $c_6$ and $c_H$ considered to be non-vanishing, and the SM case are presented for comparison. The dependence of the gauge-Higgs coupling on the sensitivity of  $HHH$ coupling is clear from the plots. The $HH$ invariant mass, as well as the missing invariant mass distributions,  also indicate a similar dependence, as presented in Fig.\ref{fig:ctHH_Mhh_Mnn} (second row) and (third row). On the other hand, the missing transverse energy distribution does not show much influence of the Higgs-gauge couplings on the sensitivity of $c_6$ and $c_H$. 

\section{Summary and Conclusions}\label{sec:summary}

The recent discovery of the Higgs boson at the LHC has established the Higgs mechanism as the way to have electroweak symmetry breaking, thus generating masses to all the particles. While the mass of the particle is more or less precisely measured, details like the strengths of its self interactions, its couplings with other particles like the gauge bosons, etc. need to be known precisely to understand and pinpoint the exact mechanism of electroweak symmetry breaking. Precise knowledge of the trilinear Higgs self-coupling, which is typically probed directly through processes involving two Higgs production, play a vital role in reconstructing the Higgs potential. Typically, such processes also involve other couplings from the Higgs sector, like the Higgs-gauge boson couplings. We consider the $ZHH$ and $\nu \bar\nu HH$ productions at the ILC to understand the influence of the $ZZH$ and $ZZHH$ couplings, in the first process, and $WWH$ and $WWHH$ couplings, on the second process, on the sensitivity of $HHH$ coupling on this process. Single and two parameter limits on the $c_6$ and $c_H$ couplings, which are related to the $HHH$ couplings, are considered in the case of the ILC with $\sqrt{s}=500$ GeV and integrated luminosity of 1000 fb$^{-1}$, to see how the other parameters, $c_W$, $c_{HW}$ and $c_{HB}$ influence the limits. It is seen that these latter parameters have significant influence of the reach of $c_6$ and $c_H$, indicating that prior, and somewhat precise knowledge of the Higgs-gauge coupling is necessary to draw any conclusion on the influence of trilinear couplings on the process considered. The kinematic distributions also indicate a strong influence of Higgs-gauge couplings, showing that, in the presence of very moderate Higgs-gauge couplings, it is difficult to extract reliable information regarding $c_6$ and $c_H$. A similar story is unfolded by considerations of $e^+e^-\rightarrow \nu \bar \nu HH$, where the influence of $WWH$ and $WWHH$ on the sensitivity of the trilinear Higgs self-coupling is explored. Concluding, one may need to rely on knowledge of the  Higgs gauge couplings from elsewhere, or consider clever observables eliminating or subduing their effects, in order to extract meaningful information regarding the trilinear Higgs couplings.

\vskip 5mm
\noindent
{\bf Acknowledgements} S.~K. would like to acknowledge the financial support from the SERB-DST, India, under the National Post-doctoral Fellowship programme, Grant No. PDF/2015/000167. R.~R. thanks Department of Science 
and Technology, Government of India for support through DST-INSPIRE Fellowship 
for doctoral program, INSPIRE CODE IF140075.
\\[2mm]

%%%%%%%%%%%%%%%%%%%%%%%%%%%%%%%%%%%%%%%%%%%%%%%%%%%%%%%
\bibliography{Bibliography}

\providecommand{\href}[2]{#2}\begingroup\raggedright\begin{thebibliography}{10}

\bibitem{Chatrchyan:2012xdj}
{\bfseries CMS} Collaboration, S.~Chatrchyan {\em et~al.}, {Observation of a
  new boson at a mass of 125 GeV with the CMS experiment at the LHC},
  \href{http://dx.doi.org/10.1016/j.physletb.2012.08.021}{{\em Phys. Lett.}
  {\bfseries B716} (2012) 30--61} CMS-HIG-12-028, CERN-PH-EP-2012-220,
\href{http://arxiv.org/abs/1207.7235}{{\ttfamily arXiv:1207.7235 [hep-ex]}}.
%%CITATION = ARXIV:1207.7235;%%.

\bibitem{Aad:2012tfa}
{\bfseries ATLAS} Collaboration, G.~Aad {\em et~al.}, {Observation of a new
  particle in the search for the Standard Model Higgs boson with the ATLAS
  detector at the LHC},
  \href{http://dx.doi.org/10.1016/j.physletb.2012.08.020}{{\em Phys. Lett.}
  {\bfseries B716} (2012) 1--29} CERN-PH-EP-2012-218,
\href{http://arxiv.org/abs/1207.7214}{{\ttfamily arXiv:1207.7214 [hep-ex]}}.
%%CITATION = ARXIV:1207.7214;%%.

\bibitem{ATLAS:2013xla}
{\bfseries ATLAS} Collaboration, {Study of the spin of the Higgs-like boson in
  the two photon decay channel using 20.7 fb$^{-1}$ of pp collisions collected
  at $\sqrt{s}$ = 8 TeV with the ATLAS detector}, {\em
  \href{http://cds.cern.ch/record/1527124/files/ATLAS-CONF-2013-029.pdf}{http://cds.cern.ch/record/1527124/files/ATLAS-CONF-2013-029.pdf}}
  (2013)
ATLAS-CONF-2013-029.
%%CITATION = ATLAS-CONF-2013-029;%%.

\bibitem{ATLAS:2013nma}
{\bfseries ATLAS} Collaboration, {Measurements of the properties of the
  Higgs-like boson in the four lepton decay channel with the ATLAS detector
  using 25 fb$^{-1}$ of proton-proton collision data}, {\em
  \href{http://cds.cern.ch/record/1523699/files/ATLAS-CONF-2013-013.pdf}{http://cds.cern.ch/record/1523699/files/ATLAS-CONF-2013-013.pdf}}
  (2013)
ATLAS-CONF-2013-013.
%%CITATION = ATLAS-CONF-2013-013;%%.

\bibitem{ATLAS:2013vla}
{\bfseries ATLAS} Collaboration, {Study of the spin properties of the
  Higgs-like particle in the $\boldsymbol{H\to WW^{(\ast)}\to e\nu\mu\nu}$
  channel with 21 fb$^{-1}$ of $\sqrt{s} = 8$ TeV data collected with the ATLAS
  detector.}, {\em
  \href{http://cds.cern.ch/record/1527127/files/ATLAS-CONF-2013-031.pdf}{http://cds.cern.ch/record/1527127/files/ATLAS-CONF-2013-031.pdf}}
  (2013)
ATLAS-CONF-2013-031.
%%CITATION = ATLAS-CONF-2013-031;%%.

\bibitem{CMS:xwa}
{\bfseries CMS} Collaboration, {Properties of the Higgs-like boson in the decay
  $H \to ZZ \to 4l$ in $pp$ collisions at $\sqrt{s}=7$ and $8$ TeV}, {\em
  \href{http://cds.cern.ch/record/1523767/files/HIG-13-002-pas.pdf}{http://cds.cern.ch/record/1523767/files/HIG-13-002-pas.pdf}}
  (2013)
CMS-PAS-HIG-13-002.
%%CITATION = CMS-PAS-HIG-13-002;%%.

\bibitem{CMS:bxa}
{\bfseries CMS} Collaboration, {Update on the search for the standard model
  Higgs boson in pp collisions at the LHC decaying to $W^+W^-$ in the fully
  leptonic final state}, {\em
  \href{http://cds.cern.ch/record/1523673/files/HIG-13-003-pas.pdf}{http://cds.cern.ch/record/1523673/files/HIG-13-003-pas.pdf}}
  (2013)
CMS-PAS-HIG-13-003.
%%CITATION = CMS-PAS-HIG-13-003;%%.

\bibitem{Aad:2013wqa}
{\bfseries ATLAS} Collaboration, G.~Aad {\em et~al.}, {Measurements of Higgs
  boson production and couplings in diboson final states with the ATLAS
  detector at the LHC}, \href{http://dx.doi.org/10.1016/j.physletb.2014.05.011,
  10.1016/j.physletb.2013.08.010}{{\em Phys. Lett.} {\bfseries B726} (2013)
  88--119} CERN-PH-EP-2013-103,
  \href{http://arxiv.org/abs/1307.1427}{{\ttfamily arXiv:1307.1427 [hep-ex]}}.
[Erratum: Phys. Lett.B734,406(2014)].
%%CITATION = ARXIV:1307.1427;%%.

\bibitem{Chatrchyan:2013lba}
{\bfseries CMS} Collaboration, S.~Chatrchyan {\em et~al.}, {Observation of a
  New Boson with Mass Near 125 GeV in $pp$ Collisions at $\sqrt{s}$ = 7 and 8
  TeV}, \href{http://dx.doi.org/10.1007/JHEP06(2013)081}{{\em JHEP} {\bfseries
  06} (2013) 081} CMS-HIG-12-036, CERN-PH-EP-2013-035,
\href{http://arxiv.org/abs/1303.4571}{{\ttfamily arXiv:1303.4571 [hep-ex]}}.
%%CITATION = ARXIV:1303.4571;%%.

\bibitem{ATLAS:2016oum}
{\bfseries ATLAS} Collaboration, {Study of the Higgs boson properties and
  search for high-mass scalar resonances in the $H \rightarrow ZZ^* \rightarrow
  4\ell$ decay channel at $\sqrt{s}$ = 13 TeV with the ATLAS detector}, {\em
  \href{https://atlas.web.cern.ch/Atlas/GROUPS/PHYSICS/CONFNOTES/ATLAS-CONF-2016-079}{https://atlas.web.cern.ch/Atlas/GROUPS/PHYSICS/CONFNOTES/ATLAS-CONF-2016-079}}
  (2016)
ATLAS-CONF-2016-079.
%%CITATION = ATLAS-CONF-2016-079;%%.

\bibitem{ATLAS:2016pkl}
{\bfseries ATLAS} Collaboration, {Search for the Standard Model Higgs boson
  produced in association with a vector boson and decaying to a $b\bar{b}$ pair
  in $pp$ collisions at 13 TeV using the ATLAS detector}, {\em
  \href{https://atlas.web.cern.ch/Atlas/GROUPS/PHYSICS/CONFNOTES/ATLAS-CONF-2016-091}{https://atlas.web.cern.ch/Atlas/GROUPS/PHYSICS/CONFNOTES/ATLAS-CONF-2016-091}}
  (2016)
ATLAS-CONF-2016-091.
%%CITATION = ATLAS-CONF-2016-091;%%.

\bibitem{ATLAS:2016zzs}
{\bfseries ATLAS} Collaboration, {Search for Higgs bosons decaying into di-muon
  in $pp$ collisions at $\sqrt{s}$ = 13 TeV with the ATLAS detector}, {\em
  \href{https://atlas.web.cern.ch/Atlas/GROUPS/PHYSICS/CONFNOTES/ATLAS-CONF-2016-041}{https://atlas.web.cern.ch/Atlas/GROUPS/PHYSICS/CONFNOTES/ATLAS-CONF-2016-041}}
  (2016)
ATLAS-CONF-2016-041.
%%CITATION = ATLAS-CONF-2016-041;%%.

\bibitem{Aad:2015vsa}
{\bfseries ATLAS} Collaboration, G.~Aad {\em et~al.}, {Evidence for the
  Higgs-boson Yukawa coupling to tau leptons with the ATLAS detector},
  \href{http://dx.doi.org/10.1007/JHEP04(2015)117}{{\em JHEP} {\bfseries 04}
  (2015) 117} CERN-PH-EP-2014-262,
\href{http://arxiv.org/abs/1501.04943}{{\ttfamily arXiv:1501.04943 [hep-ex]}}.
%%CITATION = ARXIV:1501.04943;%%.

\bibitem{ATLAS:2016hru}
{\bfseries ATLAS} Collaboration, {Combined measurements of the Higgs boson
  production and decay rates in $H\to ZZ^*\to 4\ell$ and $H\to\gamma\gamma$
  final states using $pp$ collision data at $\sqrt{s}=$ 13 TeV in the ATLAS
  experiment}, {\em
  \href{https://atlas.web.cern.ch/Atlas/GROUPS/PHYSICS/CONFNOTES/ATLAS-CONF-2016-081}{https://atlas.web.cern.ch/Atlas/GROUPS/PHYSICS/CONFNOTES/ATLAS-CONF-2016-081}}
  (2016)
ATLAS-CONF-2016-081.
%%CITATION = ATLAS-CONF-2016-081;%%.

\bibitem{Aad:2015mxa}
{\bfseries ATLAS} Collaboration, G.~Aad {\em et~al.}, {Study of the spin and
  parity of the Higgs boson in diboson decays with the ATLAS detector},
  \href{http://dx.doi.org/10.1140/epjc/s10052-015-3685-1,
  10.1140/epjc/s10052-016-3934-y}{{\em Eur. Phys. J.} {\bfseries C75} no.~10,
  (2015) 476} CERN-PH-EP-2015-114,
  \href{http://arxiv.org/abs/1506.05669}{{\ttfamily arXiv:1506.05669
  [hep-ex]}}.
[Erratum: Eur. Phys. J.C76,no.3,152(2016)].
%%CITATION = ARXIV:1506.05669;%%.

\bibitem{Degrassi:2012ry}
G.~Degrassi, S.~Di~Vita, J.~Elias-Miro, J.~R. Espinosa, G.~F. Giudice,
  G.~Isidori, and A.~Strumia, {Higgs mass and vacuum stability in the Standard
  Model at NNLO}, \href{http://dx.doi.org/10.1007/JHEP08(2012)098}{{\em JHEP}
  {\bfseries 08} (2012) 098} CERN-PH-TH-2012-134, RM3-TH-12-9,
\href{http://arxiv.org/abs/1205.6497}{{\ttfamily arXiv:1205.6497 [hep-ph]}}.
%%CITATION = ARXIV:1205.6497;%%.

\bibitem{BrauJames:2007aa}
{\bfseries ILC} Collaboration, G.~Aarons {\em et~al.}, {ILC Reference Design
  Report Volume 1 - Executive Summary}, FERMILAB-DESIGN-2007-03,
  FERMILAB-PUB-07-794-E,
\href{http://arxiv.org/abs/0712.1950}{{\ttfamily arXiv:0712.1950
  [physics.acc-ph]}}.
%%CITATION = ARXIV:0712.1950;%%.

\bibitem{Djouadi:2007ik}
{\bfseries ILC} Collaboration, G.~Aarons {\em et~al.}, {International Linear
  Collider Reference Design Report Volume 2: Physics at the ILC}, SLAC-R-975,
  FERMILAB-DESIGN-2007-04, FERMILAB-PUB-07-795-E,
\href{http://arxiv.org/abs/0709.1893}{{\ttfamily arXiv:0709.1893 [hep-ph]}}.
%%CITATION = ARXIV:0709.1893;%%.

\bibitem{MoortgatPick:2005cw}
G.~Moortgat-Pick {\em et~al.}, {The Role of polarized positrons and electrons
  in revealing fundamental interactions at the linear collider},
  \href{http://dx.doi.org/10.1016/j.physrep.2007.12.003}{{\em Phys. Rept.}
  {\bfseries 460} (2008) 131--243} CERN-PH-TH-2005-036, DCPT-04-100,
  DESY-05-059, FERMILAB-PUB-05-060-T, IPPP-04-50, KEK-2005-16, PRL-TH-05-01,
  SHEP-05-03, SLAC-PUB-11087,
\href{http://arxiv.org/abs/hep-ph/0507011}{{\ttfamily arXiv:hep-ph/0507011
  [hep-ph]}}.
%%CITATION = HEP-PH/0507011;%%.

\bibitem{Ananthanarayan:2013cia}
B.~Ananthanarayan, S.~K. Garg, J.~Lahiri, and P.~Poulose, {Probing the
  indefinite CP nature of the Higgs boson through decay distributions in the
  process $e^+e^- \to t\overline{t}\Phi$},
  \href{http://dx.doi.org/10.1103/PhysRevD.87.114002}{{\em Phys. Rev.}
  {\bfseries D87} no.~11, (2013) 114002},
\href{http://arxiv.org/abs/1304.4414}{{\ttfamily arXiv:1304.4414 [hep-ph]}}.
%%CITATION = ARXIV:1304.4414;%%.

\bibitem{Ananthanarayan:2014eea}
B.~Ananthanarayan, S.~K. Garg, C.~S. Kim, J.~Lahiri, and P.~Poulose, {Top
  Yukawa coupling measurement with indefinite CP Higgs in $e^+e^-\to
  t\bar{t}\Phi$}, \href{http://dx.doi.org/10.1103/PhysRevD.90.014016}{{\em
  Phys. Rev.} {\bfseries D90} no.~1, (2014) 014016},
\href{http://arxiv.org/abs/1405.6465}{{\ttfamily arXiv:1405.6465 [hep-ph]}}.
%%CITATION = ARXIV:1405.6465;%%.

\bibitem{Muhlleitner:2012jy}
M.~Muhlleitner, R.~M. Godbole, C.~Hangst, S.~D. Rindani, and P.~Sharma,
  {Analysis of Higgs spin and CP properties in a model-independent way in $e^+
  e^- \to t\bar{t} \Phi$},
{\em Frascati Phys. Ser.} {\bfseries 54} (2012) 188--197.
%%CITATION = 00309,54,188;%%.

\bibitem{Godbole:2011hw}
R.~M. Godbole, C.~Hangst, M.~Muhlleitner, S.~D. Rindani, and P.~Sharma,
  {Model-independent analysis of Higgs spin and CP properties in the process
  $e^+ e^- \to t \bar{t} \Phi$},
  \href{http://dx.doi.org/10.1140/epjc/s10052-011-1681-7}{{\em Eur. Phys. J.}
  {\bfseries C71} (2011) 1681},
\href{http://arxiv.org/abs/1103.5404}{{\ttfamily arXiv:1103.5404 [hep-ph]}}.
%%CITATION = ARXIV:1103.5404;%%.

\bibitem{Weinberg:1978kz}
S.~Weinberg, {Phenomenological Lagrangians},
  \href{http://dx.doi.org/10.1016/0378-4371(79)90223-1}{{\em Physica}
  {\bfseries A96} no.~1-2, (1979) 327--340}
HUTP-78-A051A.
%%CITATION = PHYSA,A96,327;%%.

\bibitem{Weinberg:1980wa}
S.~Weinberg, {Effective Gauge Theories},
  \href{http://dx.doi.org/10.1016/0370-2693(80)90660-7}{{\em Phys. Lett.}
  {\bfseries 91B} (1980) 51--55}
HUTP-80/A001.
%%CITATION = PHLTA,91B,51;%%.

\bibitem{Georgi:1994qn}
H.~Georgi, {Effective field theory},
\href{http://dx.doi.org/10.1146/annurev.ns.43.120193.001233}{{\em Ann. Rev.
  Nucl. Part. Sci.} {\bfseries 43} (1993) 209--252}.
%%CITATION = ARNUA,43,209;%%.

\bibitem{Buchmuller:1985jz}
W.~Buchmuller and D.~Wyler, {Effective Lagrangian Analysis of New Interactions
  and Flavor Conservation},
  \href{http://dx.doi.org/10.1016/0550-3213(86)90262-2}{{\em Nucl. Phys.}
  {\bfseries B268} (1986) 621--653}
CERN-TH-4254/85.
%%CITATION = NUPHA,B268,621;%%.

\bibitem{Hagiwara:1993ck}
K.~Hagiwara, S.~Ishihara, R.~Szalapski, and D.~Zeppenfeld, {Low-energy effects
  of new interactions in the electroweak boson sector},
  \href{http://dx.doi.org/10.1103/PhysRevD.48.2182}{{\em Phys. Rev.} {\bfseries
  D48} (1993) 2182--2203}
MAD-PH-737, UT-635, KEK-TH-356, KEK-PREPRINT-92-214.
%%CITATION = PHRVA,D48,2182;%%.

\bibitem{Hagiwara:1993qt}
K.~Hagiwara, R.~Szalapski, and D.~Zeppenfeld, {Anomalous Higgs boson production
  and decay}, \href{http://dx.doi.org/10.1016/0370-2693(93)91799-S}{{\em Phys.
  Lett.} {\bfseries B318} (1993) 155--162} MAD-PH-783, KEK-TH-370,
\href{http://arxiv.org/abs/hep-ph/9308347}{{\ttfamily arXiv:hep-ph/9308347
  [hep-ph]}}.
%%CITATION = HEP-PH/9308347;%%.

\bibitem{Alam:1997nk}
S.~Alam, S.~Dawson, and R.~Szalapski, {Low-energy constraints on new physics
  revisited}, \href{http://dx.doi.org/10.1103/PhysRevD.57.1577}{{\em Phys.
  Rev.} {\bfseries D57} (1998) 1577--1590} KEK-TH-519, KEK-PREPRINT-97-88,
  BNL-HET-SD-97-003,
\href{http://arxiv.org/abs/hep-ph/9706542}{{\ttfamily arXiv:hep-ph/9706542
  [hep-ph]}}.
%%CITATION = HEP-PH/9706542;%%.

\bibitem{Barger:2003rs}
V.~Barger, T.~Han, P.~Langacker, B.~McElrath, and P.~Zerwas, {Effects of
  genuine dimension-six Higgs operators},
  \href{http://dx.doi.org/10.1103/PhysRevD.67.115001}{{\em Phys. Rev.}
  {\bfseries D67} (2003) 115001} MADPH-02-1303, UPR-1007-T, DESY-02-222,
\href{http://arxiv.org/abs/hep-ph/0301097}{{\ttfamily arXiv:hep-ph/0301097
  [hep-ph]}}.
%%CITATION = HEP-PH/0301097;%%.

\bibitem{Giudice:2007fh}
G.~F. Giudice, C.~Grojean, A.~Pomarol, and R.~Rattazzi, {The
  Strongly-Interacting Light Higgs},
  \href{http://dx.doi.org/10.1088/1126-6708/2007/06/045}{{\em JHEP} {\bfseries
  06} (2007) 045} CERN-PH-TH-2007-47,
\href{http://arxiv.org/abs/hep-ph/0703164}{{\ttfamily arXiv:hep-ph/0703164
  [hep-ph]}}.
%%CITATION = HEP-PH/0703164;%%.

\bibitem{Grzadkowski:2010es}
B.~Grzadkowski, M.~Iskrzynski, M.~Misiak, and J.~Rosiek, {Dimension-Six Terms
  in the Standard Model Lagrangian},
  \href{http://dx.doi.org/10.1007/JHEP10(2010)085}{{\em JHEP} {\bfseries 10}
  (2010) 085} IFT-9-2010, TTP10-35,
\href{http://arxiv.org/abs/1008.4884}{{\ttfamily arXiv:1008.4884 [hep-ph]}}.
%%CITATION = ARXIV:1008.4884;%%.

\bibitem{Contino:2010rs}
R.~Contino, \href{http://dx.doi.org/10.1142/9789814327183_0005}{{The Higgs as a
  Composite Nambu-Goldstone Boson},} in {\em {Physics of the large and the
  small, TASI 09, proceedings of the Theoretical Advanced Study Institute in
  Elementary Particle Physics, Boulder, Colorado, USA, 1-26 June 2009}},
  pp.~235--306.
\newblock 2011.
\newblock
\href{http://arxiv.org/abs/1005.4269}{{\ttfamily arXiv:1005.4269 [hep-ph]}}.
\newblock
%%CITATION = ARXIV:1005.4269;%%.

\bibitem{GutierrezRodriguez:2011gi}
A.~Gutierrez-Rodriguez, J.~Peressutti, and O.~A. Sampayo, {Higgs Boson
  Self-Coupling at High Energy $\gamma \gamma$ Collider},
  \href{http://dx.doi.org/10.1088/0954-3899/38/9/095002}{{\em J. Phys.}
  {\bfseries G38} (2011) 095002},
\href{http://arxiv.org/abs/1107.0245}{{\ttfamily arXiv:1107.0245 [hep-ph]}}.
%%CITATION = ARXIV:1107.0245;%%.

\bibitem{GutierrezRodriguez:2009uz}
A.~Gutierrez-Rodriguez, M.~A. Hernandez-Ruiz, and O.~A. Sampayo, {Neutral Higgs
  Boson Pair-Production and Trilinear Self-Couplings in the MSSM at ILC and
  CLIC Energies}, \href{http://dx.doi.org/10.1142/S0217751X09044784}{{\em Int.
  J. Mod. Phys.} {\bfseries A24} (2009) 5299--5318},
\href{http://arxiv.org/abs/0903.1383}{{\ttfamily arXiv:0903.1383 [hep-ph]}}.
%%CITATION = ARXIV:0903.1383;%%.

\bibitem{GutierrezRodriguez:2005fe}
A.~Gutierrez-Rodriguez, M.~A. Hernandez-Ruiz, and O.~A. Sampayo,
  {Pairs-production of higgs in association with bottom quarks pairs at
  $e^+e^-$ colliders}, \href{http://dx.doi.org/10.1142/S0217732305017548}{{\em
  Mod. Phys. Lett.} {\bfseries A20} (2005) 2629--2638},
\href{http://arxiv.org/abs/hep-ph/0504266}{{\ttfamily arXiv:hep-ph/0504266
  [hep-ph]}}.
%%CITATION = HEP-PH/0504266;%%.

\bibitem{Rindani:2010pi}
S.~D. Rindani and P.~Sharma, {Decay-lepton correlations as probes of anomalous
  ZZH and gammaZH interactions in $e^+e^- \to ZH$ with polarized beams},
  \href{http://dx.doi.org/10.1016/j.physletb.2010.08.027}{{\em Phys. Lett.}
  {\bfseries B693} (2010) 134--139},
\href{http://arxiv.org/abs/1001.4931}{{\ttfamily arXiv:1001.4931 [hep-ph]}}.
%%CITATION = ARXIV:1001.4931;%%.

\bibitem{Rindani:2009pb}
S.~D. Rindani and P.~Sharma, {Angular distributions as a probe of anomalous ZZH
  and gammaZH interactions at a linear collider with polarized beams},
  \href{http://dx.doi.org/10.1103/PhysRevD.79.075007}{{\em Phys. Rev.}
  {\bfseries D79} (2009) 075007},
\href{http://arxiv.org/abs/0901.2821}{{\ttfamily arXiv:0901.2821 [hep-ph]}}.
%%CITATION = ARXIV:0901.2821;%%.

\bibitem{Baak:2012kk}
M.~Baak, M.~Goebel, J.~Haller, A.~Hoecker, D.~Kennedy, R.~Kogler, K.~Moenig,
  M.~Schott, and J.~Stelzer, {The Electroweak Fit of the Standard Model after
  the Discovery of a New Boson at the LHC},
  \href{http://dx.doi.org/10.1140/epjc/s10052-012-2205-9}{{\em Eur. Phys. J.}
  {\bfseries C72} (2012) 2205} DESY-12-154,
\href{http://arxiv.org/abs/1209.2716}{{\ttfamily arXiv:1209.2716 [hep-ph]}}.
%%CITATION = ARXIV:1209.2716;%%.

\bibitem{Einhorn:2013kja}
M.~B. Einhorn and J.~Wudka, {The Bases of Effective Field Theories},
  \href{http://dx.doi.org/10.1016/j.nuclphysb.2013.08.023}{{\em Nucl. Phys.}
  {\bfseries B876} (2013) 556--574} UCRHEP-T529, NSF-ITP-13-115, MCTP-13-18,
\href{http://arxiv.org/abs/1307.0478}{{\ttfamily arXiv:1307.0478 [hep-ph]}}.
%%CITATION = ARXIV:1307.0478;%%.

\bibitem{Contino:2013kra}
R.~Contino, M.~Ghezzi, C.~Grojean, M.~Muhlleitner, and M.~Spira, {Effective
  Lagrangian for a light Higgs-like scalar},
  \href{http://dx.doi.org/10.1007/JHEP07(2013)035}{{\em JHEP} {\bfseries 07}
  (2013) 035} CERN-PH-TH-2013-047, KA-TP-06-2013, PSI-PR-13-04,
\href{http://arxiv.org/abs/1303.3876}{{\ttfamily arXiv:1303.3876 [hep-ph]}}.
%%CITATION = ARXIV:1303.3876;%%.

\bibitem{Amar:2014fpa}
G.~Amar, S.~Banerjee, S.~von Buddenbrock, A.~S. Cornell, T.~Mandal, B.~Mellado,
  and B.~Mukhopadhyaya, {Exploration of the tensor structure of the Higgs boson
  coupling to weak bosons in $e^+e^-$ collisions},
  \href{http://dx.doi.org/10.1007/JHEP02(2015)128}{{\em JHEP} {\bfseries 02}
  (2015) 128} HRI-RECAPP-2014-011, WITS-CTP-135,
\href{http://arxiv.org/abs/1405.3957}{{\ttfamily arXiv:1405.3957 [hep-ph]}}.
%%CITATION = ARXIV:1405.3957;%%.

\bibitem{Masso:2014xra}
E.~Masso, {An Effective Guide to Beyond the Standard Model Physics},
  \href{http://dx.doi.org/10.1007/JHEP10(2014)128}{{\em JHEP} {\bfseries 10}
  (2014) 128},
\href{http://arxiv.org/abs/1406.6376}{{\ttfamily arXiv:1406.6376 [hep-ph]}}.
%%CITATION = ARXIV:1406.6376;%%.

\bibitem{Biekoetter:2014jwa}
A.~Biekötter, A.~Knochel, M.~Krämer, D.~Liu, and F.~Riva, {Vices and virtues
  of Higgs effective field theories at large energy},
  \href{http://dx.doi.org/10.1103/PhysRevD.91.055029}{{\em Phys. Rev.}
  {\bfseries D91} (2015) 055029},
\href{http://arxiv.org/abs/1406.7320}{{\ttfamily arXiv:1406.7320 [hep-ph]}}.
%%CITATION = ARXIV:1406.7320;%%.

\bibitem{Willenbrock:2014bja}
S.~Willenbrock and C.~Zhang, {Effective Field Theory Beyond the Standard
  Model}, \href{http://dx.doi.org/10.1146/annurev-nucl-102313-025623}{{\em Ann.
  Rev. Nucl. Part. Sci.} {\bfseries 64} (2014) 83--100} CP3-14-02,
\href{http://arxiv.org/abs/1401.0470}{{\ttfamily arXiv:1401.0470 [hep-ph]}}.
%%CITATION = ARXIV:1401.0470;%%.

\bibitem{Bonnet:2011yx}
F.~Bonnet, M.~B. Gavela, T.~Ota, and W.~Winter, {Anomalous Higgs couplings at
  the LHC, and their theoretical interpretation},
  \href{http://dx.doi.org/10.1103/PhysRevD.85.035016}{{\em Phys. Rev.}
  {\bfseries D85} (2012) 035016} FTUAM-11-40, IFT-UAM-CSIC-11-10, MPP-2011-54,
\href{http://arxiv.org/abs/1105.5140}{{\ttfamily arXiv:1105.5140 [hep-ph]}}.
%%CITATION = ARXIV:1105.5140;%%.

\bibitem{Corbett:2012dm}
T.~Corbett, O.~J.~P. Eboli, J.~Gonzalez-Fraile, and M.~C. Gonzalez-Garcia,
  {Constraining anomalous Higgs interactions},
  \href{http://dx.doi.org/10.1103/PhysRevD.86.075013}{{\em Phys. Rev.}
  {\bfseries D86} (2012) 075013},
\href{http://arxiv.org/abs/1207.1344}{{\ttfamily arXiv:1207.1344 [hep-ph]}}.
%%CITATION = ARXIV:1207.1344;%%.

\bibitem{Chang:2013cia}
W.-F. Chang, W.-P. Pan, and F.~Xu, {Effective gauge-Higgs operators analysis of
  new physics associated with the Higgs boson},
  \href{http://dx.doi.org/10.1103/PhysRevD.88.033004}{{\em Phys. Rev.}
  {\bfseries D88} no.~3, (2013) 033004},
\href{http://arxiv.org/abs/1303.7035}{{\ttfamily arXiv:1303.7035 [hep-ph]}}.
%%CITATION = ARXIV:1303.7035;%%.

\bibitem{Elias-Miro:2013mua}
J.~Elias-Miro, J.~R. Espinosa, E.~Masso, and A.~Pomarol, {Higgs windows to new
  physics through d=6 operators: constraints and one-loop anomalous
  dimensions}, \href{http://dx.doi.org/10.1007/JHEP11(2013)066}{{\em JHEP}
  {\bfseries 11} (2013) 066},
\href{http://arxiv.org/abs/1308.1879}{{\ttfamily arXiv:1308.1879 [hep-ph]}}.
%%CITATION = ARXIV:1308.1879;%%.

\bibitem{Banerjee:2013apa}
S.~Banerjee, S.~Mukhopadhyay, and B.~Mukhopadhyaya, {Higher dimensional
  operators and the LHC Higgs data: The role of modified kinematics},
  \href{http://dx.doi.org/10.1103/PhysRevD.89.053010}{{\em Phys. Rev.}
  {\bfseries D89} no.~5, (2014) 053010} IPMU13-0160, RECAPP-HRI-2013-018,
\href{http://arxiv.org/abs/1308.4860}{{\ttfamily arXiv:1308.4860 [hep-ph]}}.
%%CITATION = ARXIV:1308.4860;%%.

\bibitem{Boos:2013mqa}
E.~Boos, V.~Bunichev, M.~Dubinin, and Y.~Kurihara, {Higgs boson signal at
  complete tree level in the SM extension by dimension-six operators},
  \href{http://dx.doi.org/10.1103/PhysRevD.89.035001}{{\em Phys. Rev.}
  {\bfseries D89} (2014) 035001} SINP-MSU-2013-2-885,
\href{http://arxiv.org/abs/1309.5410}{{\ttfamily arXiv:1309.5410 [hep-ph]}}.
%%CITATION = ARXIV:1309.5410;%%.

\bibitem{Masso:2012eq}
E.~Massó and V.~Sanz, {Limits on anomalous couplings of the Higgs boson to
  electroweak gauge bosons from LEP and the LHC},
  \href{http://dx.doi.org/10.1103/PhysRevD.87.033001}{{\em Phys. Rev.}
  {\bfseries D87} no.~3, (2013) 033001} CERN-PH-TH-2012-298,
\href{http://arxiv.org/abs/1211.1320}{{\ttfamily arXiv:1211.1320 [hep-ph]}}.
%%CITATION = ARXIV:1211.1320;%%.

\bibitem{Han:2004az}
Z.~Han and W.~Skiba, {Effective theory analysis of precision electroweak data},
  \href{http://dx.doi.org/10.1103/PhysRevD.71.075009}{{\em Phys. Rev.}
  {\bfseries D71} (2005) 075009},
\href{http://arxiv.org/abs/hep-ph/0412166}{{\ttfamily arXiv:hep-ph/0412166
  [hep-ph]}}.
%%CITATION = HEP-PH/0412166;%%.

\bibitem{Corbett:2012ja}
T.~Corbett, O.~J.~P. Eboli, J.~Gonzalez-Fraile, and M.~C. Gonzalez-Garcia,
  {Robust Determination of the Higgs Couplings: Power to the Data},
  \href{http://dx.doi.org/10.1103/PhysRevD.87.015022}{{\em Phys. Rev.}
  {\bfseries D87} (2013) 015022} YITP-SB-12-42,
\href{http://arxiv.org/abs/1211.4580}{{\ttfamily arXiv:1211.4580 [hep-ph]}}.
%%CITATION = ARXIV:1211.4580;%%.

\bibitem{Dumont:2013wma}
B.~Dumont, S.~Fichet, and G.~von Gersdorff, {A Bayesian view of the Higgs
  sector with higher dimensional operators},
  \href{http://dx.doi.org/10.1007/JHEP07(2013)065}{{\em JHEP} {\bfseries 07}
  (2013) 065} CPHT-RR025.0413, ICTP-SAIFR-2013-005, LPSC13097,
\href{http://arxiv.org/abs/1304.3369}{{\ttfamily arXiv:1304.3369 [hep-ph]}}.
%%CITATION = ARXIV:1304.3369;%%.

\bibitem{Pomarol:2013zra}
A.~Pomarol and F.~Riva, {Towards the Ultimate SM Fit to Close in on Higgs
  Physics}, \href{http://dx.doi.org/10.1007/JHEP01(2014)151}{{\em JHEP}
  {\bfseries 01} (2014) 151},
\href{http://arxiv.org/abs/1308.2803}{{\ttfamily arXiv:1308.2803 [hep-ph]}}.
%%CITATION = ARXIV:1308.2803;%%.

\bibitem{Ellis:2014dva}
J.~Ellis, V.~Sanz, and T.~You, {Complete Higgs Sector Constraints on
  Dimension-6 Operators}, \href{http://dx.doi.org/10.1007/JHEP07(2014)036}{{\em
  JHEP} {\bfseries 07} (2014) 036} KCL-PH-TH-2014-15, LCTS-2014-14,
  CERN-PH-TH-2014-061,
\href{http://arxiv.org/abs/1404.3667}{{\ttfamily arXiv:1404.3667 [hep-ph]}}.
%%CITATION = ARXIV:1404.3667;%%.

\bibitem{Belusca-Maito:2014dpa}
H.~Belusca-Maito, {Effective Higgs Lagrangian and Constraints on Higgs
  Couplings}, LPT-Orsay-14-22,
\href{http://arxiv.org/abs/1404.5343}{{\ttfamily arXiv:1404.5343 [hep-ph]}}.
%%CITATION = ARXIV:1404.5343;%%.

\bibitem{Gupta:2014rxa}
R.~S. Gupta, A.~Pomarol, and F.~Riva, {BSM Primary Effects},
  \href{http://dx.doi.org/10.1103/PhysRevD.91.035001}{{\em Phys. Rev.}
  {\bfseries D91} no.~3, (2015) 035001},
\href{http://arxiv.org/abs/1405.0181}{{\ttfamily arXiv:1405.0181 [hep-ph]}}.
%%CITATION = ARXIV:1405.0181;%%.

\bibitem{Ellis:2017kfi}
J.~Ellis, P.~Roloff, V.~Sanz, and T.~You, {Dimension-6 Operator Analysis of the
  CLIC Sensitivity to New Physics},
  \href{http://dx.doi.org/10.1007/JHEP05(2017)096}{{\em JHEP} {\bfseries 05}
  (2017) 096} KCL-PH-TH-2017-04, CERN-TH-2017-009, CAVENDISH-HEP-17-01,
  CERN-PH-TH-2017-009, DAMTP-2017-01,
\href{http://arxiv.org/abs/1701.04804}{{\ttfamily arXiv:1701.04804 [hep-ph]}}.
%%CITATION = ARXIV:1701.04804;%%.

\bibitem{Denizli:2017pyu}
H.~Denizli and A.~Senol, {Constraints on Higgs effective couplings in $H\nu
  \bar{\nu}$ production of CLIC at 380 GeV},
  \href{http://dx.doi.org/10.1155/2018/1627051}{{\em Adv. High Energy Phys.}
  {\bfseries 2018} (2018) 1627051},
\href{http://arxiv.org/abs/1707.03890}{{\ttfamily arXiv:1707.03890 [hep-ph]}}.
%%CITATION = ARXIV:1707.03890;%%.

\bibitem{DiVita:2017vrr}
S.~Di~Vita, G.~Durieux, C.~Grojean, J.~Gu, Z.~Liu, G.~Panico, M.~Riembau, and
  T.~Vantalon, {A global view on the Higgs self-coupling at lepton colliders},
  \href{http://dx.doi.org/10.1007/JHEP02(2018)178}{{\em JHEP} {\bfseries 02}
  (2018) 178} DESY-17-131, FERMILAB-PUB-17-462-T,
\href{http://arxiv.org/abs/1711.03978}{{\ttfamily arXiv:1711.03978 [hep-ph]}}.
%%CITATION = ARXIV:1711.03978;%%.

\bibitem{Ellis:2018gqa}
J.~Ellis, C.~W. Murphy, V.~Sanz, and T.~You, {Updated Global SMEFT Fit to
  Higgs, Diboson and Electroweak Data},
  \href{http://dx.doi.org/10.1007/JHEP06(2018)146}{{\em JHEP} {\bfseries 06}
  (2018) 146} Cavendish-HEP-2018-06, DAMTP-2018-12, KCL-PH-TH/2018-12,
  CERN-PH-TH/2018-042, CERN-TH-2018-042,
\href{http://arxiv.org/abs/1803.03252}{{\ttfamily arXiv:1803.03252 [hep-ph]}}.
%%CITATION = ARXIV:1803.03252;%%.

\bibitem{Liu:2018peg}
T.~Liu, K.-F. Lyu, J.~Ren, and H.~X. Zhu, {Probing the quartic Higgs boson
  self-interaction}, \href{http://dx.doi.org/10.1103/PhysRevD.98.093004}{{\em
  Phys. Rev.} {\bfseries D98} no.~9, (2018) 093004},
\href{http://arxiv.org/abs/1803.04359}{{\ttfamily arXiv:1803.04359 [hep-ph]}}.
%%CITATION = ARXIV:1803.04359;%%.

\bibitem{Rindani:2018ubx}
S.~D. Rindani and B.~Singh, {Indirect measurement of triple-Higgs coupling at
  an electron-positron collider with polarized beams},
\href{http://arxiv.org/abs/1805.03417}{{\ttfamily arXiv:1805.03417 [hep-ph]}}.
%%CITATION = ARXIV:1805.03417;%%.

\bibitem{Hesari:2018ssq}
H.~Hesari, H.~Khanpour, and M.~Mohammadi~Najafabadi, {Study of Higgs Effective
  Couplings at Electron-Proton Colliders},
  \href{http://dx.doi.org/10.1103/PhysRevD.97.095041}{{\em Phys. Rev.}
  {\bfseries D97} no.~9, (2018) 095041},
\href{http://arxiv.org/abs/1805.04697}{{\ttfamily arXiv:1805.04697 [hep-ph]}}.
%%CITATION = ARXIV:1805.04697;%%.

\bibitem{Teyssier:2014hta}
{\bfseries ATLAS, CMS} Collaboration, D.~Teyssier, {LHC results and prospects:
  Beyond Standard Model}, in {\em {International Workshop on Future Linear
  Colliders (LCWS13) Tokyo, Japan, November 11-15, 2013}}.
\newblock 2014.
\newblock
\href{http://arxiv.org/abs/1404.7311}{{\ttfamily arXiv:1404.7311 [hep-ex]}}.
\newblock
%%CITATION = ARXIV:1404.7311;%%.

\bibitem{ATLAS:2018otd}
{\bfseries ATLAS} Collaboration, {Combination of searches for Higgs boson pairs
  in $pp$ collisions at 13 TeV with the ATLAS experiment.}, {\em
  \href{https://atlas.web.cern.ch/Atlas/GROUPS/PHYSICS/CONFNOTES/ATLAS-CONF-2018-043}{https://atlas.web.cern.ch/Atlas/GROUPS/PHYSICS/CONFNOTES/ATLAS-CONF-2018-043}}
  (2018)
ATLAS-CONF-2018-043.
%%CITATION = ATLAS-CONF-2018-043;%%.

\bibitem{Khachatryan:2016vau}
{\bfseries ATLAS, CMS} Collaboration, G.~Aad {\em et~al.}, {Measurements of the
  Higgs boson production and decay rates and constraints on its couplings from
  a combined ATLAS and CMS analysis of the LHC pp collision data at $
  \sqrt{s}=7 $ and 8 TeV},
  \href{http://dx.doi.org/10.1007/JHEP08(2016)045}{{\em JHEP} {\bfseries 08}
  (2016) 045} CERN-EP-2016-100, ATLAS-HIGG-2015-07, CMS-HIG-15-002,
\href{http://arxiv.org/abs/1606.02266}{{\ttfamily arXiv:1606.02266 [hep-ex]}}.
%%CITATION = ARXIV:1606.02266;%%.

\bibitem{CMS:2013xfa}
{\bfseries CMS} Collaboration, {Projected Performance of an Upgraded CMS
  Detector at the LHC and HL-LHC: Contribution to the Snowmass Process}, in
  {\em {Proceedings, 2013 Community Summer Study on the Future of U.S. Particle
  Physics: Snowmass on the Mississippi (CSS2013): Minneapolis, MN, USA, July
  29-August 6, 2013}}.
\newblock 2013.
\newblock \href{http://arxiv.org/abs/1307.7135}{{\ttfamily arXiv:1307.7135
  [hep-ex]}}.
\newblock
\url{http://www.slac.stanford.edu/econf/C1307292/docs/submittedArxivFiles/1307.7135.pdf}.
\newblock
%%CITATION = ARXIV:1307.7135;%%.

\bibitem{DeRujula:1991ufe}
A.~De~Rujula, M.~B. Gavela, P.~Hernandez, and E.~Masso, {The Selfcouplings of
  vector bosons: Does LEP-1 obviate LEP-2?},
  \href{http://dx.doi.org/10.1016/0550-3213(92)90460-S}{{\em Nucl. Phys.}
  {\bfseries B384} (1992) 3--58}
CERN-TH-6272-91, FTUAM-91-31.
%%CITATION = NUPHA,B384,3;%%.

\bibitem{GutierrezRodriguez:2008nk}
A.~Gutierrez-Rodriguez, M.~A. Hernandez-Ruiz, O.~A. Sampayo, A.~Chubykalo, and
  A.~Espinoza-Garrido, {The Triple Higgs Boson Self-Coupling at Future Linear
  $e^+e^-$ Colliders Energies: ILC and CLIC},
  \href{http://dx.doi.org/10.1143/JPSJ.77.094101}{{\em J. Phys. Soc. Jap.}
  {\bfseries 77} (2008) 094101},
\href{http://arxiv.org/abs/0807.0663}{{\ttfamily arXiv:0807.0663 [hep-ph]}}.
%%CITATION = ARXIV:0807.0663;%%.

\bibitem{Takubo:2009ws}
Y.~Takubo, {Analysis of Higgs Self-coupling with ZHH at ILC}, in {\em {8th
  General Meeting of the ILC Physics Subgroup Tsukuba, Japan, January 21,
  2009}}.
\newblock 2009.
\newblock
\href{http://arxiv.org/abs/0907.0524}{{\ttfamily arXiv:0907.0524 [hep-ph]}}.
\newblock
%%CITATION = ARXIV:0907.0524;%%.

\bibitem{Tian:2010np}
J.~Tian, K.~Fujii, and Y.~Gao, {Study of Higgs Self-coupling at ILC},
\href{http://arxiv.org/abs/1008.0921}{{\ttfamily arXiv:1008.0921 [hep-ex]}}.
%%CITATION = ARXIV:1008.0921;%%.

\bibitem{Battaglia:2001nn}
M.~Battaglia, E.~Boos, and W.-M. Yao, {Studying the Higgs potential at the e+
  e- linear collider}, {\em eConf} {\bfseries C010630} (2001) E3016
  SNOWMASS-2001-E3016,
\href{http://arxiv.org/abs/hep-ph/0111276}{{\ttfamily arXiv:hep-ph/0111276
  [hep-ph]}}.
%%CITATION = HEP-PH/0111276;%%.

\bibitem{Killick:2013mya}
R.~Killick, K.~Kumar, and H.~E. Logan, {Learning what the Higgs boson is mixed
  with}, \href{http://dx.doi.org/10.1103/PhysRevD.88.033015}{{\em Phys. Rev.}
  {\bfseries D88} (2013) 033015},
\href{http://arxiv.org/abs/1305.7236}{{\ttfamily arXiv:1305.7236 [hep-ph]}}.
%%CITATION = ARXIV:1305.7236;%%.

\bibitem{Djouadi:1999gv}
A.~Djouadi, W.~Kilian, M.~Muhlleitner, and P.~M. Zerwas, {Testing Higgs
  selfcouplings at $e^+ e^-$ linear colliders},
  \href{http://dx.doi.org/10.1007/s100529900082}{{\em Eur. Phys. J.} {\bfseries
  C10} (1999) 27--43} DESY-99-001, TTP-99-02, PM-99-01,
\href{http://arxiv.org/abs/hep-ph/9903229}{{\ttfamily arXiv:hep-ph/9903229
  [hep-ph]}}.
%%CITATION = HEP-PH/9903229;%%.

\bibitem{Baer:2013cma}
H.~Baer, T.~Barklow, K.~Fujii, Y.~Gao, A.~Hoang, S.~Kanemura, J.~List, H.~E.
  Logan, A.~Nomerotski, M.~Perelstein, {\em et~al.}, {The International Linear
  Collider Technical Design Report - Volume 2: Physics}, ILC-REPORT-2013-040,
  ANL-HEP-TR-13-20, BNL-100603-2013-IR, IRFU-13-59, CERN-ATS-2013-037,
  COCKCROFT-13-10, CLNS-13-2085, DESY-13-062, FERMILAB-TM-2554,
  IHEP-AC-ILC-2013-001, INFN-13-04-LNF, JAI-2013-001, JINR-E9-2013-35,
  JLAB-R-2013-01, KEK-REPORT-2013-1, KNU-CHEP-ILC-2013-1, LLNL-TR-635539,
  SLAC-R-1004, ILC-HIGRADE-REPORT-2013-003,
\href{http://arxiv.org/abs/1306.6352}{{\ttfamily arXiv:1306.6352 [hep-ph]}}.
%%CITATION = ARXIV:1306.6352;%%.

\bibitem{Castanier:2001sf}
C.~Castanier, P.~Gay, P.~Lutz, and J.~Orloff, {Higgs self coupling measurement
  in $e^+ e^-$ collisions at center-of-mass energy of 500-GeV},
  LC-PHSM-2000-061,
\href{http://arxiv.org/abs/hep-ex/0101028}{{\ttfamily arXiv:hep-ex/0101028
  [hep-ex]}}.
%%CITATION = HEP-EX/0101028;%%.

\bibitem{Alloul:2013naa}
A.~Alloul, B.~Fuks, and V.~Sanz, {Phenomenology of the Higgs Effective
  Lagrangian via FEYNRULES},
  \href{http://dx.doi.org/10.1007/JHEP04(2014)110}{{\em JHEP} {\bfseries 04}
  (2014) 110} CERN-PH-TH-2013-248,
\href{http://arxiv.org/abs/1310.5150}{{\ttfamily arXiv:1310.5150 [hep-ph]}}.
%%CITATION = ARXIV:1310.5150;%%.

\bibitem{Alwall:2014hca}
J.~Alwall, R.~Frederix, S.~Frixione, V.~Hirschi, F.~Maltoni, O.~Mattelaer,
  H.~S. Shao, T.~Stelzer, P.~Torrielli, and M.~Zaro, {The automated computation
  of tree-level and next-to-leading order differential cross sections, and
  their matching to parton shower simulations},
  \href{http://dx.doi.org/10.1007/JHEP07(2014)079}{{\em JHEP} {\bfseries 07}
  (2014) 079} CERN-PH-TH-2014-064, CP3-14-18, LPN14-066, MCNET-14-09,
  ZU-TH-14-14,
\href{http://arxiv.org/abs/1405.0301}{{\ttfamily arXiv:1405.0301 [hep-ph]}}.
%%CITATION = ARXIV:1405.0301;%%.

\bibitem{Alloul:2013bka}
A.~Alloul, N.~D. Christensen, C.~Degrande, C.~Duhr, and B.~Fuks, {FeynRules 2.0
  - A complete toolbox for tree-level phenomenology},
  \href{http://dx.doi.org/10.1016/j.cpc.2014.04.012}{{\em Comput. Phys.
  Commun.} {\bfseries 185} (2014) 2250--2300} CERN-PH-TH-2013-239, MCNET-13-14,
  IPPP-13-71, DCPT-13-142, PITT-PACC-1308,
\href{http://arxiv.org/abs/1310.1921}{{\ttfamily arXiv:1310.1921 [hep-ph]}}.
%%CITATION = ARXIV:1310.1921;%%.

\bibitem{Antony:GetDist}
A.~Lewis, {GetDist: Kernel Density Estimation,
  \href{http://cosmologist.info/notes/GetDist.pdf}{url::.http://cosmologist.info/notes/GetDist.pdf}},
  {\em Homepage
  \href{http://getdist.readthedocs.org/en/latest/index.html}{http://getdist.readthedocs.org/en/latest/index.html}}
  .

\bibitem{Behnke:2013xla}
T.~Behnke, J.~E. Brau, B.~Foster, J.~Fuster, M.~Harrison, J.~M. Paterson,
  M.~Peskin, M.~Stanitzki, N.~Walker, and H.~Yamamoto, {The International
  Linear Collider Technical Design Report - Volume 1: Executive Summary},
  ILC-REPORT-2013-040, ANL-HEP-TR-13-20, BNL-100603-2013-IR, IRFU-13-59,
  CERN-ATS-2013-037, COCKCROFT-13-10, CLNS-13-2085, DESY-13-062,
  FERMILAB-TM-2554, IHEP-AC-ILC-2013-001, INFN-13-04-LNF, JAI-2013-001,
  JINR-E9-2013-35, JLAB-R-2013-01, KEK-REPORT-2013-1, KNU-CHEP-ILC-2013-1,
  LLNL-TR-635539, SLAC-R-1004, ILC-HIGRADE-REPORT-2013-003,
\href{http://arxiv.org/abs/1306.6327}{{\ttfamily arXiv:1306.6327
  [physics.acc-ph]}}.
%%CITATION = ARXIV:1306.6327;%%.

\end{thebibliography}\endgroup
\bibliographystyle{utphys}

\end{document}